\documentclass[nofootinbib,superscriptaddress,a4paper,twocolumn,floatfix,longbibliography]{revtex4-2}
\usepackage{array}[=2016-10-06]
\usepackage[english]{babel}
\usepackage[utf8]{inputenc}

\usepackage{amsthm}
\usepackage{amssymb}
\usepackage{mathtools}

\usepackage{subfigure}
\usepackage{siunitx}
\usepackage[dvipsnames]{xcolor}
\usepackage[ruled,vlined]{algorithm2e}
\usepackage{amsmath, amssymb} 
\usepackage{makecell}

\usepackage{enumitem}

\usepackage{tikz}
\usetikzlibrary{calc}
\usetikzlibrary{automata,positioning}
\usetikzlibrary{arrows.meta, positioning, shapes, fit}
\usetikzlibrary{backgrounds}

\definecolor{unia}{HTML}{3C2673}
\definecolor{unia2}{HTML}{8c358b}
\usepackage[pdftex, pdftitle={PauliEngine 1.0: A High-Performance Backend for Quantum Operator Algebra}, 
pdfauthor={Leon Müller, Adelina Bärligea, Alexander Knapp, Jakob S. Kottmann}]{hyperref}
\hypersetup{
    colorlinks,
    linkcolor={unia},
    citecolor={blue!50!black},
    urlcolor={unia2}
}

\usepackage{booktabs}
\usepackage{multirow}
\usepackage{listings}

\usepackage{orcidlink}
\usepackage[capitalize]{cleveref}
\usepackage{xcolor}
\usepackage{braket}

\newcommand{\revtexseparator}{%
    \par
    \onecolumngrid
    \vspace{1em}%
    \noindent\makebox[\textwidth][c]{%
        \makebox[0pt]{\rule[-0.25pt]{245.5pt}{0.5pt}}%
        \makebox[0pt]{\rule[-0.375pt]{184.125pt}{0.75pt}}%
        \makebox[0pt]{\rule[-0.5pt]{122.75pt}{1pt}}%
        \makebox[0pt]{\rule[-0.625pt]{61.375pt}{1.25pt}}%
    }\par
    \vspace{0.5em}%
    \twocolumngrid
}

\usepackage{xcolor}
\usepackage[most]{tcolorbox}

\definecolor{ExampleBG}{HTML}{EAF3FC}
\definecolor{SoftwareBG}{HTML}{E8F6F3}
\definecolor{InsightBG}{HTML}{F1EDFA}

\definecolor{ExampleAccent}{HTML}{5B8FC4}
\definecolor{SoftwareAccent}{HTML}{4A9B91}
\definecolor{InsightAccent}{HTML}{8068B3}

\definecolor{part0}{RGB}{127,119,221}
\definecolor{part1}{RGB}{216,90,48}
\definecolor{part00}{RGB}{29,158,117}
\definecolor{part01}{RGB}{212,83,126}

\definecolor{pX}{RGB}{29,158,117}
\definecolor{pY}{RGB}{212,83,126}
\definecolor{pZ}{RGB}{127,119,221}

\tcbset{
    PauliEngineBox/.style={
        enhanced,
        breakable,
        coltitle=white,
        fonttitle=\normalfont\normalsize,
        fontupper=\small,
        halign title=flush left,
        arc=1.5mm,
        boxrule=0.4mm,
        boxsep=1mm,
        left=3mm,
        right=3mm,
        top=2.5mm,
        bottom=2.5mm,
        lefttitle=3mm,
        righttitle=3mm,
        toptitle=0mm,
        bottomtitle=0mm,
        before skip=3mm,
        after skip=3mm
    }
}

\newtcolorbox[auto counter]{ExampleBox}[2][]{%
    PauliEngineBox,
    colback=ExampleBG,
    colframe=ExampleAccent,
    colbacktitle=ExampleAccent,
    title={Example~\thetcbcounter: #2},
    #1
}

\usepackage{listings}
\usepackage{upquote}
\tcbuselibrary{skins,breakable,listings}

\providecolor{SoftwareKeyword}{HTML}{235A9F}
\providecolor{SoftwareString}{HTML}{825180}
\providecolor{SoftwareMuted}{gray}{0.45}

\tcbset{
    PauliEngineSoftware/.style={
        PauliEngineBox,
        colback=SoftwareBG,
        colframe=SoftwareAccent,
        colbacktitle=SoftwareAccent
    }
}

\newtcolorbox[auto counter]{SoftwareBox}[2][]{%
    PauliEngineSoftware,
    title={Software~\thetcbcounter: #2},
    #1
}

\lstdefinestyle{paulienginepython}{
    language=Python,
    basicstyle=\ttfamily\footnotesize,
    keywordstyle=\color{SoftwareKeyword},
    stringstyle=\color{SoftwareString},
    commentstyle=\color{SoftwareMuted},
    numbers=none,
    frame=none,
    xleftmargin=0pt,
    xrightmargin=0pt,
    columns=fullflexible,
    keepspaces=true,
    tabsize=4,
    showstringspaces=false,
    showspaces=false,
    showtabs=false,
    upquote=false,
    mathescape=false,
    texcl=false,
    breaklines=true,
    breakatwhitespace=false,
    breakautoindent=true,
    breakindent=1em,
    postbreak=\mbox{\textcolor{SoftwareMuted}{$\hookrightarrow$}\space}
}

\newtcblisting[use counter from=SoftwareBox]{SoftwareCodeBox}[2][]{%
    PauliEngineSoftware,
    title={Software~\thetcbcounter: #2},
    listing only,
    listing engine=listings,
    listing options={
        style=paulienginepython,
        basicstyle=\ttfamily\footnotesize,
        numbers=none,
        backgroundcolor={},
        aboveskip=0pt,
        belowskip=0pt
    },
    #1
}

\newtcolorbox[auto counter]{InsightBox}[2][]{%
    PauliEngineBox,
    colback=InsightBG,
    colframe=InsightAccent,
    colbacktitle=InsightAccent,
    title={Algorithmic Insight~\thetcbcounter: #2},
    #1
}

\makeatletter
\newcommand{\maintoc}{%
  \begingroup
    \let\l@subsubsection\@gobbletwo
    \tableofcontents
  \endgroup}
\newcommand{\stopmaintoc}{%
  \addtocontents{toc}{\protect\let\protect\contentsline\protect\@gobblefour}}
\newcommand{\appendixtoc}{%
  \let\apx@addcontentsline\addcontentsline
  \def\addcontentsline##1##2##3{%
    \apx@addcontentsline{##1}{##2}{##3}%
    \def\apx@tmp{##1}%
    \ifx\apx@tmp\apx@toc\apx@addcontentsline{apx}{##2}{##3}\fi}%
  \begingroup
    \let\l@subsubsection\@gobbletwo
    \let\toc@post\relax 
    \@starttoc{apx}%
  \endgroup}
\def\apx@toc{toc}
\makeatother

\newcommand{\PE}{\textsf{PauliEngine}}
\newcommand{\g}{\mathfrak{g}}
\def\gsim{\g{\text - }{\rm sim}} 

\begin{document}

\title{PauliEngine 1.0:  A High-Performance Backend for Quantum Operator Algebra}

\author{Leon Müller\,\orcidlink{0009-0002-6754-9324}}
\email{These authors contributed equally.}
\affiliation{{Institute for Computer Science, University of Augsburg, Germany }}

\author{Adelina Bärligea\,\orcidlink{0009-0008-5497-1941}}
\email{These authors contributed equally.}
\affiliation{{Institute for Computer Science, University of Augsburg, Germany }}

\author{Alexander Knapp\,\orcidlink{0000-0002-4050-3249}}
\affiliation{{Institute for Computer Science, University of Augsburg, Germany }}

\author{Jakob~S.~Kottmann\,\orcidlink{0000-0002-4156-2048}}
\email{jakob.kottmann@uni-a.de}
\affiliation{{Institute for Computer Science, University of Augsburg, Germany }}
\affiliation{{Center for Advanced Analytics and Predictive Sciences, University of Augsburg, Germany }}


\begin{abstract}
Efficient classical simulation of quantum systems often relies on exploiting their algebraic structure to identify compact operator representations and perform arithmetic directly within them. We introduce \PE, a high-performance C++ backend with a unified Python interface for numerical and symbolic quantum operator arithmetic. We develop adaptive storage and multiplication algorithms that exploit both the binary structure and sparsity of Pauli operators, together with specialized routines for local commutators and the direct evaluation of individual commutator coefficients. To exploit symmetries throughout operator calculations, we extend these arithmetic capabilities to translation-invariant Pauli cycles and permutation-invariant Pauli orbits, enabling polynomial-time arithmetic between elementary orbit operators even for exponentially large Pauli expansions. Extensive benchmarks demonstrate leading performance across most investigated workloads, with runtime improvements of several orders of magnitude over many existing implementations and particularly pronounced advantages for large operators and registers. We demonstrate how these advances extend the computational reach of Lie-algebraic simulation, observable propagation, symmetric quantum dynamics, and variational quantum chemistry. With its combination of computational efficiency, accessibility, and extensibility, \PE\ establishes a reusable arithmetic foundation for quantum software, enabling researchers to exploit the algebraic structure of quantum systems in the development of increasingly scalable classical simulation methods.
\end{abstract}

\maketitle

\let\savedaddcontentsline\addcontentsline
\renewcommand{\addcontentsline}[3]{}
\section{Introduction}
\label{sec:introduction}
\let\addcontentsline\savedaddcontentsline

The exponential growth of the Hilbert-space dimension with the number of qubits constitutes a fundamental challenge in the classical simulation of quantum systems. A recurring principle underlying efficient classical simulation methods is to exploit the algebraic or physical structure of a system to identify a compact representation of its states or observables, in which the relevant operations can be performed efficiently. An early example of this approach is the stabilizer formalism, which enables the efficient classical simulation of Clifford circuits~\cite{Gottesman.1997-StabilizerCodesQuantum,Gottesman.1998-HeisenbergRepresentationQuantum,Aaronson.2004-ImprovedSimulationStabilizer}. Other examples include fermionic linear optics, where the evolution of fermionic Gaussian states can be described through covariance matrices of polynomial dimension~\cite{Valiant.2002-QuantumCircuitsThat,Knill.2001-FermionicLinearOptics,Terhal.2002-ClassicalSimulationNoninteractingfermion,Bravyi.2005-LagrangianRepresentationFermionic}, and Lie-algebraic classical simulation methods, which describe the evolution of observables through the adjoint representation of a low-dimensional dynamical Lie algebra~\cite{Somma.2006-EfficientSolvabilityHamiltonians,Goh.2025-LiealgebraicClassicalSimulations,Barligea.2026-EnablingLieAlgebraicClassical}. The computational efficiency of these methods relies on both the availability of compact representations and the ability to perform arithmetic directly within them.

Pauli strings play a central role in many such representations. Their algebraic structure allows us to encode an $n$-qubit Pauli operator, whose matrix representation has dimension $2^n\times 2^n$, through two $n$-bit binary vectors and a scalar phase~\cite{Calderbank.1998-QuantumErrorCorrection,Dehaene.2003-CliffordGroupStabilizer}. Products and commutation relations can then be evaluated through elementary bitwise operations. Since Pauli strings form a basis of the space of linear operators on $n$ qubits, this arithmetic extends naturally to arbitrary quantum operators through their Pauli expansions. Whenever these expansions remain sufficiently sparse, we can perform operator calculations without constructing the underlying exponentially large matrices. This principle is exploited, for example, in Pauli propagation methods for the classical simulation of quantum circuits and many-body dynamics~\cite{Rall.2019-SimulationQubitQuantum,Begusic.2023-SimulatingQuantumCircuit,Begusic.2024-FastConvergedClassical,Fontana.2023-ClassicalSimulationsNoisy,Angrisani.2025-ClassicallyEstimatingObservables,Begusic.2025-RealtimeOperatorEvolution,Rudolph.2025-PauliPropagationComputational}, in related bit-encoded algebraic approaches to many-body dynamics~\cite{Broers.2026-ExclusiveorEncodedAlgebraic,Broers.2026-ScalableSimulationQuantum}, and in Majorana propagation for fermionic systems~\cite{Miller.2025-SimulationFermionicCircuits,Facelli.2026-FastConvergenceMajorana}.

The widespread use of Pauli operators has motivated the development of numerous software implementations of Pauli arithmetic, ranging from the operator classes of established quantum software frameworks~\cite{JavadiAbhari.2024-QuantumComputingQiskit,Bergholm.2018-PennyLaneAutomaticDifferentiation,Cirq.2025-Cirq,Sivarajah.2021-TketRetargetableCompiler,Suzuki.2021-QulacsFastVersatile,Kim.2023-CUDAQuantumPlatform,McClean.2020-OpenFermionElectronicStructure,HQS.2026-Struqture,Ralli.2026-Symmer} to specialized libraries for the efficient manipulation of Pauli expansions~\cite{Dion.2024-EfficientlyManipulatingPauli,Loizeau.2025-QuantumManybodySimulations,Krotz.2026-PauLIBHighperformanceLibrary,Li.2026-HighperformancePaulialgebraFramework,Qognitive.2024-FastPauli,Quantinuum.2026-Zixy,Fawzi.2026-FastFermion} and dedicated hardware accelerators~\cite{Le.2026-PACOXFPGAbasedPauli}. Many of the simulation methods outlined above require the same elementary operator calculations, including products, commutators, inner products, and the evaluation of individual expansion coefficients. This motivates the development of a general-purpose computational backend that provides these operations across different operator representations and exploits the particular algebraic structure of each representation.

In this work, we introduce \PE~1.0, an open-source, high-performance computational backend for quantum operator arithmetic, implemented in C++ with a unified Python interface. Starting from the binary representation of Pauli strings, we develop optimized algorithms for arithmetic on sparse linear combinations of Pauli operators with complex and symbolic coefficients. We place particular emphasis on the multiplication of Pauli sums, where the efficient evaluation of individual string products must be complemented by an efficient accumulation of the resulting contributions. To this end, we introduce an adaptive prefix-partitioned multiplication algorithm that groups pairs of input terms according to their possible output signatures, allowing their contributions to be accumulated independently and in parallel. For low-weight Pauli operators on large qubit registers, we introduce a complementary sparse storage format and adaptively select the representation and arithmetic algorithms according to the structure of the operators. We further exploit the sparsity of individual Pauli strings by developing a commutator algorithm that restricts the arithmetic to pairs of strings with overlapping supports, substantially reducing the computational cost of commutators between operators arising from local interactions.

The advantages of exploiting algebraic structure extend beyond the representation of individual Pauli strings. Motivated by recent developments in the classical simulation of symmetric quantum systems~\cite{Anschuetz.2023-EfficientClassicalAlgorithms,Barligea.2026-EnablingLieAlgebraicClassical,Teng.2025-LeveragingSymmetryMerging,Chang.2026-PracticalFrameworkSimulating,Bhandari.2026-ReadoutawareTheoryApproximate}, we incorporate two symmetry-adapted operator representations into \PE: translation-invariant Pauli cycles and permutation-invariant Pauli orbits. Individual elements of these bases can represent entire families of symmetry-equivalent Pauli strings, whose number can grow exponentially with system size in the case of Pauli orbits. Building on the commutator formulas derived in Ref.~\cite{Barligea.2026-EnablingLieAlgebraicClassical}, we develop and implement arithmetic routines that operate directly on these representations. We extend the arithmetic to products and inner products and develop targeted commutator routines for evaluating individual expansion coefficients without constructing the full commutator, reducing the asymptotic computational cost for ordinary Pauli sums and Pauli orbits. The common mathematical and computational structure of our implementations provides a framework for extending \PE\ to further operator bases by specifying their algebraic relations and developing the corresponding arithmetic algorithms.

Through extensive numerical benchmarks, we compare \PE\ against a comprehensive selection of existing software packages providing related arithmetic capabilities, covering a broad range of system sizes, Pauli weights, and operator expansion sizes. \PE\ achieves the shortest runtimes in most of the considered comparisons of Pauli arithmetic, with improvements of several orders of magnitude in many configurations. For symmetry-adapted operators, our benchmarks further demonstrate the predicted improvements in asymptotic scaling and substantial reductions in memory requirements. We illustrate how these advantages translate to application-level calculations, including the construction of dynamical Lie algebras on hundreds of qubits, exact observable propagation through near-Clifford circuits, the simulation of permutation-invariant quantum dynamics, and the screening of excitation pools in variational quantum chemistry. By establishing a common computational foundation for efficient quantum operator arithmetic, we aim to extend the computational reach of existing classical simulation methods and enable the development of new quantum software that exploits the algebraic structure of quantum systems.

The source code, documentation, and usage examples of \PE\ are available at~\url{https://github.com/tequilahub/pauliengine}. In~\cref{sec:framework}, we introduce the mathematical and computational framework underlying the software. We develop the representation-specific arithmetic algorithms and analyze their computational complexities in~\cref{sec:operatorrepresentations}, before describing their software implementation, the Python interface, and the extensibility of the framework in~\cref{sec:software}. In~\cref{sec:benchmarks}, we benchmark our implementations against existing software, and in~\cref{sec:applications}, we discuss their applicability to various classical simulation methods. The appendices provide detailed derivations of the arithmetic algorithms, implementation specifications, numerical validation, supplementary benchmarks, and application-level numerical demonstrations.

\maintoc

\begin{figure*}[htbp]
    \centering
    \includegraphics[width=0.9\linewidth]{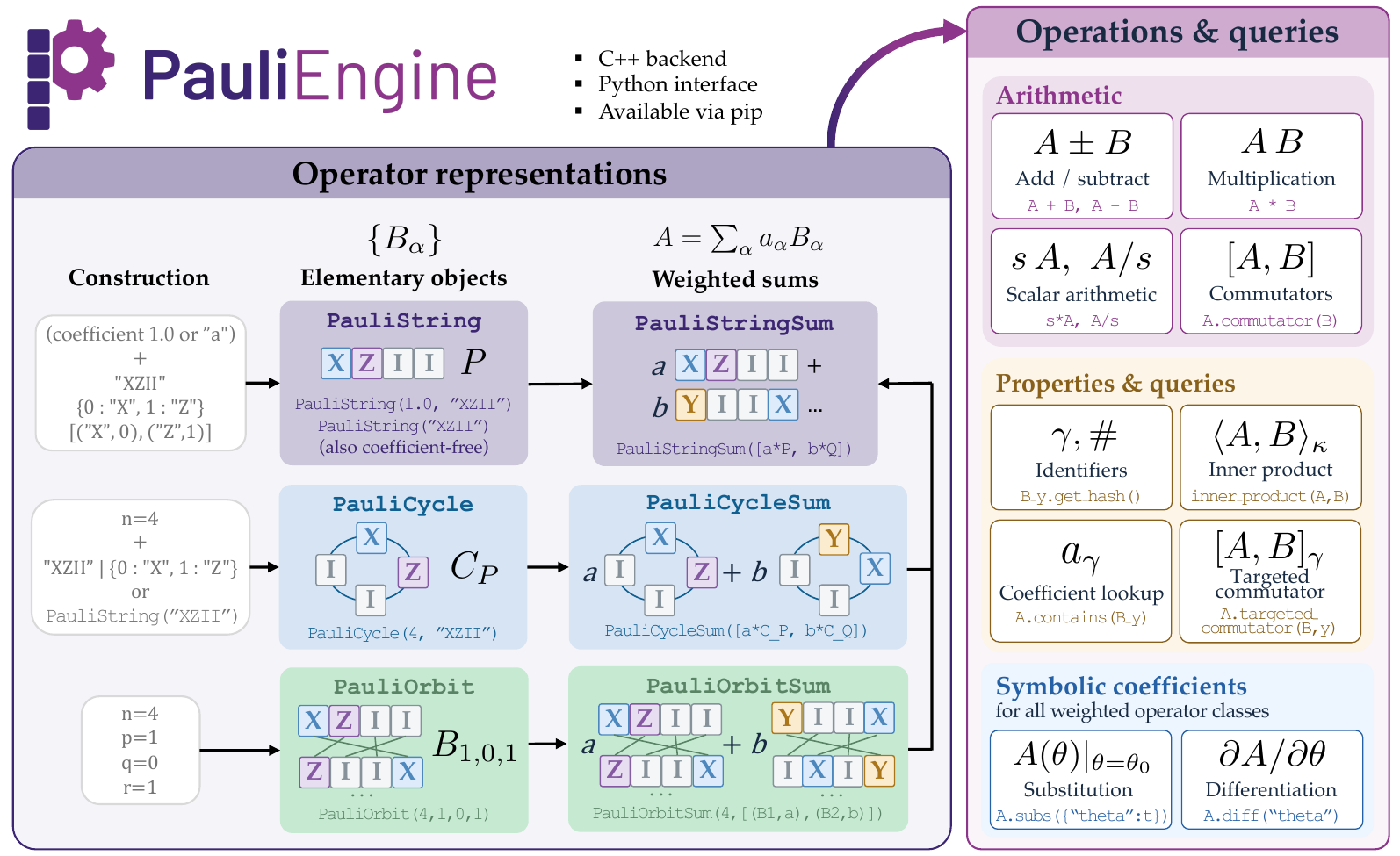}
\caption{Overview of structured operator arithmetic in \PE. The three implemented operator families, Pauli strings, Pauli cycles, and Pauli orbits, exploit different algebraic structures to represent quantum operators through compact elementary labels and weighted linear combinations. Their mathematical definitions and representation-specific arithmetic algorithms are developed in~\cref{sec:operatorrepresentations}. All three families share a common interface for arithmetic, inner products, and coefficient queries, including the direct evaluation of individual commutator coefficients (\cref{sec:shared_operations}). Arithmetic retains the native representation wherever possible, while mixed-family operations are supported through conversion to ordinary Pauli sums, i.e., \texttt{PauliStringSum}s. Numerical and symbolic coefficients can be combined within the same framework. The software architecture, Python interface, symbolic functionality, and integration of further operator families are discussed in~\cref{sec:software}. The constructors follow the software normalization conventions specified in Appendix~\ref{app:software_conventions}.}
    \label{fig:overview}
\end{figure*}

\section{A framework for structured operator arithmetic}
\label{sec:framework}

The operator representations implemented in \PE\ comprise ordinary Pauli strings and two symmetry-adapted operator families motivated by Ref.~\cite{Barligea.2026-EnablingLieAlgebraicClassical}: Pauli cycles for translation-invariant systems and Pauli orbits for permutation-invariant systems. Each family has a corresponding container for weighted linear combinations, as illustrated in Fig.~\ref{fig:overview}. Their elementary labels encode different algebraic structures, which we exploit through the representation-specific arithmetic algorithms developed in~\cref{sec:operatorrepresentations}.

The framework provides a common interface for operator arithmetic and queries, including addition, multiplication, commutators, inner products, coefficient lookup, and targeted commutator evaluation. Arithmetic is performed directly in the native representation whenever possible, avoiding explicit Hilbert-space matrices or expanded Pauli sums. Weighted operators support numerical and symbolic coefficients, while ordinary Pauli strings additionally admit coefficient-free representations for computations involving only operator labels and phases. Arithmetic between different operator families can be evaluated through conversion to the ordinary Pauli basis, although the resulting expansions can become arbitrarily large. Preserving native representations wherever possible is therefore a central design principle of \PE, whose separation of a common arithmetic interface from representation-specific implementations also facilitates the integration of new operator families with their own storage and algorithms~(\cref{sec:software,ssec:Extensibility}).

\subsection{Operators, representations, and linear combinations}

Consider a finite-dimensional Hilbert space $\mathcal H$ and a real vector space of Hermitian operators $\mathcal V\subseteq\operatorname{Herm}(\mathcal H)$. We describe this space through a basis of elementary operators,
\begin{equation}
    \mathcal B=\{B_\alpha:\alpha\in\mathcal I\},
    \qquad
    \mathcal V=\operatorname{span}_{\mathbb R}\mathcal B,
\end{equation}
where $\mathcal I$ denotes the set of basis labels. Allowing complex coefficients extends the represented space to $\mathcal V_{\mathbb C}=\operatorname{span}_{\mathbb C}\mathcal B$, in which every operator admits a unique expansion
\begin{equation}
    A=\sum_{\alpha\in S_A}a_\alpha B_\alpha,
    \qquad a_\alpha\in\mathbb C,
    \label{eq:general_operator_expansion}
\end{equation}
where $S_A\subseteq\mathcal I$ contains the labels with nonzero coefficients. Since the basis elements are Hermitian, the expansion represents a Hermitian operator precisely when all its coefficients are real, i.e., $a_\alpha\in\mathbb R$. 
Pauli strings form such a basis for operators on qubit systems and constitute the standard operator representation in \PE. The computational representation of $\mathcal B$ is determined by how its elementary operators are identified, stored, and manipulated. We encode ordinary Pauli strings through packed binary signatures or sparse site lists (cf.~\cref{ssec:PauliStrings}), identify Pauli cycles through cyclic equivalence classes (cf.~\cref{ssec:PauliCycles}), and characterize Pauli orbits through the multiplicities of their single-qubit Pauli factors (cf.~\cref{ssec:PauliOrbits}). The corresponding sum containers associate these labels with the coefficients of their linear combinations. This allows arithmetic to be carried out without constructing dense operator matrices, while symmetry-adapted labels can additionally avoid large Pauli expansions.

For computations involving only operator labels and commutation relations, \PE\ also provides coefficient-free Pauli strings, i.e., Pauli-group elements $i^kP$ with tracked phase $k\in\{0,1,2,3\}$~\cite{Gottesman.1997-StabilizerCodesQuantum,Nielsen.2016-QuantumComputationQuantum}, and ordered collections of distinct labels. Together, the weighted and coefficient-free representations supply the primitives for operator-based computational approaches to describing quantum systems, including the applications discussed in~\cref{sec:applications}.

\subsection{Generalized operator arithmetic}

Arithmetic on linear combinations follows from the products and coefficient accumulation of their elementary basis operators. When $\mathcal V_{\mathbb C}$ is closed under multiplication, elementary products and commutators take the form
\begin{equation}
\begin{aligned}
    B_\alpha B_\beta
    &=\sum_{\gamma\in\mathcal I}
        m_{\alpha\beta}^{\gamma}B_\gamma,\\
    [B_\alpha,B_\beta]
    &=\sum_{\gamma\in\mathcal I}
        c_{\alpha\beta}^{\gamma}B_\gamma,
\end{aligned}
\label{eq:general_arithmetic}
\end{equation}
where $[A,B]=AB-BA$ and $c_{\alpha\beta}^{\gamma}=m_{\alpha\beta}^{\gamma}-m_{\beta\alpha}^{\gamma}$. For Hermitian bases, we additionally use $f_{\alpha\beta}^{\gamma}=ic_{\alpha\beta}^{\gamma}$ to denote the \emph{structure constants} of the Hermitian bracket $i[B_\alpha,B_\beta]$. If $\mathcal V$ is closed under this bracket, it forms a real Lie algebra, with real structure constants $f_{\alpha\beta}^{\gamma}$~\cite{DAlessandro.2021-IntroductionQuantumControl}.

These rules extend bilinearly to linear combinations. For $A=\sum_\alpha a_\alpha B_\alpha$ and $B=\sum_\beta b_\beta B_\beta$, a coefficient $c_\gamma$, in the expansion of the product $(AB)=\sum_\gamma c_\gamma B_\gamma$ is then
\begin{align}
    \operatorname{coeff}_{\mathcal B}(AB;\gamma)
    &=\sum_{\alpha,\beta}
        a_\alpha b_\beta m_{\alpha\beta}^{\gamma},\label{eq:general_targeted_arithmetic} 
\end{align}
and, analogously, for the commutator
\begin{align}
    \operatorname{coeff}_{\mathcal B}([A,B];\gamma)
    &=\sum_{\alpha,\beta}
        a_\alpha b_\beta c_{\alpha\beta}^{\gamma}.
\end{align}
These expressions distinguish constructing a complete output expansion from evaluating only selected coefficients. Depending on the representation, the latter can avoid generating and accumulating many irrelevant terms. The corresponding algorithms and complexity differences are examined in~\cref{sec:shared_operations}.

The framework also supports operator inner products, defined through the normalized Hilbert--Schmidt product
\begin{equation}
    \langle A,B\rangle_\kappa
    =\kappa\operatorname{Tr}(A^\dagger B),
    \qquad
    \kappa=\frac{1}{d_{\mathcal H}},
    \label{eq:general_inner_product}
\end{equation}
where $d_{\mathcal H}=\dim_{\mathbb C}\mathcal H$. Together with coefficient lookup, inner products provide access to individual operator coordinates without necessarily constructing a full arithmetic result; for orthogonal bases, the coordinate is the normalized overlap given in~\cref{eq:coefficient_extraction}. When the represented space is not closed under a requested operation, the output is expressed in a suitable larger space. For qubit operators, \PE\ provides this fallback through the ordinary Pauli basis.

\subsection{Computational model and notation}\label{sec:comp-model-and-notation}

The computational cost of operator arithmetic depends on both the representation of elementary operators and the number of terms in their linear combinations as in Eq.~\eqref{eq:general_operator_expansion}.
For input operators $A$ and $B$, we denote by $M_A=|S_A|$ and $M_B=|S_B|$ the numbers of nonzero input terms, by $K_{\mathrm{raw}}$ the number of generated nonzero terms before accumulation, and by $K_{\mathrm{out}}$ the number of distinct nonzero output terms remaining after accumulation. For addition or subtraction, we have $K_{\mathrm{out}}\leq M_A+M_B$. Products and commutators may generate several output terms from each pair of elementary operators, so their output sizes depend on the multiplication rules of the chosen representation.

For binary Pauli signatures, we use $w$ to denote the width of a machine word, a fixed-size group of bits processed together by the processor, typically $w=64$. Each binary signature vector is stored in
$ L_{\mathrm{word}}
    =\left\lceil\frac{n}{w}\right\rceil$
machine words for an $n$-qubit operator. Unless stated otherwise, complexity estimates count machine-word operations, as in the word-RAM model~\cite{Fredman.1993-SurpassingInformationTheoretic,Hagerup.1998-SortingSearchingWord}, and assume constant-time arithmetic on fixed-width numerical coefficients. Symbolic coefficient manipulation, combinatorial precomputation, and output accumulation can introduce additional costs and are distinguished where relevant. Hash-based algorithms are analyzed under the assumption of expected constant-time table access (cf.~Ref.~\cite[Ch.~11]{Cormen.2022-IntroductionAlgorithms}), with the construction and processing of operator labels accounted for explicitly.

These conventions allow us to compare the cost of elementary arithmetic, full operator products, and targeted coefficient evaluation across the representations developed in~\cref{sec:operatorrepresentations}.

\section{Operator representations and arithmetic}
\label{sec:operatorrepresentations}

\subsection{Pauli strings and sparse linear combinations}
\label{ssec:PauliStrings}

Pauli strings form a natural operator basis for quantum computation. On a single qubit, the Pauli matrices are
\begin{equation}
    X = \begin{pmatrix}0 & 1 \\ 1 & 0\end{pmatrix}, \quad
    Y = \begin{pmatrix}0 & -i \\ i & 0\end{pmatrix}, \quad
    Z = \begin{pmatrix}1 & 0 \\ 0 & -1\end{pmatrix}.
    \label{eq:Pmatrices}
\end{equation}
Together with the identity $\sigma_0=I$, and writing $\sigma_1=X$, $\sigma_2=Y$, $\sigma_3=Z$, they define the set of $n$-qubit Pauli strings
\begin{equation}
    \mathcal P_n
    = \left\{
        P_{\boldsymbol\alpha}
        = \bigotimes_{j=1}^{n}\sigma_{\alpha_j}
        \;\middle|\;
        \boldsymbol\alpha\in\{0,1,2,3\}^n
      \right\}.
    \label{eq:Paulistring_def}
\end{equation}
These $4^n$ operators form an orthonormal basis of the complex operator space $\operatorname{End}(\mathcal H_n)$, containing the maps from $\mathcal H_n=(\mathbb C^2)^{\otimes n}$ to itself, under the normalized Hilbert--Schmidt inner product $\langle A,B\rangle_\kappa$ of~\cref{eq:general_inner_product} with $\kappa=2^{-n}$. Consequently, every operator $A\in\operatorname{End}(\mathcal H_n)$ admits a unique expansion
\begin{equation}
    A=\sum_{P\in\mathcal P_n}a_P P,
    \qquad
    a_P=2^{-n}\operatorname{Tr}(PA).
\label{eq:PSlinercomb}
\end{equation}
In particular, $A$ is Hermitian precisely when all coefficients $a_P$ are real. For operators with only a small number of nonzero coefficients, the expansion can be stored sparsely by associating each Pauli string with its coefficient.

The computational efficiency of this approach depends on how the individual Pauli strings are represented. In \PE, we provide two complementary storage formats for weighted Pauli strings, exploiting either their binary algebraic structure or the sparsity of their nonidentity factors. The first is based on the \emph{binary symplectic representation}, which originates from the stabilizer formalism~\cite{Gottesman.1997-StabilizerCodesQuantum,Calderbank.1998-QuantumErrorCorrection,Aaronson.2004-ImprovedSimulationStabilizer} and underlies most high-performance Pauli-string software~\cite{Gidney.2021-StimFastStabilizer,JavadiAbhari.2024-QuantumComputingQiskit,Dion.2024-EfficientlyManipulatingPauli,Loizeau.2025-QuantumManybodySimulations,Krotz.2026-PauLIBHighperformanceLibrary,Li.2026-HighperformancePaulialgebraFramework}.
In it, each Pauli string $P\in\mathcal P_n$ is identified by two binary vectors\footnote{This differs from the conventional $X/Z$ signature $(\boldsymbol\xi,\boldsymbol\zeta)$~\cite{Aaronson.2004-ImprovedSimulationStabilizer,Dehaene.2003-CliffordGroupStabilizer}, in which $Y$ carries both bits instead of $Z$. The two are related by the invertible linear map $(\boldsymbol\xi,\boldsymbol\zeta)=(\boldsymbol x\oplus\boldsymbol y,\boldsymbol y)$, which preserves both the XOR product rule and the symplectic parity introduced below.},
\begin{equation}
    v(P)=(\boldsymbol x,\boldsymbol y)
    \in\mathbb F_2^{2n},
    \label{eq:binsymprep}
\end{equation}
whose entries encode the local Pauli factors according to
\begin{equation}
    (x_j,y_j)=
    \begin{cases}
        (0,0), & \sigma_{\alpha_j}=I,\\
        (1,0), & \sigma_{\alpha_j}=X,\\
        (0,1), & \sigma_{\alpha_j}=Y,\\
        (1,1), & \sigma_{\alpha_j}=Z.
    \end{cases}
\end{equation}
The encoding is invertible and requires only $2n$ bits to represent an unweighted Pauli string, excluding machine-word alignment and storage overhead. By comparison, its explicit matrix representation contains $4^n$ entries ($2^n$ are nonzero). Complex coefficients, including global phases, are stored separately from the binary signature. This compact representation enables Pauli string arithmetic to be performed directly through bitwise operations.

For two strings $P,Q\in\mathcal P_n$ with signatures $v(P)=(\boldsymbol x,\boldsymbol y)$ and $v(Q)=(\boldsymbol x',\boldsymbol y')$, their product is
\begin{equation}
    PQ=\omega(P,Q)R,
    \qquad
    v(R)=v(P)\oplus v(Q),
    \label{eq:pauli_multiplication}
\end{equation}
where $\oplus$ denotes bitwise XOR and $\omega(P,Q)\in\{1,-1,i,-i\}$ is the phase arising from the local Pauli multiplication rules~\cite{Dehaene.2003-CliffordGroupStabilizer,Aaronson.2004-ImprovedSimulationStabilizer,Gargiulo.2026-PauliStringsQuantum}. 
This phase is determined by the numbers $\nu_+$ of single-qubit products $XY$, $YZ$, and $ZX$ contributing factors of $+i$, and the number $\nu_-$ of products $YX$, $ZY$, and $XZ$ contributing a factor of $-i$,
\begin{equation}
    \omega(P,Q)=i^{\nu_+(P,Q)-\nu_-(P,Q)}.
    \label{eq:pauli_phase}
\end{equation}
Our phase reconstruction algorithm, adapted from Stim~\cite{Gidney.2021-StimFastStabilizer}, requires only two population counts across the accumulated bitwise counters. The corresponding update rules and phase reconstruction are derived in Appendix~\ref{app:pauli_phase_reconstruction}.

The same representation allows commutators to be evaluated without forming both products. Defining the binary symplectic parity
\begin{equation}
    \chi(P,Q)
    =\boldsymbol x\cdot\boldsymbol y'
    +\boldsymbol y\cdot\boldsymbol x'
    \pmod 2,
    \label{eq:pauli_symplectic_parity}
\end{equation}
the commutation relation and commutator follow as~\cite{Dehaene.2003-CliffordGroupStabilizer,Gargiulo.2026-PauliStringsQuantum}
\begin{equation}
\begin{aligned}
    PQ&=(-1)^{\chi(P,Q)}QP,
    \\
    [P,Q]&=
    \begin{cases}
        0, & \chi(P,Q)=0,\\
        2\omega(P,Q)R, & \chi(P,Q)=1.
    \end{cases}
\end{aligned}
    \label{eq:pauli_commutator}
\end{equation}
The parity can be evaluated directly from the binary signatures. In \PE, the commutator algorithm first tests this parity and returns zero for commuting strings; otherwise, it evaluates a single Pauli multiplication. For example, $P=XI$ and $Q=YZ$ yield $PQ=iZZ$ and $[P,Q]=2iZZ$. The corresponding derivation and implementation details are given in Appendix~\ref{app:pauli_commutation}.
As each of the two $n$-bit vectors $\boldsymbol x$ and $\boldsymbol y$ in \cref{eq:binsymprep} is stored in $L_{\mathrm{word}}=\lceil n/w\rceil$ machine words, both multiplication~\eqref{eq:pauli_multiplication} and commutator evaluation~\eqref{eq:pauli_commutator} therefore require $\mathcal O(L_{\mathrm{word}})$ word operations, assuming constant-time bitwise instructions and population counts. 
Their computational cost thus scales linearly with the number of stored words, independently of the exponentially large Hilbert-space dimension.

For Pauli strings acting nontrivially on only a small fraction of the qubits, we can further reduce the storage and arithmetic costs by retaining only their nonidentity factors. Let
\begin{equation}
    \operatorname{supp}(P)=\{j:P_j\neq I\},
    \qquad
    \operatorname{wt}(P)=|\operatorname{supp}(P)|
    \label{eq:pauli_support_weight}
\end{equation}
denote the support and Pauli weight of a string $P$, respectively. We represent such strings through ordered lists of occupied sites and their corresponding Pauli factors, encoding each site index and local Pauli label in a single integer.

The product of two strings in this representation can be obtained by simultaneously traversing their ordered lists. Factors acting on distinct sites are retained, while factors acting on the same site are combined according to the local Pauli multiplication rules, accumulating their contributions to the global phase. The corresponding multiplication and commutation algorithms require
\begin{equation}
    \mathcal O\bigl(1+\operatorname{wt}(P)+\operatorname{wt}(Q)\bigr)
    \label{eq:sparse_pauli_elementary_complexity}
\end{equation}
operations, independently of the separation of the occupied sites. Elementary arithmetic between strings of bounded Pauli weight thus has constant computational cost in this representation. We detail the storage format and arithmetic algorithms in Appendix~\ref{app:sparse_pauli_arithmetic}.

We represent sparse linear combinations of Pauli strings, such as~\cref{eq:PSlinercomb}, by the \texttt{PauliStringSum} class, which stores distinct Pauli labels and their numerical or symbolic coefficients. The labels can be stored either as packed binary signatures or as lists of their nonidentity factors.
For two operators $A=\sum_{P\in S_A}a_PP$ and $B=\sum_{Q\in S_B}b_QQ$, multiplication and commutation follow from
\begin{equation}
\begin{aligned}
    AB&=\sum_{P\in S_A}\sum_{Q\in S_B}a_P\,b_Q\,PQ,\\
    [A,B]&=\sum_{P\in S_A}\sum_{Q\in S_B}a_P\,b_Q\,[P,Q],
\end{aligned}
    \label{eq:pauli_sum_arithmetic}
\end{equation}
where $S_A$ and $S_B$ contain the nonzero Pauli terms of the respective expansions. The computational cost of these operations arises both from evaluating individual Pauli products and from accumulating contributions with identical output labels. We employ different accumulation algorithms for the two storage formats and further exploit the structure of the input operators and the requested output to reduce the number of elementary operations.

\subsubsection{Prefix-partitioned arithmetic.}

For Pauli sums stored through packed binary signatures, we implement an adaptive output-prefix partitioning algorithm for multiplying Pauli sums with complex coefficients. The underlying observation is that contributions to the same output Pauli string can be identified through their binary signatures before constructing the corresponding products. Fixing an order of the signature bits, let $h_d(P)$ denote the first $d$ coordinates that vary within at least one input sum. For $PQ=\omega(P,Q)R$, the XOR rule of \cref{eq:pauli_multiplication} gives
\begin{equation}
    h_d(R)=h_d(P)\oplus h_d(Q).
    \label{eq:pauli_prefix_routing}
\end{equation}
The XOR relation allows us to partition the product space without explicitly evaluating the individual Pauli products. For inputs sorted by signature, pairs of input ranges with fixed prefixes are grouped according to their common output prefix. We recursively refine these partitions according to their computational workload, thereby controlling the number of contributions that need to be accumulated simultaneously.

Since all contributions to a given Pauli string belong to the same partition, coefficient accumulation can be performed independently and in parallel. Within each partition, we either sort and merge compact product keys or accumulate coefficients directly in a table, depending on the remaining key space. Full Pauli signatures are reconstructed only for the surviving terms, and the locally sorted results are concatenated to obtain the final operator without a global merge, as illustrated in~\cref{fig:prefix-partitioned}. While related adaptive $\mathbb F_2$-linear partitioning has been developed for Pauli propagation through local gates~\cite{orfi2026scalable}, our approach exploits this structure for the multiplication of arbitrary sparse Pauli sums.

\begin{figure}[t]
  \centering
  \begingroup
  \definecolor{part0}{HTML}{4C88BD}
  \definecolor{part00}{HTML}{4C88BD}
  \definecolor{part01}{HTML}{2D9688}
  \definecolor{part1}{HTML}{7B5CB2}
  \begin{tikzpicture}[
  show background rectangle,
    background rectangle/.style={
        fill=black!3,
        rounded corners=5pt
    },
      font=\footnotesize,
      cell/.style={draw=#1!75!black, fill=#1!25, line width=0.4pt, rounded corners=1pt},
      lab/.style={font=\scriptsize, text=black!65},
      stage/.style={font=\footnotesize\bfseries, align=center, anchor=south},
      arr/.style={-{Stealth[length=1.8mm]}, line width=0.5pt, black!60},
    ]
    \node[stage] at (1.0, 0.45) {1.\ Partition};
    \foreach \c/\lbl in {0/0, 1/1} \node[lab] at (\c*1.0+0.5, 0.22) {$B_{\lbl}$};
    \foreach \r/\lbl in {0/0, 1/1} \node[lab] at (-0.28, -\r*1.0-0.5) {$A_{\lbl}$};
    \foreach \r/\c/\p in {0/0/0, 0/1/1, 1/0/1, 1/1/0} {
      \draw[cell=part\p] (\c*1.0+0.03, -\r*1.0-0.03) rectangle (\c*1.0+0.97, -\r*1.0-0.97);
      \node at (\c*1.0+0.5, -\r*1.0-0.5) {\p};
    }
    \draw[arr] (2.08, -1.0) -- (2.42, -1.0);

    \begin{scope}[xshift=2.95cm]
      \node[stage] at (1.0, 0.45) {2.\ Refine};
      \foreach \c/\lbl in {0/00, 1/01, 2/10, 3/11} \node[lab] at (\c*0.5+0.25, 0.22) {$B_{\lbl}$};
      \foreach \r/\lbl in {0/00, 1/01, 2/10, 3/11} \node[lab] at (-0.28, -\r*0.5-0.25) {$A_{\lbl}$};
      \foreach \r/\c/\p in {0/0/00, 0/1/01, 1/0/01, 1/1/00, 2/2/00, 2/3/01, 3/2/01, 3/3/00} {
        \draw[cell=part\p] (\c*0.5+0.025, -\r*0.5-0.025) rectangle (\c*0.5+0.475, -\r*0.5-0.475);
        \node[font=\scriptsize] at (\c*0.5+0.25, -\r*0.5-0.25) {\p};
      }
      \foreach \r/\c in {0/2, 2/0} {
        \draw[cell=part1] (\c*0.5+0.025, -\r*0.5-0.025) rectangle (\c*0.5+0.975, -\r*0.5-0.975);
        \node at (\c*0.5+0.5, -\r*0.5-0.5) {1};
      }
    \end{scope}
    \draw[arr] (5.03, -1.0) -- (5.37, -1.0);

    \begin{scope}[xshift=5.45cm]
      \node[stage] at (0.8, 0.45) {3.\ Accumulate};
      \draw[cell=part00] (0, -0.025) rectangle (1.6, -0.475);
      \node at (0.8, -0.25) {prefix 00};
      \draw[cell=part01] (0, -0.525) rectangle (1.6, -0.975);
      \node at (0.8, -0.75) {prefix 01};
      \draw[cell=part1] (0, -1.025) rectangle (1.6, -1.975);
      \node at (0.8, -1.5) {prefix 1};
    \end{scope}

    \node[lab, anchor=west] at (-0.05, -2.4) {result array:};
    \draw[cell=part00] (0.00, -2.58) rectangle (1.74, -3.06);
    \node[font=\scriptsize] at (0.87, -2.82) {terms 00\ldots};
    \draw[cell=part01] (1.79, -2.58) rectangle (3.53, -3.06);
    \node[font=\scriptsize] at (2.66, -2.82) {terms 01\ldots};
    \draw[cell=part1] (3.58, -2.58) rectangle (7.05, -3.06);
    \node[font=\scriptsize] at (5.315, -2.82) {terms 1\ldots};
  \end{tikzpicture}
  \endgroup
\caption{Adaptive output-prefix partitioning for the multiplication of sparse Pauli sums. 1.~Pairs of input ranges $A_\alpha$ and $B_\beta$ are grouped by their output signature prefix $\alpha\oplus\beta$, with colors distinguishing the resulting partitions. 2.~Partitions are refined according to their computational workload. Here, partition $0$ is subdivided into $00$ and $01$, while partition $1$ remains unchanged. 3.~Contributions to identical output signatures are accumulated independently within each partition. The locally sorted results are concatenated in prefix order without requiring a global merge. See~\cref{eq:pauli_prefix_routing,app:pauli_sum_product} for the routing identity and algorithmic details.}
\label{fig:prefix-partitioned}
\end{figure}

For $M_A=|S_A|$ and $M_B=|S_B|$, multiplication using packed binary signatures requires $\mathcal O(M_AM_BL_{\mathrm{word}})$ work for evaluating pairwise signatures and phases, with additional costs for partitioning and local accumulation. Our partitioning algorithm reduces temporary storage overhead and avoids global sorting of the pairwise products. It also applies to commutators by retaining only anticommuting pairs. For operator squares, the number of evaluated pairs can be further reduced by exploiting the cancellation of anticommuting cross terms. The algorithm and its complexity are detailed in Appendix~\ref{app:pauli_sum_product}.

\subsubsection{Arithmetic with sparsely supported Pauli strings}
\label{sssec:sparse_pauli_arithmetic}

For operators composed predominantly of low-weight Pauli strings, storing their nonidentity factors can substantially reduce the cost of evaluating the elementary products entering~\cref{eq:pauli_sum_arithmetic}. To characterize the sparsity of an operator $A$, we introduce its total and mean Pauli weights,
\begin{equation}
    W_A=\sum_{P\in S_A}\operatorname{wt}(P),
    \qquad
    \overline w_A=\frac{W_A}{M_A},
    \label{eq:pauli_sum_mean_weight}
\end{equation}
where $M_A>0$, and denote by $s_A$ its span, i.e., the highest occupied qubit index plus one in zero-based indexing. The storage required for the individual Pauli factors then scales with $W_A$ independently of $s_A$. We set $\overline{w}_A=0$ for the zero operator and $s_A=0$ for operators without occupied sites.

For multiplication in this representation, we generate elementary products by merging their ordered site lists and accumulate the resulting contributions in independent hash partitions. The partitions are determined by the hashes of the output labels, allowing contributions to identical labels to be accumulated together and separate partitions to be processed in parallel. Summing the cost of elementary arithmetic in~\cref{eq:sparse_pauli_elementary_complexity} over all input pairs yields the expected label-processing cost
\begin{equation}
    \mathcal O\bigl(M_AM_B(1+\overline w_A+\overline w_B)\bigr),
    \label{eq:sparse_pauli_sum_complexity}
\end{equation}
assuming expected constant hash-table probe counts. For operators of bounded mean Pauli weight, the pair-processing cost thus becomes independent of the register span.

The relative efficiency of the two storage formats depends on the number of stored machine words and occupied sites. We use this observation to automatically select sparse site storage at construction when the operator span satisfies $s_A\geq256$ and the mean Pauli weight satisfies $\overline w_A\leq s_A/256$. Since products and commutators can increase the Pauli weights of the constituent strings, we additionally estimate their weight growth before performing sparse arithmetic and reevaluate the storage choice for the accumulated result. The storage selection, hash-partitioned accumulation, and corresponding complexity estimates are detailed in Appendix~\ref{app:sparse_pauli_arithmetic}.

\subsubsection{Overlap-indexed commutators.}

Two Pauli strings $P$ and $Q$ can only anticommute if their supports overlap, i.e.,
\begin{equation}
    \operatorname{supp}(P)\cap\operatorname{supp}(Q)
    =\varnothing
    \quad\Longrightarrow\quad [P,Q]=0.
\end{equation}
We exploit this property by indexing the terms of $B$ according to their nonidentity sites, allowing us to directly identify the terms sharing at least one qubit with a given $P\in S_A$. A related site-based anticommutation index accelerates operator propagation through local gates~\cite{Miller.2026-MonopropScalingPauli}; here, we apply this idea to commutators between two arbitrary sparse Pauli sums. Denoting the number of distinct input pairs with overlapping supports by $N_{\mathrm{ov}}$, overlap indexing reduces the number of anticommutation tests from $M_AM_B$ to $N_{\mathrm{ov}}$. In particular, for local Hamiltonians with $M_A,M_B=\mathcal O(n)$, bounded Pauli weight, and a bounded number of terms acting on each qubit, we obtain $N_{\mathrm{ov}}=\mathcal O(n)$. For packed binary signatures, this reduces the pair-processing cost from $\mathcal O(n^2L_{\mathrm{word}})$ to $\mathcal O(nL_{\mathrm{word}})$, including index construction and candidate enumeration. With sparse site storage, the elementary arithmetic for each overlapping pair additionally has constant cost, yielding a total of $\mathcal O(n)$ operations for processing the identified pairs. Both implementations select overlap indexing adaptively according to an estimate of the number of overlapping pairs. See Appendix~\ref{app:pauli_overlap_commutator} for the algorithms and a detailed complexity analysis.

\subsubsection{Targeted commutator coefficients.}

When only the coefficient of a prescribed Pauli string $R$ is required, constructing the complete commutator can be avoided. For each $P\in S_A$, the XOR rule uniquely determines the only possible contributing partner $Q$ through $v(Q)=v(P)\oplus v(R)$. Our implementation traverses the smaller input sum and uses a hash index of the larger sum to identify these partners. Including the construction of the index, the expected signature-processing cost of a targeted commutator query is $\mathcal O((M_A+M_B)L_{\mathrm{word}})$ for packed binary signatures. When using sparse site storage, the partner labels are constructed through ordered-list merges. For bounded input and target Pauli weights, the expected label-processing cost then reduces to $\mathcal O(M_A+M_B)$, independently of the register span. The derivation and implementation details are given in Appendix~\ref{app:pauli_targeted}.

\begin{ExampleBox}{Arithmetic with sparse Pauli sums}
Consider the $n$-qubit operators
\begin{equation*}
    A=\sum_{j=1}^{n}X_j,
    \qquad
    B=\sum_{j=1}^{n}Y_j.
\end{equation*}
Since operators acting on different qubits commute,
\begin{equation*}
    [A,B]=2i\sum_{j=1}^{n}Z_j.
\end{equation*}
Of the $n^2$ input term pairs, only $n$ have overlapping supports. The overlap-indexed commutator algorithm (Appendix~\ref{app:pauli_overlap_commutator}) therefore reduces the number of anticommutation tests from $n^2$ to $n$. If only the coefficient of a prescribed $Z_k$ is required, the targeted commutator algorithm (\cref{eq:targeted_pauli_commutator,app:pauli_targeted}) directly returns the coefficient,
\begin{equation*}
    \operatorname{coeff}_{\mathcal P_n}([A,B];Z_k)=2i,
\end{equation*}
without constructing the complete commutator.
\end{ExampleBox}

\subsection{Pauli cycles and cycle sums}
\label{ssec:PauliCycles}
Translation-invariant operators on a periodic chain admit a compact representation through symmetrized Pauli strings. Such representations appear in Lie-algebraic classical simulation~\cite{Barligea.2026-EnablingLieAlgebraicClassical}, the analysis of dynamical Lie algebras of periodic spin chains and QAOA circuits~\cite{Wiersema.2024-ClassificationDynamicalLie,Allcock.2026-DynamicalLieAlgebras,Kazi.2025-AnalyzingQuantumApproximate}, in symmetry-merged Pauli propagation~\cite{Teng.2025-LeveragingSymmetryMerging}, and in software for operator dynamics~\cite{Loizeau.2025-QuantumManybodySimulations,Lychkovskiy.2026-QCommuteToolSymbolic}.

Let $C_n=\langle\tau\rangle$ denote the cyclic group acting on $n$ qubits, where conjugation by $U_\tau$ shifts each Pauli factor by one site. The corresponding twirling map, i.e., the Hilbert--Schmidt orthogonal projector onto the translation-invariant operators~\cite{Zeier.2011-SymmetryPrinciplesQuantum,Nguyen.2024-TheoryEquivariantQuantum}, is
\begin{equation}
    \mathcal T(A)
    =\frac{1}{n}\sum_{t=0}^{n-1}
    U_\tau^t A(U_\tau^\dagger)^t.
    \label{eq:cyclic_twirl}
\end{equation}
Following Ref.~\cite{Barligea.2026-EnablingLieAlgebraicClassical}, we define the \emph{Pauli cycle} associated with a string $P\in\mathcal P_n$ as
\begin{equation}
    C_P=\mathcal T(P)
    =\frac{1}{d_P}\sum_{Q\in\Omega_P^{\mathrm{cyc}}}Q,
    \label{eq:pauli_cycle_def}
\end{equation}
where $\Omega_P^{\mathrm{cyc}}$ is the set of distinct cyclic translations of $P$. Its cardinality $d_P=|\Omega_P^{\mathrm{cyc}}|$ is the cyclic period of $P$, i.e., the smallest positive number of sites by which the string can be shifted to recover itself. For example, the four-qubit string $XIXI$ has period $d_P=2$, since its only distinct translations are $XIXI$ and $IXIX$. In general, $d_P$ divides $n$, and each distinct translation occurs $n/d_P$ times in~\cref{eq:cyclic_twirl}.

Two Pauli strings are called cyclically equivalent, denoted by $P\sim_{\mathrm{cyc}}Q$, if one can be obtained from the other by a cyclic translation. All strings related in this way generate the same Pauli cycle. Choosing one representative from each such set of strings yields an orthogonal basis of the translation-invariant operator space, including the identity~\cite{Barligea.2026-EnablingLieAlgebraicClassical}, the size of which scales as $\Theta(4^n/n)$~\cite{Mansky.2025-ScalingSymmetryrestrictedQuantum}. From the orthogonality of the Pauli strings, we obtain
\begin{equation}
    \langle C_P,C_Q\rangle_\kappa
    =\begin{cases}
        1/d_P, & P\sim_{\mathrm{cyc}}Q,\\
        0, & \text{otherwise},
    \end{cases}
    \label{eq:pauli_cycle_inner_product}
\end{equation}
with $\kappa=2^{-n}$.

Our \texttt{PauliCycle} class identifies equivalent strings by assigning them a common canonical representative. The \texttt{unify()} routine exploits the cyclic distances between nonidentity Pauli factors and selects the lexicographically least rotation of the resulting sequence, a classic string-canonization problem~\cite{Booth.1980-LexicographicallyLeastCircular,Shiloach.1981-FastCanonizationCircular}, to construct this representative in $\mathcal O(L_{\mathrm{word}}+\mathrm{wt}(P))$ operations, avoiding the explicit comparison of all translations (see Appendix~\ref{app:cycle_unification}).

The software convention differs from~\cref{eq:pauli_cycle_def} by a factor of $n$: the coefficient $a$ of a \texttt{PauliCycle} multiplies the sum over all $n$ cyclic translations, such that the represented operator is $a\,n\,C_P$. To construct the averaged cycle $C_P$, one therefore supplies the coefficient $a=1/n$.

Cyclic symmetry also simplifies operator arithmetic. Extending the notation $C_A=\mathcal T(A)$ to arbitrary operators, we obtain
\begin{equation}
\begin{aligned}
    C_PC_Q
    &=\frac{1}{n}\sum_{t=0}^{n-1}C_{PQ^{(t)}},\\
    [C_P,C_Q]
    &=\frac{1}{n}\sum_{t=0}^{n-1}C_{[P,Q^{(t)}]},
\end{aligned}
    \label{eq:pauli_cycle_arithmetic}
\end{equation}
where $Q^{(t)}=U_\tau^tQ(U_\tau^\dagger)^t$. The double sum over translations reduces to a single sum over relative shifts, as illustrated in~\cref{fig:cycle_relative_shifts}. Evaluating the corresponding Pauli products and identifying their canonical cycle representatives requires $\mathcal O(n^2)$ operations at fixed machine-word width, assuming expected constant-time indexed accumulation.

For commutators, a further reduction follows from the fact that Pauli strings with disjoint supports commute. If the input strings have Pauli weights $p=\mathrm{wt}(P)$ and $q=\mathrm{wt}(Q)$, at most $pq$ shifts need to be considered. For fixed $p$ and $q$, this reduces the commutator cost to $\mathcal O(n)$~\cite{Barligea.2026-EnablingLieAlgebraicClassical}. The multiplication rule follows from the same cyclic averaging identity, but requires all relative shifts in general, since products of disjoint-support strings remain nonzero. The corresponding derivations are given in Appendix~\ref{app:cycle_arithmetic}.

Our \texttt{PauliCycleSum} class represents linear combinations of cycles with numerical or symbolic coefficients. Products and commutators are evaluated directly in this basis by accumulating contributions with identical canonical representatives. Targeted commutator evaluation further allows individual cycle coefficients to be computed without constructing the complete output expansion (see~\cref{eq:cycle_targeted_coefficient,sec:shared_operations}).

\begin{ExampleBox}{Arithmetic with translation-invariant operators}
Consider the four-qubit Pauli cycles
\begin{equation*}
    C_{XIII}=\frac14\sum_{j=1}^{4}X_j, \quad C_{YIII}=\frac14\sum_{j=1}^{4}Y_j,
\end{equation*}
as defined in~\cref{eq:pauli_cycle_def}. Since only Pauli factors acting on the same qubit anticommute, the relative-shift formula in~\cref{eq:pauli_cycle_arithmetic} contains only one nonzero commutator contribution,
\begin{equation*}
\begin{aligned}
    [C_{XIII},C_{YIII}]
    &=\frac14 C_{[XIII,YIII]}\\
    &=\frac{i}{2}C_{ZIII}.
\end{aligned}
\end{equation*}
The commutator is thus obtained from a single elementary Pauli-string commutator without expanding either cycle into its constituent strings. The reduction to relative shifts is illustrated in~\cref{fig:cycle_relative_shifts}.
\end{ExampleBox}

\subsection{Pauli orbits and orbit sums}
\label{ssec:PauliOrbits}
Permutation symmetry arises whenever all qubits play equivalent roles, as in collective spin models such as the Lipkin--Meshkov--Glick model~\cite{Lipkin.1965-ValidityManybodyApproximation}, collective open-system dynamics~\cite{Shammah.2018-OpenQuantumSystems}, the tomography of permutationally invariant states~\cite{Toth.2010-PermutationallyInvariantQuantum}, QAOA on complete graphs~\cite{Allcock.2026-DynamicalLieAlgebras}, and permutation-equivariant quantum neural networks~\cite{Schatzki.2024-TheoreticalGuaranteesPermutationequivariant,Nguyen.2024-TheoryEquivariantQuantum}.

Permutation-invariant operators admit a particularly compact representation, since arbitrary permutations of the qubits preserve only the multiplicities of the individual Pauli factors. Following Ref.~\cite{Barligea.2026-EnablingLieAlgebraicClassical}, we define the permutation twirl
\begin{equation}
    \mathcal S(A)=\frac{1}{n!}\sum_{\pi\in S_n}
    U_\pi A U_\pi^\dagger,
    \label{eq:permutation_twirl}
\end{equation}
where $S_n$ denotes the symmetric group acting on the qubits through permutation operators $U_\pi$. The resulting Pauli orbit is uniquely characterized by the label $\alpha=(p,q,r)$, specifying the numbers of $X$, $Y$, and $Z$ factors, respectively, with $p+q+r\leq n$. For any Pauli string $P$ with these multiplicities, we obtain
\begin{equation}
\begin{aligned}
    B_{p,q,r}
    &=\mathcal S(P)
     =\frac{1}{N_{p,q,r}}
       \sum_{Q\in\Omega_{p,q,r}}Q,\\
    N_{p,q,r}
    &=\frac{n!}{p!\,q!\,r!\,(n-p-q-r)!},
\end{aligned}
\label{eq:pauli_orbit_def}
\end{equation}
where $\Omega_{p,q,r}$ contains all distinct permutations of $P$ and $N_{p,q,r}$ denotes their number. The permutation-invariant operator space thus admits a basis of only $\binom{n+3}{3}$ elements~\cite{Albertini.2018-ControllabilitySymmetricSpin,Anschuetz.2023-EfficientClassicalAlgorithms,Schatzki.2024-TheoreticalGuaranteesPermutationequivariant,Kazi.2025-AnalyzingQuantumApproximate,Barligea.2026-EnablingLieAlgebraicClassical}, compared with the $4^n$ elements of the ordinary Pauli basis.
As for Pauli cycles, the disjointness of the underlying sets of Pauli strings yields an orthogonal basis with
\begin{equation}
    \langle B_\alpha,B_\beta\rangle_\kappa
    =\frac{\delta_{\alpha\beta}}{N_\alpha}.
    \label{eq:pauli_orbit_inner_product}
\end{equation}
Note that similar orbit sums of Pauli strings as in \cref{eq:pauli_orbit_def} appear in related literature~\cite{Allcock.2026-DynamicalLieAlgebras,Teng.2025-LeveragingSymmetryMerging}.

The arithmetic of Pauli orbits can be understood analogously to that of Pauli cycles (\cref{eq:pauli_cycle_arithmetic}), by considering the relative permutations of the generating strings. Since explicitly enumerating these permutations remains factorial in $n$, we instead group their contributions according to the overlap patterns of their local Pauli factors. Each pattern is characterized by a $4\times4$ contingency table $\eta$, whose entry $\eta_{ab}$ counts the sites carrying the Pauli factors $a$ and $b$ in the first and second string, respectively. The input orbit labels $\alpha$ and $\beta$ determine the row and column sums of the table, thereby defining the set $\mathcal T_{\alpha\beta}$ of feasible overlap patterns.

Each table determines both the output orbit label $\gamma(\eta)$ and the corresponding multiplication phase. Let $D(\eta)$ count the sites with distinct nonidentity Pauli factors, and let $E(\eta)=\eta_{YX}+\eta_{ZY}+\eta_{XZ}$ count those contributing a phase $-i$. The product of two Pauli orbits is then given by
\begin{equation}
    B_\alpha B_\beta
    =\sum_\gamma m_{\alpha\beta}^{\gamma}B_\gamma,
    \label{eq:pauli_orbit_product}
\end{equation}
with coefficients
\begin{equation}
\begin{aligned}
    m_{\alpha\beta}^{\gamma}
    &=\frac{1}{N_\alpha N_\beta}
      \sum_{\substack{\eta\in\mathcal T_{\alpha\beta}\\
                      \gamma(\eta)=\gamma}}
      W(\eta)(-1)^{E(\eta)}i^{D(\eta)},\\
    W(\eta)&=\frac{n!}{\prod_{a,b}\eta_{ab}!}.
\end{aligned}
\label{eq:pauli_orbit_product_coeff}
\end{equation}
Here, the multinomial factor $W(\eta)$ counts all ordered pairs of Pauli strings realizing the same overlap pattern. Their contributions can therefore be evaluated collectively through a single contingency table, as illustrated in~\cref{fig:orbit_overlap_patterns}.

The commutator follows from the same construction by retaining only tables with odd $D(\eta)$ and multiplying their contributions by two, recovering the commutator algorithm of Ref.~\cite{Barligea.2026-EnablingLieAlgebraicClassical}. Both products and commutators require at most $\mathcal O(n^9)$ arithmetic operations after combinatorial precomputation and yield at most $\mathcal O(n^3)$ distinct output labels. Despite the potentially exponential sizes of the underlying Pauli expansions, arithmetic in the orbit basis thus remains polynomial in the number of qubits.

For targeted commutator evaluation, prescribing the output orbit label imposes three additional constraints on the contingency table. These reduce the enumeration cost to $\mathcal O(n^6)$ in general and to $\mathcal O(1)$ for targets of fixed bounded Pauli weight, assuming precomputed combinatorial factors~\cite{Barligea.2026-EnablingLieAlgebraicClassical}. The derivations and complexity estimates for these elementary orbit operations are given in Appendix~\ref{app:pauli_orbit_product}.

Our \texttt{PauliOrbit} class stores individual orbit labels, while \texttt{PauliOrbitSum} represents their linear combinations with numerical or symbolic coefficients. Products, commutators, and targeted commutator coefficients are evaluated directly in this representation. \PE\ uses the anti-Hermitian basis $iB_{p,q,r}$, such that representing a Hermitian orbit $B_{p,q,r}$ requires a coefficient $-i$ in the corresponding orbit sum (see Appendix~\ref{app:software_conventions}).

\begin{ExampleBox}{Arithmetic with permutation-invariant operators}
Consider the Pauli orbits
\begin{equation*}
    B_{1,0,0}=\frac{1}{n}\sum_{j=1}^{n}X_j,
    \qquad
    B_{0,1,0}=\frac{1}{n}\sum_{j=1}^{n}Y_j,
\end{equation*}
as defined in~\cref{eq:pauli_orbit_def}. The $n^2$ pairs of Pauli strings belong to only two overlap patterns: $n$ pairs with both nonidentity factors acting on the same qubit and $n(n-1)$ pairs with disjoint supports. Applying the contingency-table multiplication rule in~\cref{eq:pauli_orbit_product_coeff} gives
\begin{equation*}
    B_{1,0,0}B_{0,1,0}
    =\frac{i}{n}B_{0,0,1}
     +\frac{n-1}{n}B_{1,1,0}.
\end{equation*}
Only the first contribution changes sign upon reversing the operands, yielding
\begin{equation*}
    [B_{1,0,0},B_{0,1,0}]
    =\frac{2i}{n}B_{0,0,1}.
\end{equation*}
Both operations are thus evaluated using only two overlap patterns, independently of the number of qubits. The corresponding contingency tables for $n=4$ are illustrated in~\cref{fig:orbit_overlap_patterns}.
\end{ExampleBox}

\subsection{Shared operations and targeted coefficients}
\label{sec:shared_operations}

The operator representations introduced above share a common set of arithmetic and scalar operations. In particular, the orthogonality of their respective bases $\mathcal B=\{B_\alpha\}$ allows us to extract individual operator coefficients through the normalized Hilbert--Schmidt inner product introduced in~\cref{eq:general_inner_product},
\begin{equation}
    \operatorname{coeff}_{\mathcal B}(A;\alpha)
    =\frac{\langle B_\alpha,A\rangle_\kappa}
           {\langle B_\alpha,B_\alpha\rangle_\kappa},
    \label{eq:coefficient_extraction}
\end{equation}
for any $A\in\operatorname{span}_{\mathbb C}\mathcal B$. The squared norms of Pauli strings, averaged cycles, and averaged permutation orbits are given by
\begin{equation}
    \nu_P=1,\qquad
    \nu_{[P]}=\frac{1}{d_P},\qquad
    \nu_{(p,q,r)}=\frac{1}{N_{p,q,r}},
    \label{eq:representation_norms}
\end{equation}
respectively. Inner products can thus be evaluated directly from the operator labels and coefficients, including between different representations, without constructing dense operator matrices (see Appendix~\ref{app:inner_product_evaluation}).

When only the coefficient of a prescribed basis element $B_\gamma$ in a commutator $[A,B]$ is required, targeted evaluation avoids constructing the complete output expansion. While this offers no asymptotic advantage for individual Pauli strings, it reduces the expected pair-processing cost for ordinary Pauli sums from quadratic to linear in the input sizes when both sums grow comparably (Appendix~\ref{app:pauli_targeted}). For Pauli cycles, targeting reduces the number of output terms that need to be stored and accumulated without changing the asymptotic bounds of the relative-shift algorithm (Appendix~\ref{app:cycle_arithmetic}). For Pauli orbits, the additional constraints on the contingency tables reduce the enumeration cost from $\mathcal O(n^9)$ to $\mathcal O(n^6)$, and even to $\mathcal O(1)$ for targets of fixed bounded Pauli weight after combinatorial precomputation (Appendix~\ref{app:pauli_orbit_product}).

The resulting computational costs are summarized in~\cref{tab:representations_operations}. Despite the different algorithms underlying these operations, all three families are accessible through a common software interface, preserving their native representations wherever possible. We discuss this interface, mixed-representation arithmetic, and the integration of additional operator families in~\cref{sec:software}.

\begin{table*}[htbp]
    \centering
    \small
    \setlength{\tabcolsep}{5pt}
    \renewcommand{\arraystretch}{1.35}
    \begin{tabular}{
        p{0.18\textwidth}
        p{0.18\textwidth}
        p{0.29\textwidth}
        p{0.27\textwidth}
    }
        \toprule
        \textbf{Representation}
        & \textbf{Arithmetic output}
        & \textbf{Product and commutator cost}
        & \textbf{Targeted commutator cost} \\
        \midrule

\texttt{PauliString}
& \texttt{PauliString} or zero
& & \\
{\footnotesize \textit{Packed signatures}}
&
& $\mathcal O(L_{\mathrm{word}})$
& $\mathcal O(L_{\mathrm{word}})$ \\
{\footnotesize \textit{Sparse sites}}
&
& $\mathcal O(1)$ bounded weights
& $\mathcal O(1)$ bounded weights \\
\midrule

\texttt{PauliStringSum}
& \texttt{PauliStringSum}
& & \\
{\footnotesize \textit{Packed signatures}}
&
& $\mathcal O(M_AM_BL_{\mathrm{word}})$;
  \newline
  $\mathcal O(nL_{\mathrm{word}})$ local commutators
& $\mathcal O((M_A+M_B)L_{\mathrm{word}})$ \\
{\footnotesize \textit{Sparse sites}}
&
& $\mathcal O(M_AM_B)$ bounded mean weights;
  \newline
  $\mathcal O(n)$ local candidate-pair processing
& $\mathcal O(M_A+M_B)$ bounded weights \\
\midrule

        \texttt{PauliCycle},
        \newline
        \texttt{PauliCycleSum}
        & \texttt{PauliCycleSum}
        & $\mathcal O(n^2)$;
          \newline
          $\mathcal O(n)$ bounded-weight commutators
        & $\mathcal O(n^2)$;
          \newline
          $\mathcal O(n)$ bounded input weights \\
        \midrule

        \texttt{PauliOrbit},
        \newline
        \texttt{PauliOrbitSum}
        & \texttt{PauliOrbitSum}
        & $\mathcal O(n^9)$
        & $\mathcal O(n^6)$;
          \newline
          $\mathcal O(1)$ fixed bounded-weight targets \\
        \bottomrule
    \end{tabular}
    \caption{Output representations and asymptotic costs of multiplication, commutators, and targeted commutator evaluation for the weighted operator families implemented in \PE. We use $n$ for the number of qubits and $L_{\mathrm{word}}=\lceil n/w\rceil$ for the number of machine words per binary signature vector (see~\cref{sec:comp-model-and-notation}). For ordinary Pauli sums, the product and commutator estimates describe pair processing before coefficient accumulation. The local commutator bound assumes $M_A,M_B=\mathcal O(n)$, bounded Pauli weights, and bounded term incidence per qubit (Appendix~\ref{app:pauli_overlap_commutator}), while the targeted estimate includes hash-index construction (Appendix~\ref{app:pauli_targeted}). Cycle and orbit estimates refer to elementary operator pairs; arithmetic on their linear combinations additionally depends on the numbers of input terms. All complexity estimates assume fixed-cost numerical coefficient arithmetic, expected constant-time hash-table access wherever applicable, and precomputed combinatorial factors for orbit arithmetic.}
    \label{tab:representations_operations}
\end{table*}

\section{Software interface and additional capabilities}
\label{sec:software}

\subsection{Software architecture}
\label{sec:software_architecture}

\PE\ is implemented as a C++20 library with Python bindings provided through \texttt{nanobind}~\cite{Jakob.2022-Nanobind}. The arithmetic algorithms developed in~\cref{sec:operatorrepresentations} are implemented in C++, while the Python interface provides access to the different operator representations through a common set of operations, as illustrated in~\cref{fig:overview}.

The internal data structures exploit the compact operator labels introduced above, using flat storage for ordinary Pauli sums with either packed binary signatures or sparse site lists, and retaining the symmetry-adapted labels of cycles and orbits. For weighted ordinary Pauli operators, we adaptively select the storage format and corresponding arithmetic kernels according to the span and mean Pauli weight of the operators (Appendix~\ref{app:sparse_pauli_arithmetic}). Independently of the storage format, the coefficient type is a template parameter of the weighted C++ representations, allowing us to specialize the arithmetic for complex floating-point numbers and symbolic polynomial coefficients. Selected arithmetic operations, including the multiplication of large Pauli sums, can be parallelized using OpenMP~\cite{Dagum.1998-OpenMPIndustryStandard}.

The representation-specific data structures and arithmetic algorithms are detailed in Appendices~\ref{app:pauli_arithmetic} and \ref{app:symmetry_arithmetic}, and the coefficient backends and software normalization conventions are described in Appendix~\ref{app:software_conventions}. The source code, documentation, and installation instructions are available on \href{https://github.com/tequilahub/pauliengine}{GitHub}~\cite{PE_repo}. The Python package can be installed from source or via \texttt{pip} from the \href{https://pypi.org/project/pauliengine/}{Python Package Index (PyPI)}.

\subsection{A unified Python interface}
\label{sec:software_interface}

The Python interface provides constructors for elementary Pauli strings, cycles, and orbits, and their corresponding sum representations \texttt{PauliStringSum}
\footnote{The name \texttt{QubitHamiltonian}, used in earlier versions of \PE, is retained as a compatibility alias in Python and remains the class name in the C++ implementation.}, \texttt{PauliCycleSum}, and \texttt{PauliOrbitSum}, as illustrated in~\cref{fig:overview}. Numerical or symbolic coefficients supplied at construction determine the underlying coefficient type. Addition, subtraction, multiplication, and scalar arithmetic use standard Python operators, while commutators, inner products, and coefficient queries are exposed through dedicated methods.

For weighted ordinary Pauli operators, the constructors \texttt{PauliString} and \texttt{PauliStringSum} additionally select between packed binary signatures and sparse site storage according to the criteria described in Appendix~\ref{app:sparse_pauli_arithmetic}. Both storage formats support the same arithmetic interface and can be combined in arithmetic expressions, with the necessary conversions performed internally. The selection is independent of the input syntax and can be configured through \texttt{set\_sparse\_threshold()}.

Products of individual Pauli strings retain the elementary Pauli-string representation, while products and commutators within the cycle and orbit families retain their respective sum representations. Arithmetic between different families generally returns an ordinary Pauli sum, requiring an expansion into the ordinary Pauli basis. Numerical coefficients are promoted to symbolic coefficients when necessary. The mixed-representation dispatch rules are detailed in Appendix~\ref{app:mixed_representation_dispatch}, while specialized inner-product algorithms can avoid expanding symmetry-adapted operators (see Appendix~\ref{app:inner_product_evaluation}).

\begin{SoftwareCodeBox}{Operator arithmetic and coefficient queries}
import pauliengine as pe

X = pe.PauliString(1.0, "X")
Y = pe.PauliString(1.0, "Y")
Z = pe.PauliString(1.0, "Z")
H = pe.PauliStringSum([X, Y])

XY = X * Y
K = X.commutator(Y)
a_X = H.contains(X)                # 1
c_Z = H.targeted_commutator(X, Z)  # -2j

C = pe.PauliCycle(4, "XZ", 0.25)
B = pe.PauliOrbit(4, 1, 0, 1)

CC = C * C  # PauliCycleSum
BB = B * B  # PauliOrbitSum
CB = C * B  # PauliStringSum

overlap = pe.inner_product(C, B)
\end{SoftwareCodeBox}

The method \texttt{contains(target)} retrieves the stored coefficient associated with a target label, whereas \texttt{targeted\_commutator(other,target)} computes its coefficient in a commutator without constructing the complete result (\cref{sec:shared_operations}). Both queries disregard the scalar coefficient attached to the target object. Elementary operator labels can also be identified through \texttt{get\_hash()}.

Omitting the coefficient when constructing an ordinary Pauli string, e.g., \texttt{PauliString("X")}, selects its coefficient-free representation, which retains the operator label and exact phase. Collections of such strings form ordered sets of distinct labels, supporting Boolean membership queries and stable indexing through \texttt{index()}. Their commutator method returns the product of anticommuting strings without the factor of two in the weighted commutator. Numerical scaling or arithmetic with weighted operators promotes coefficient-free objects to weighted representations (see Appendix~\ref{app:software_conventions}). These set operations are particularly useful for constructing operator bases, as illustrated in~\cref{sec:applications}.

\subsection{Symbolic coefficients and interoperability}
\label{sec:software_additional}

Symbolic coefficients allow us to perform arithmetic on parameter-dependent operators while retaining their algebraic dependence on the parameters. The operator labels follow the arithmetic rules of their respective representations, while the coefficients are manipulated symbolically. Parameter-dependent operators in each weighted family support coefficient differentiation through \texttt{diff()} and parameter substitution through \texttt{subs()}, preserving the underlying operator representation. For Pauli orbits, parameter-dependent coefficients are stored in \texttt{PauliOrbitSum}. We motivate and describe the implementation of our symbolic coefficient backend in Appendix~\ref{app:software_conventions}.

\begin{SoftwareCodeBox}{Parameter-dependent operator arithmetic}
import pauliengine as pe

H = pe.PauliStringSum([
    ("theta", {0: "X"}),
    (1.0,     {0: "Z"})
])
Y = pe.PauliString(1.0, "Y")

K = H.commutator(H.commutator(Y))
dK = K.diff("theta")
K_half = K.subs({"theta": 0.5})
\end{SoftwareCodeBox}

For $H(\theta)=\theta X+Z$, the above example computes the nested commutator
\begin{equation}
    K(\theta)=[H(\theta),[H(\theta),Y]]
    =4(1+\theta^2)Y,
\end{equation}
from which we obtain the derivative $\partial_\theta K=8\theta Y$ and, for example, the substituted operator $K(1/2)=5Y$. The symbolic representation thus allows us to evaluate operator expressions and their parameter derivatives at different parameter values without repeating the underlying operator arithmetic.

The ordinary Pauli interface also provides Hermiticity queries, adjoints, complex conjugation, transposition, and qubit remapping. Dictionary-based representations and conversion to and from OpenFermion operators facilitate the exchange of operators with other software. Operators using sparse site storage can also be explicitly converted to packed binary signatures through \texttt{to\_dense()}, preserving their Pauli expansion.

Numerical matrices can be decomposed into Pauli sums through \texttt{PauliStringSum.from\_matrix()}, using the algorithm of Georges \emph{et al.}~\cite{Georges.2025-PauliDecompositionFast}, while \texttt{to\_matrix()} provides dense matrix export. The corresponding qubit-indexing and register conventions are specified in Appendix~\ref{app:software_conventions}.

\subsection{Extensibility}
\label{ssec:Extensibility}

The separation of operator representations, their internal storage formats, and their arithmetic interface allows the software architecture described in~\cref{sec:software_architecture} to accommodate further operator families. These may be adapted to different symmetries, particle statistics, local Hilbert-space dimensions, or restricted Hilbert-space sectors. The matrix-unit example below illustrates how the mathematical properties of an operator basis determine the corresponding data structures and arithmetic algorithms.

\begin{InsightBox}{Constructing a new operator representation}
Consider the matrix-unit basis $E_{ij}=|i\rangle\langle j|$ of a restricted Hilbert-space sector spanned by $d$ orthonormal states. Its implementation requires four ingredients:

\begin{enumerate}[leftmargin=*, labelindent=0pt]
    \item \textbf{Basis and labels.} Specify the elementary operators, their unique labels, and normalization (\cref{sec:operatorrepresentations}).
    \begin{equation*}
        E_{ij}\longleftrightarrow(i,j),
        \qquad 1\leq i,j\leq d.
    \end{equation*}

    \item \textbf{Data structures.} Store elementary labels and coefficients, and accumulate contributions with identical labels (\cref{sec:software_architecture}).
    \begin{equation*}
        A=\sum_{i,j}a_{ij}E_{ij}
        \longleftrightarrow
        \{(i,j)\mapsto a_{ij}\}.
    \end{equation*}

    \item \textbf{Arithmetic.} Implement the elementary multiplication rules and determine the resulting representation (\cref{sec:operatorrepresentations,app:mixed_representation_dispatch}).
    \begin{equation*}
    \begin{aligned}
        E_{ij}E_{kl}&=\delta_{jk}E_{il},\\
        [E_{ij},E_{kl}]
            &=\delta_{jk}E_{il}-\delta_{li}E_{kj}.
    \end{aligned}
    \end{equation*}

    \item \textbf{Scalar operations.} Implement inner products and direct coefficient queries (\cref{sec:shared_operations}).
    \begin{equation*}
    \begin{aligned}
        \langle E_{ij},E_{kl}\rangle_\kappa
            &=\kappa\delta_{ik}\delta_{jl},\\
        \operatorname{coeff}(A;ij)
            &=\kappa^{-1}\langle E_{ij},A\rangle_\kappa.
    \end{aligned}
    \end{equation*}
\end{enumerate}
\end{InsightBox}

Integrating a representation into \PE\ requires implementing the corresponding C++ types and arithmetic routines detailed above, extending the dispatch mechanisms and Python bindings, and validating the supported operations. Operations that leave the represented operator space require suitable conversion rules, while coefficient extraction in nonorthogonal bases requires the Gram matrix or a dual basis. Representation-specific algorithms for targeted arithmetic can further reduce computational costs.

One prospective extension is the modified generalized Gell-Mann (MGGM) basis for operators acting within fixed-excitation sectors~\cite{Barligea.2026-EnablingLieAlgebraicClassical}, which would allow arithmetic to be performed directly within these restricted operator spaces. More generally, the framework can be extended to fermionic and bosonic operator algebras by exploiting their respective canonical anticommutation and commutation relations, as in existing Majorana-basis arithmetic~\cite{Miller.2025-SimulationFermionicCircuits,Fawzi.2026-FastFermion}, as well as to operators on systems with higher local Hilbert-space dimensions through suitable generalizations of the Pauli basis, such as the Weyl--Heisenberg operators or generalized Gell-Mann matrices~\cite{Gottesman.1999-FaulttolerantQuantumComputation,Bertlmann.2008-BlochVectorsQudits}.

\section{Performance benchmarks}
\label{sec:benchmarks}


In this section, we benchmark the operator arithmetic developed in~\cref{sec:operatorrepresentations}, comparing its performance with existing software implementations and assessing the computational advantages of symmetry-adapted representations. We investigate runtime scaling with the number of qubits, operator terms, and Pauli weights, as well as the effect of parallel execution. The benchmark methodology and numerical validation are detailed in Appendices~\ref{app:benchmark_methodology} and \ref{app:numerical_validation}. The complete benchmark suite, including the underlying measurement data, reproducible figure-generation scripts, and supplementary numerical results, is available at \url{https://github.com/adelina-b/pauliengine-benchmarks}.

\subsection{Related software and benchmarking scope}
\label{ssec:benchmark_software}

Efficient Pauli arithmetic is implemented in general-purpose quantum software such as Qiskit~\cite{JavadiAbhari.2024-QuantumComputingQiskit}, OpenFermion~\cite{McClean.2020-OpenFermionElectronicStructure}, and PennyLane~\cite{Bergholm.2018-PennyLaneAutomaticDifferentiation}, as well as specialized libraries including PauliArray~\cite{Dion.2024-EfficientlyManipulatingPauli}, PauliStrings.jl~\cite{Loizeau.2025-QuantumManybodySimulations}, PauLIB~\cite{Krotz.2026-PauLIBHighperformanceLibrary}, BinSim~\cite{Li.2026-HighperformancePaulialgebraFramework}, zixy~\cite{Quantinuum.2026-Zixy}, fast-pauli~\cite{Qognitive.2024-FastPauli}, and fastfermion~\cite{Fawzi.2026-FastFermion}. These implementations differ in their Pauli-string encodings, data structures for operator sums, supported coefficient types, and parallelization strategies.

Our Pauli-arithmetic benchmarks compare operations through the Python interfaces and, where available, directly through the native C++, Rust, or Julia implementations. For elementary Pauli-string operations, we additionally consider specialized stabilizer and Lie-algebra tools such as Stim~\cite{Gidney.2021-StimFastStabilizer}, QuantumClifford.jl~\cite{Krastanov.2026-QuantumCliffordJl}, and PauLie~\cite{Shaya.2026-PauLieFastClassification}. The symmetry-adapted benchmarks include comparisons with the translation-symmetric arithmetic of PauliStrings.jl and the earlier cycle and orbit implementations developed for Lie-algebraic simulation ($\gsim$)~\cite{Barligea.2026-EnablingLieAlgebraicClassical}, as well as the symmetry-merging approach of Ref.~\cite{Teng.2025-LeveragingSymmetryMerging}.

A detailed discussion of the considered packages, their arithmetic implementations, supported capabilities, and versions used in our experiments is provided in Appendix~\ref{app:benchmark_software}, with a comparative overview in~\cref{tab:related_software}.

\subsection{Performance of Pauli-string and Pauli-sum arithmetic}
\label{ssec:benchmark_pauli}

We first investigate the performance of ordinary Pauli arithmetic as a function of the register size, the number of terms in the input operators, and the Pauli weights of these terms. For each configuration, we generate ten independent pairs of operators and measure the runtime of the respective arithmetic operation ten times for each pair, using identical inputs across the considered software implementations. We report arithmetic mean runtimes with error bars indicating the standard error of the mean across the ten input instances. The experimental configurations, software interfaces, and measurement procedures are detailed in Appendix~\ref{app:benchmark_methodology}.

\Cref{fig:frontend-comm-t1} compares single-threaded commutator runtimes through the Python interfaces of the respective packages. We consider both dense Pauli strings, with an expected weight of $\tfrac{3}{4}n$, and randomly distributed Pauli strings of fixed weight four. For both distributions, we vary the number of terms $M$ in each operand at fixed register size $n=64$ and the number of qubits $n$ at fixed $M=316$.

\begin{figure*}[htbp]
    \centering
    \includegraphics[width=1\linewidth]{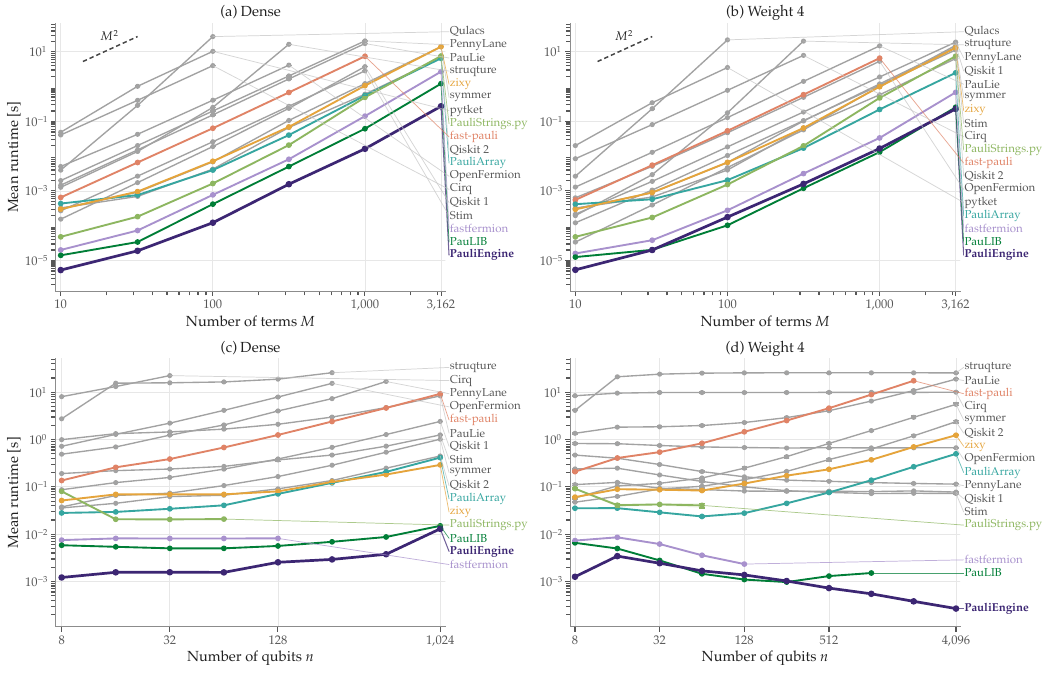}
    \caption{Single-threaded commutator runtimes of Pauli sums through the Python interfaces of the considered software packages. Both operands contain $M$ distinct Pauli strings with randomly sampled numerical coefficients. Panels (a) and (b) show the dependence on the number of terms $M$ at fixed register size $n=64$, while panels (c) and (d) show the dependence on the register size $n$ at fixed $M=316$. The Pauli strings are sampled uniformly from $\mathcal{P}_n$ in (a) and (c), and from strings of weight four in (b) and (d). Dashed lines in (a) and (b) indicate quadratic scaling with $M$. Each data point represents ten timing measurements for each of ten independently generated input pairs, with error bars indicating the standard error of the mean across the input instances. Qiskit~1 and Qiskit~2 refer to the \texttt{SparseObservable} and \texttt{SparsePauliOp} implementations, respectively. The experimental methodology and software-specific configurations are detailed in Appendix~\ref{app:benchmark_methodology}.}
    \label{fig:frontend-comm-t1}
\end{figure*}

When increasing the number of terms in the input operators, we observe approximately quadratic runtime scaling for most implementations once the arithmetic dominates the cost of invoking the corresponding routines (see \cref{fig:frontend-comm-t1}(a,b)). This agrees with the pairwise arithmetic underlying the product and commutator algorithms discussed in~\cref{sec:operatorrepresentations}. Deviations from quadratic scaling arise from the additional computational costs of accumulating and simplifying the generated terms, as well as from differences in how the respective implementations exploit the commutation relations of the input strings. Despite the similar scaling behavior, the absolute runtimes differ by several orders of magnitude. The implementations based on packed binary signatures generally achieve lower runtimes, with \PE\ and PauLIB exhibiting the lowest commutator runtimes across most of the considered configurations.

The dependence on the register size is particularly sensitive to the support of the individual Pauli strings. For dense operators (see \cref{fig:frontend-comm-t1}(c)), increasing $n$ generally increases the runtime, although the packed representations exhibit a comparatively weak dependence on the register size over substantial parts of the considered range. In the fixed-weight case (see \cref{fig:frontend-comm-t1}(d)), increasing the register size also changes the fraction of input pairs with overlapping supports. For two independently sampled strings of weight four, the probability of overlapping supports is
\begin{equation}
    p_{\mathrm{ov}}(n)
    =1-\frac{\binom{n-4}{4}}{\binom{n}{4}}
    =\frac{16}{n}+\mathcal{O}(n^{-2}),
\end{equation}
for $n\geq 8$. Consequently, at fixed $M$, the expected number of input pairs that can contribute to the commutator decreases with increasing register size. Several implementations benefit from the resulting reduction in nonvanishing arithmetic contributions. In \PE, the overlap-indexed algorithm developed in Appendix~\ref{app:pauli_overlap_commutator} additionally reduces the number of input pairs for which commutation relations need to be evaluated, leading to decreasing runtimes over a substantial range of register sizes. For sufficiently large registers, the sparse site representation introduced in~\cref{sssec:sparse_pauli_arithmetic} further reduces the cost of processing the individual Pauli labels.

\begin{figure*}[htbp]
    \centering
    \includegraphics[width=1\linewidth]{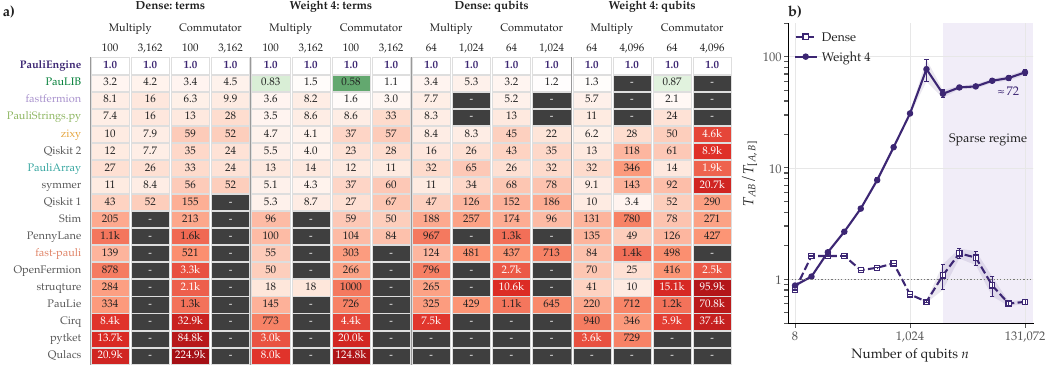}
    \caption{Relative runtimes of single-threaded Pauli-sum arithmetic through the Python interfaces. (a) Ratio of the mean runtime of each implementation to that of \PE\ for selected configurations of the product and commutator benchmarks. The term-count scans use $n=64$, while the register-size scans use $M=316$. Values greater (smaller) than one indicate longer (shorter) runtimes than those of \PE. Dark-gray cells indicate configurations for which no complete runtime measurement is available. The implementations are ordered by the geometric mean of their available runtime ratios. (b) Ratio of the product runtime $T_{AB}$ to the commutator runtime $T_{[A,B]}$ in \PE\ as a function of the register size at fixed $M=316$, for dense Pauli strings and strings of weight four. The horizontal dotted line indicates equal runtimes for both operations. All ratios are computed from the arithmetic mean runtimes obtained using the experimental protocol described in Appendix~\ref{app:benchmark_methodology}.}
    \label{fig:frontend-factors}
\end{figure*}

To quantify the differences between the implementations, \cref{fig:frontend-factors}(a) shows the ratios of their mean runtimes to the corresponding mean runtimes of \PE\ for selected configurations of product and commutator arithmetic. PauLIB achieves runtimes closest to those of \PE, outperforming our implementation in some configurations involving low-weight Pauli strings. For the remaining packages, we observe runtime differences ranging from factors of a few to several orders of magnitude, depending on the operation and the structure of the input operators.

The advantages of our adaptive storage and arithmetic algorithms become particularly pronounced for low-weight operators on large registers. At $n=4096$, where the sparse site representation is employed, the multiplication runtimes of the remaining implementations with complete measurements exceed those of \PE\ by factors ranging from $3.4$ to more than three orders of magnitude. The smallest factor is obtained for Qiskit's \texttt{SparseObservable}, which similarly exploits the sparsity of individual Pauli strings. The dependence of multiplication runtimes on the register size and the transition between the two storage formats are investigated further in Appendix~\ref{app:pauli_benchmarks}.

The computational advantages of sparse site arithmetic are also reflected in the ratio between product and commutator runtimes in~\cref{fig:frontend-factors}(b). For dense operators, both operations exhibit comparable runtimes, while for operators of fixed weight their runtimes increasingly separate as the number of overlapping input pairs decreases. Upon switching to sparse site arithmetic, the ratio decreases and subsequently exhibits a considerably weaker dependence on the register size, reaching approximately $72$ at $n=131072$. The remaining difference reflects the computational advantages of overlap indexing, while both operations benefit from the reduced cost of elementary sparse arithmetic.

Further comparisons of product arithmetic, including native implementations and multithreaded execution, as well as benchmarks of elementary Pauli-string products, are presented in Appendix~\ref{app:pauli_benchmarks}.

\subsection{Performance of symmetry-adapted arithmetic}
\label{ssec:benchmark_symmetry}

To investigate the computational advantages of the symmetry-adapted representations introduced in~\cref{ssec:PauliCycles,ssec:PauliOrbits}, we compare arithmetic performed directly in the respective symmetry-adapted bases with arithmetic on the equivalent ordinary Pauli expansions using \texttt{PauliStringSum}. We further compare with the original implementations of cycle and orbit commutators developed for Lie-algebraic simulation in $\gsim$~\cite{Barligea.2026-EnablingLieAlgebraicClassical}. For translation-invariant operators, we additionally consider the dedicated arithmetic routines of PauliStrings.jl~\cite{Loizeau.2025-QuantumManybodySimulations}. A further approach to exploiting symmetry is provided by the symmetry-merging procedure proposed in Ref.~\cite{Teng.2025-LeveragingSymmetryMerging} and implemented in the accompanying SymMerge code~\cite{Teng.2026-SymMerge}, which combines the coefficients of symmetry-equivalent Pauli strings during operator propagation. For our comparison, we use the arithmetic routines of the underlying PauliPropagation.jl library~\cite{Rudolph.2025-PauliPropagationComputational} and subsequently merge the resulting terms into their respective symmetry classes.

For both Pauli cycles and orbits, we consider arithmetic between two individual basis elements and investigate the dependence of the runtime on the number of qubits $n$ and the Pauli weights of the operands. All measurements are performed on a single thread, following the experimental protocol of Appendix~\ref{app:benchmark_methodology}, with ten timing repetitions for each of ten sampled input pairs.

\begin{figure}[htbp]
    \centering
    \includegraphics[width=\linewidth]{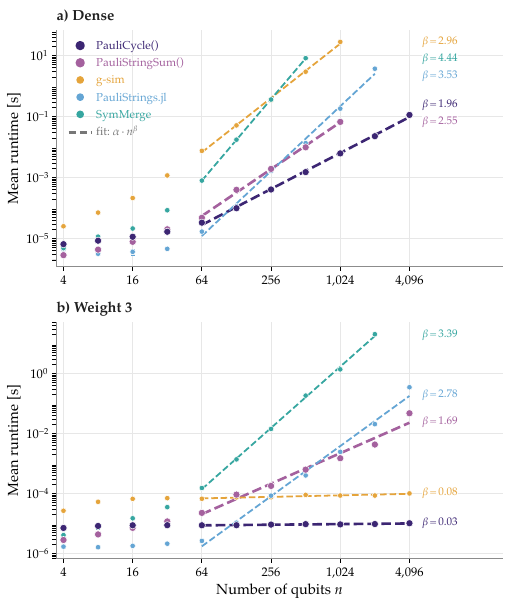}
    \caption{Runtime of commutators between two Pauli cycles on a periodic register of $n$ qubits, evaluated using different implementations of translation-invariant operator arithmetic. We sample the representative Pauli strings (a) uniformly from $\{X,Y,Z\}^{\otimes n}$ and (b) from Pauli strings of weight three with randomly distributed supports. The ordinary Pauli-sum representation is obtained by expanding the cycles into their constituent Pauli strings. Dashed lines indicate power-law fits of the form $\alpha n^\beta$, with the fitted exponents $\beta$ given in the respective colors. Each data point represents the mean runtime across ten input pairs with ten timing repetitions each. Error bars indicate the standard error of the mean across the input pairs. All measurements use a single thread, with the arithmetic of \PE\ and $\gsim$ invoked through Python and the Julia implementations measured natively. The software normalization conventions of Appendix~\ref{app:software_conventions} apply.}
    \label{fig:cycles-commutators-t1}
\end{figure}

\Cref{fig:cycles-commutators-t1} shows the commutator runtimes for Pauli cycles generated by dense and fixed-weight Pauli strings. In the dense case, the quadratic number of pairs of ordinary Pauli strings can be reduced to $n$ relative translations using the arithmetic identity derived in Appendix~\ref{app:cycle_arithmetic}. Taking into account the cost of elementary Pauli arithmetic, which is linear in $n$, and the identification of equivalent output cycles, we observe approximately quadratic runtime scaling for our implementation, with a fitted exponent of $\beta=1.96$. The ordinary Pauli-sum implementation exhibits a steeper increase in runtime, as do the other considered implementations over their respective fitting ranges.

The advantage of performing the arithmetic directly in the cyclic basis becomes particularly pronounced for Pauli strings of bounded weight. As discussed in~\cref{ssec:PauliCycles}, only relative translations for which the supports of the two representative strings overlap can contribute to the commutator. For two strings of weight three, there are at most nine such translations, independently of $n$. Accordingly, both our implementation and the earlier $\gsim$ implementation exhibit nearly constant runtimes over the considered range of register sizes (see \cref{fig:cycles-commutators-t1}(b)), with our implementation reducing the absolute runtime by approximately one order of magnitude. The arithmetic through ordinary Pauli expansions, PauliStrings.jl, and SymMerge exhibits a stronger dependence on the register size. At $n=4096$, the native cycle arithmetic of \PE\ evaluates the commutator within approximately $10\,\mu\mathrm{s}$.

\begin{figure*}[htbp]
    \centering
    \includegraphics[width=\linewidth]{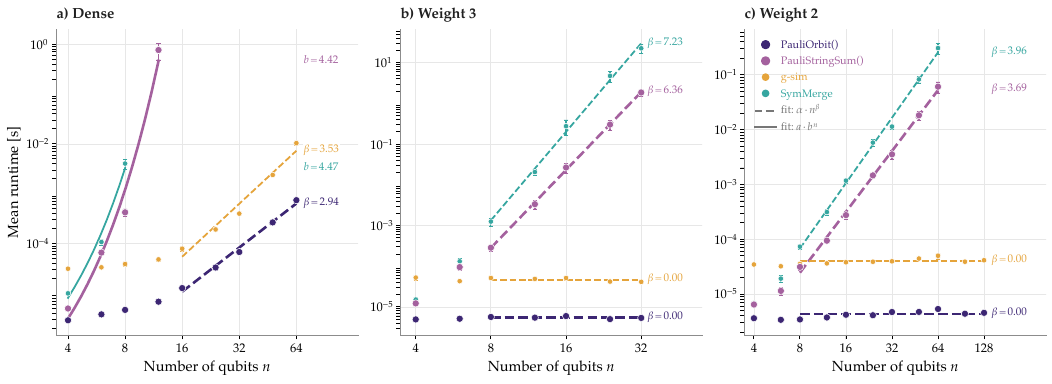}
    \caption{Runtime of commutators between two permutation-invariant Pauli-orbit basis elements on $n$ qubits. The input orbits are generated from (a) fully occupied Pauli strings and orbit labels of fixed Pauli weight (b) three and (c) two. We compare arithmetic performed directly in the orbit basis with arithmetic on the corresponding ordinary Pauli expansions, the earlier $\gsim$ implementation, and the arithmetic and symmetry-merging routines associated with SymMerge. Dashed and solid lines indicate fits of the forms $\alpha n^\beta$ and $ab^n$, respectively, with the fitted exponents $\beta$ and exponential bases $b$ given in the respective colors. Each data point represents the mean runtime across ten input pairs with ten timing repetitions each, with error bars indicating the standard error of the mean across the input pairs. All measurements use a single thread, with the arithmetic of \PE\ and $\gsim$ invoked through Python and the Julia implementations measured natively. The software normalization conventions of Appendix~\ref{app:software_conventions} apply.}
    \label{fig:orbits-commutators-t1}
\end{figure*}

For permutation-invariant operators, the number of Pauli strings represented by an individual orbit can grow exponentially with the register size, even though the orbit itself is specified by only three integers. The consequences of this distinction are evident in~\cref{fig:orbits-commutators-t1}. For dense orbits, the arithmetic on ordinary Pauli expansions exhibits exponential growth in runtime, which is also observed for the SymMerge implementation, where symmetry-equivalent terms are combined after the arithmetic. By performing the arithmetic directly in the orbit basis, we replace the enumeration of exponentially many pairs of Pauli strings with the enumeration of contingency tables, for which we derived a general upper bound of $\mathcal{O}(n^9)$ arithmetic operations in Appendix~\ref{app:pauli_orbit_product} (see also~\cref{tab:representations_operations}). For the fully occupied input orbits considered here, the absence of identity factors reduces the contingency table to a $3\times3$ table with prescribed row and column sums, leaving only four independent entries and hence at most $\mathcal{O}(n^4)$ admissible configurations. Both our implementation and $\gsim$ exhibit polynomial runtime growth, with fitted exponents of $\beta=2.94$ and $\beta=3.53$, respectively, consistent with this reduced complexity.

For orbits of fixed Pauli weight $k$, the number of strings in an ordinary Pauli expansion grows as $\mathcal{O}(n^k)$, as follows from the orbit cardinality introduced in~\cref{ssec:PauliOrbits}. Consequently, pairwise arithmetic between two such expansions can involve $\mathcal{O}(n^{2k})$ pairs of strings. In the orbit representation, however, the marginal constraints in~\cref{eq:orbit_contingency_constraints} bound the number of admissible contingency tables independently of $n$, yielding an arithmetic complexity of $\mathcal{O}(1)$ after precomputation of the combinatorial factors (see Appendix~\ref{app:pauli_orbit_product}). This agrees with the nearly constant commutator runtimes observed for both weight-three and weight-two orbits in~\cref{fig:orbits-commutators-t1}(b,c). Our implementation further reduces the runtime relative to the original $\gsim$ implementation, evaluating these commutators within a few microseconds throughout the considered range of register sizes.

The corresponding multiplication benchmarks exhibit similar advantages of retaining the symmetry-adapted representations, in accordance with the arithmetic complexities summarized in~\cref{tab:representations_operations}. In Appendix~\ref{app:symmetry_benchmarks}, we additionally investigate the memory requirements of arithmetic in the different representations and the computational benefit of directly evaluating individual commutator coefficients in the Pauli-orbit basis.

\section{Applications}
\label{sec:applications}

Many approaches to the classical simulation and analysis of quantum systems rely on repeated operator arithmetic. In the following, we discuss how \PE\ can serve as a computational backend for Lie-algebraic simulation methods (\cref{ssec:applications_lie}), observable propagation through quantum circuits and Hamiltonian dynamics (\cref{ssec:applications_propagation}), and operator-based calculations in variational quantum chemistry and quantum control (\cref{ssec:applications_variational}). By exploiting the sparsity, locality, and symmetries of the operators involved, the arithmetic algorithms developed in~\cref{sec:operatorrepresentations} can substantially extend the range of systems accessible to these methods.

Earlier versions of \PE\ have already been employed in the preprocessing of Lie-algebraic classical simulations ($\gsim$)~\cite{Barligea.2026-EnablingLieAlgebraicClassical} and included in independent comparisons of Pauli arithmetic conducted in the context of PauLie~\cite{QPauLie.2026-PaulistringComparison} and PauLIB~\cite{Krotz.2026-PauLIBHighperformanceLibrary}. Further application contexts in which \PE\ has been discussed include Pauli propagation for ground-state calculations~\cite{Shrikhande.2025-PauliPropagationEnables}, fault-tolerant circuit simulation~\cite{Fang.2026-SymFTUniversalFaulttolerant}, operator-centric eigensolvers~\cite{Utama.2026-OperatorcentricCliffordAlgebra,Utama.2026-CliffordqcPythonToolkit}, quantum chemistry~\cite{Scali.2026-OpenQARPModularFramework}, quantum optimal control~\cite{Mishra.2026-AnalyticalSeriesExpansion}, and the analysis of Pauli commutation relations~\cite{Cha.2026-CountingAnticommutingPauli}. We also envisage integrating \PE\ into the quantum chemistry framework Tequila~\cite{Kottmann.2021-TequilaPlatformRapid} as an alternative backend for the operator arithmetic currently provided by OpenFermion~\cite{McClean.2020-OpenFermionElectronicStructure}.

\subsection{Lie-algebraic classical simulation methods}
\label{ssec:applications_lie}

Lie-algebraic simulation methods exploit the fact that the evolution of certain observables can be described within an operator space of considerably smaller dimension than the full operator space~\cite{Somma.2006-EfficientSolvabilityHamiltonians,Goh.2025-LiealgebraicClassicalSimulations,Barligea.2026-EnablingLieAlgebraicClassical}. A natural starting point for identifying such a space is the dynamical Lie algebra (DLA) generated by the Hamiltonians $\mathcal G=\{H_1,\ldots,H_m\}$ governing the evolution,
\begin{equation}
    \mathfrak g
    =\operatorname{Lie}_{\mathbb R}
    \langle iH_1,\ldots,iH_m\rangle.
    \label{eq:applications_dla}
\end{equation}
We can construct a basis of $\mathfrak g$ by successively evaluating commutators between the generators and previously obtained basis elements, adding linearly independent operators until the space is closed under commutation~\cite{Zeier.2011-SymmetryPrinciplesQuantum,Schirmer.2001-CompleteControllabilityQuantum,Iiyama.2025-FastNumericalGeneration}. Both the commutator arithmetic and the inner products required for testing linear independence can be evaluated using \PE.

When the generators are individual Pauli strings, every nonvanishing commutator is proportional to another Pauli string (see~\cref{eq:pauli_commutator}). The resulting Pauli Lie algebras have been studied and classified extensively~\cite{Wiersema.2024-ClassificationDynamicalLie,Kokcu.2024-ClassificationDynamicalLie,Aguilar.2024-FullClassificationPauli,Smith.2025-OptimallyGeneratingSu,Gargiulo.2026-PauliStringsQuantum}, and can be identified directly from their generators using dedicated tools such as PauLie~\cite{Shaya.2026-PauLieFastClassification}. Since distinct Pauli strings are linearly independent, constructing the Lie closure reduces to generating and identifying new Pauli labels, which form the orbit of the generators under the associated group of Clifford transvections~\cite{Gargiulo.2026-PauliStringsQuantum}. The coefficient-free representation introduced in~\cref{sec:software_interface} allows us to construct such Lie-algebra bases without the overhead of performing scalar coefficient arithmetic.

To simulate the evolution of observables, we express the action of a Hamiltonian $H$ on the basis $\mathcal B=\{B_\alpha\}_{\alpha=1}^{d}$ of an invariant Hermitian operator space through the matrix $\mathsf A_H$,
\begin{equation}
    i[H,B_\alpha]
    =\sum_{\beta=1}^{d}
    (\mathsf A_H)_{\alpha\beta}B_\beta.
    \label{eq:applications_adjoint_action}
\end{equation}
The evolution of an observable can then be computed through matrix exponentiation in this $d$-dimensional space. If $d$ is sufficiently small and the initial-state expectation values of the basis operators are accessible, this enables efficient classical simulation~\cite{Somma.2006-EfficientSolvabilityHamiltonians,Goh.2025-LiealgebraicClassicalSimulations,Barligea.2026-EnablingLieAlgebraicClassical}. In particular, each entry of $\mathsf A_H$ is a commutator coefficient, which, for the operator bases supported by \PE, can be evaluated individually using the targeted arithmetic introduced in~\cref{sec:shared_operations}.

The computational advantages become particularly pronounced for systems with symmetries, as demonstrated in Ref.~\cite{Barligea.2026-EnablingLieAlgebraicClassical} (see also Refs.~\cite{Anschuetz.2023-EfficientClassicalAlgorithms,Chang.2026-PracticalFrameworkSimulating} for permutation-symmetric systems). Using Pauli cycles and Pauli orbits, we can perform the Lie-algebraic preprocessing directly in symmetry-adapted representations, without constructing the potentially exponentially large ordinary Pauli expansions of the basis operators. Together with the reduced arithmetic complexities for low-weight operators (see~\cref{tab:representations_operations}), this makes \PE\ particularly suited to the Lie-algebraic simulation of large interacting quantum systems. 
In Appendix~\ref{app:applications_lie}, we demonstrate the computational advantages of our coefficient-free arithmetic by explicitly constructing dynamical Lie algebras comprising more than 300,000 basis elements for systems of up to $400$ qubits.

\subsection{Observable propagation methods}
\label{ssec:applications_propagation}

An alternative approach to simulating quantum dynamics is to work in the Heisenberg picture, propagating an observable $O$ under a unitary evolution $U$ and subsequently evaluating its expectation value with respect to the initial state $\rho_{\mathrm{in}}$. By representing the evolving observable as a Pauli expansion, we can perform this propagation through successive operator transformations~\cite{Rall.2019-SimulationQubitQuantum,Begusic.2023-SimulatingQuantumCircuit,Begusic.2024-FastConvergedClassical,Fontana.2023-ClassicalSimulationsNoisy,Rudolph.2025-PauliPropagationComputational}. The computational feasibility of this approach depends on how efficiently we can manipulate the expansion and control its growth throughout the evolution\footnote{Dedicated implementations of this approach include PauliPropagation.jl~\cite{Rudolph.2025-PauliPropagationComputational}, pauli-prop~\cite{IBM.2026-PauliProp}, cuPauliProp~\cite{NVIDIA.2025-CuPauliProp}, spd~\cite{Begusic.2026-Spd}, and monoprop~\cite{Miller.2026-MonopropScalingPauli}.}.

\subsubsection{Exact Clifford and near-Clifford propagation}
\label{sssec:applications_clifford}

To understand the elementary operations involved in Pauli propagation, consider the conjugation of a Pauli string $Q$ by a Pauli rotation $R_P(\theta)=e^{-i\theta P/2}$. Using the commutation relations of Pauli strings, we obtain
\begin{equation}
    R_P(\theta)^\dagger Q R_P(\theta)
    =
    \begin{cases}
        Q, & [P,Q]=0,\\
        \cos(\theta)Q+i\sin(\theta)PQ,
        & \{P,Q\}=0.
    \end{cases}
    \label{eq:applications_pauli_propagation}
\end{equation}
For Clifford rotations, the transformed operator is again proportional to a single Pauli string, such that propagation does not increase the number of terms in the observable~\cite{Gottesman.1998-HeisenbergRepresentationQuantum,Aaronson.2004-ImprovedSimulationStabilizer}; in the binary picture, Pauli rotations through angles of $\pm\pi/2$ act as symplectic transvections~\cite{Gargiulo.2026-PauliStringsQuantum}. Non-Clifford rotations, such as those corresponding to $T$ gates, can instead generate two contributions from each anticommuting string. Consequently, circuits containing relatively few non-Clifford gates can often be simulated efficiently by retaining the complete Pauli expansion~\cite{Rall.2019-SimulationQubitQuantum,Begusic.2023-SimulatingQuantumCircuit}, in line with dedicated near-Clifford simulators whose costs grow exponentially only in the number of non-Clifford gates~\cite{Bravyi.2016-ImprovedClassicalSimulation,Bravyi.2019-SimulationQuantumCircuits,Kissinger.2022-SimulatingQuantumCircuits,Chase.2026-ClifftFastExact,Haenel.2026-TsimFastUniversal}.

The propagation procedure thus repeatedly employs the commutation tests and phase-aware Pauli multiplication introduced in~\cref{ssec:PauliStrings}, together with the accumulation of contributions sharing the same Pauli label. 
The computational advantages of employing \PE\ as the arithmetic backend for this procedure are illustrated in Appendix~\ref{app:applications_clifford}, where we benchmark the exact propagation of observables through 64-qubit Clifford circuits with varying numbers of non-Clifford rotations.

\subsubsection{Approximate Heisenberg propagation}
\label{sssec:applications_approximate_propagation}

For more general quantum dynamics, the Pauli expansion of an evolving observable can grow exponentially, rendering exact propagation computationally infeasible. Approximate propagation methods address this growth by retaining selected contributions, for example, according to their coefficient magnitudes or Pauli weights~\cite{Rakovszky.2020-DissipationassistedOperatorEvolution,Fontana.2023-ClassicalSimulationsNoisy,Shao.2023-SimulatingNoisyVariational,Begusic.2024-FastConvergedClassical,Schuster.2024-PolynomialtimeClassicalAlgorithm,Angrisani.2025-ClassicallyEstimatingObservables,Begusic.2025-RealtimeOperatorEvolution,Rudolph.2025-PauliPropagationComputational}. For continuous-time evolution under a time-independent Hamiltonian $H$, the observable can be expressed through a series of nested commutators~\cite{Parker.2019-UniversalOperatorGrowth,Lychkovskiy.2026-QCommuteToolSymbolic},
\begin{equation}
    O(t)=\sum_{k=0}^{\infty}
    \frac{t^k}{k!}\operatorname{ad}_H^k(O),
    \qquad
    \operatorname{ad}_H(O):=i[H,O].
    \label{eq:applications_heisenberg_series}
\end{equation}
The successive terms of this expansion can be generated using the commutator arithmetic of \PE. For Hamiltonians composed of local interactions, the overlap-indexed commutator algorithm developed in Appendix~\ref{app:pauli_overlap_commutator} is particularly relevant, as only terms with overlapping supports can contribute at each step.

Further computational advantages arise when the dynamics preserve symmetries of the observable. In particular, observables evolving under Hamiltonians that share their translation or permutation symmetry remain within the respective invariant operator spaces. Using the arithmetic of Pauli cycles and Pauli orbits, we can therefore perform the propagation directly in these compressed bases~\cite{Bhandari.2026-ReadoutawareTheoryApproximate}. This complements approaches that exploit symmetry by merging equivalent Pauli strings during propagation~\cite{Teng.2025-LeveragingSymmetryMerging}, and allows us to combine symmetry-based compression with approximation strategies for controlling the growth of the operator expansion. 
In Appendix~\ref{app:applications_symmetric_propagation}, we demonstrate the advantages of this approach by propagating a permutation-invariant observable under the dynamics of the Lipkin--Meshkov--Glick model~\cite{Lipkin.1965-ValidityManybodyApproximation} for systems of up to 100 qubits.

\subsection{Variational quantum chemistry and quantum control}
\label{ssec:applications_variational}

Beyond the simulation of quantum dynamics, operator arithmetic constitutes an important part of many variational methods~\cite{Peruzzo.2014-VariationalEigenvalueSolver,Cerezo.2021-VariationalQuantumAlgorithms,Tilly.2022-VariationalQuantumEigensolver}. In quantum chemistry, for example, the transformation of fermionic operators into Pauli expansions~\cite{Jordan.1928-UberPaulischeAquivalenzverbot,Bravyi.2002-FermionicQuantumComputation,Seeley.2012-BravyikitaevTransformationQuantum} provides a natural setting for employing our arithmetic routines~\cite{McArdle.2020-QuantumComputationalChemistry,McClean.2020-OpenFermionElectronicStructure,Kottmann.2021-TequilaPlatformRapid}. In adaptive variational algorithms~\cite{Grimsley.2019-AdaptiveVariationalAlgorithma,Tang.2021-QubitADAPTVQEAdaptiveAlgorithm,Ryabinkin.2020-IterativeQubitCoupled}, we select suitable excitation operators from a predefined pool by evaluating the energy gradients associated with the individual excitations. For an anti-Hermitian excitation generator $A_\mu$, a molecular Hamiltonian $H$, and the current variational state $\ket{\psi}$, the corresponding gradient is given by
\begin{equation}
    g_\mu
    =\left.\frac{\partial}{\partial\theta}
    \bra{\psi}e^{-\theta A_\mu}He^{\theta A_\mu}\ket{\psi}
    \right|_{\theta=0}
    =\bra{\psi}[H,A_\mu]\ket{\psi}.
    \label{eq:applications_variational_gradient}
\end{equation}
Screening a large pool of excitations thus requires the repeated evaluation of commutators, for which we can employ the sparse Pauli-sum arithmetic of \PE. 
We demonstrate the utility of \PE\ for this task in Appendix~\ref{app:applications_variational}, where we investigate the gradient screening of excitation pools containing several thousand operators and show how the structure of the reference state can be exploited to further accelerate the arithmetic.

The symbolic coefficient arithmetic introduced in~\cref{sec:software_additional} is particularly useful when the same operator calculations are required for varying parameter values. Consider, for example, a parameterized Hamiltonian $H(\boldsymbol{\theta})=H_0+\sum_\mu\theta_\mu H_\mu$. In quantum optimal control~\cite{Khaneja.2005-OptimalControlCoupled,Glaser.2015-TrainingSchrodingersCat}, expansions in nested commutators of the Hamiltonian components can be used to derive analytical gradients~\cite{Mishra.2026-AnalyticalSeriesExpansion}. By retaining the parameters symbolically throughout the commutator arithmetic, we can construct such operator expressions once, differentiate their polynomial coefficients, and subsequently evaluate them for different parameter configurations.

Further applications can be envisaged wherever the algebraic structure of quantum operators is exploited, including the construction of commutation graphs for measurement grouping~\cite{Verteletskyi.2020-MeasurementOptimizationVariational,Gokhale.2019-MeasurementCostVariational,Yen.2020-MeasuringAllCompatible,Reggio.2024-FastPartitioningPauli, Bincoletto.2026-PhysicsinformedMeasurementProtocol}, Pauli-based circuit compilation~\cite{Bravyi.2016-TradingClassicalQuantum,Litinski.2019-GameSurfaceCodes,Zhang.2019-OptimizingTGates,Cowtan.2020-PhaseGadgetSynthesis,Paykin.2023-PCOASTPaulibasedQuantum,Peres.2023-QuantumCircuitCompilation}, and the analytical manipulation of parameterized quantum models~\cite{Schuld.2021-EffectDataEncoding,Wierichs.2022-GeneralParametershiftRules,Nemkov.2023-FourierExpansionVariational,Fontana.2023-ClassicalSimulationsNoisy}.

\section{Discussion and Conclusion}
\label{sec:conclusion}

With \PE, we have developed a general-purpose, high-performance computational backend for quantum operator algebra that combines compact operator representations, representation-specific arithmetic algorithms, and a unified software interface. Our implementation provides numerical and symbolic arithmetic for ordinary Pauli operators, translation-invariant Pauli cycles, and permutation-invariant Pauli orbits, making their respective algebraic structures directly accessible to higher-level computational methods through a common Python interface.

Beyond optimizing elementary Pauli-string arithmetic, we have developed algorithms that exploit the structure of entire operator expansions and the requirements of individual arithmetic tasks. Our adaptive storage mechanism combines efficient bitwise arithmetic on packed binary signatures with sparse site arithmetic, whose computational cost for elementary operations on bounded-weight Pauli strings is independent of the register span. Our partitioned multiplication algorithms allow contributions to be accumulated efficiently in both storage formats, while overlap indexing reduces the number of candidate pairs for commutators between local Hamiltonians from quadratic to linear in the system size. For permutation-invariant operators, arithmetic between potentially exponentially large Pauli expansions reduces to polynomial-time calculations in the Pauli orbit basis, and individual coefficients of commutators between elementary orbits can be evaluated in constant time for targets of fixed bounded Pauli weight after combinatorial precomputation (see~\cref{tab:representations_operations}).

Our comprehensive numerical benchmarks against a broad range of general-purpose and specialized quantum software packages demonstrate leading performance across most of the investigated arithmetic workloads, with runtime improvements of several orders of magnitude over many existing implementations. While the exponential growth of operator expansions remains a fundamental challenge in classical simulation, our results demonstrate how exploiting sparsity, locality, and symmetries can substantially extend the range of computationally accessible systems. The significance of these advances becomes apparent in the application-level calculations presented in Appendix~\ref{app:applications}, where we construct dynamical Lie algebras comprising more than 300,000 basis elements on 400 qubits, perform exact near-Clifford observable propagation, simulate permutation-invariant dynamics on 100 qubits, and efficiently screen excitation pools containing several thousand operators. By making the underlying arithmetic available through a common software interface, these computational advances become directly accessible to a wide range of existing and prospective classical simulation methods.

The computational advantages demonstrated in this work motivate the extension of \PE\ to operator representations adapted to further symmetries, particle statistics, and restricted Hilbert-space sectors. The mathematical and computational framework developed here provides a systematic approach to integrating such representations and exploiting their algebraic structure through specialized arithmetic algorithms. Much as standardized linear-algebra libraries such as BLAS and LAPACK~\cite{Lawson.1979-BasicLinearAlgebra,Anderson.1999-LAPACKUsersGuide} provide the computational foundation for numerical scientific computing, we envision \PE\ as a reusable arithmetic foundation for quantum software, allowing researchers to concentrate on the development of new simulation methods while benefiting from continued advances in the efficiency and scalability of the underlying operator arithmetic.

\stopmaintoc

\section*{Acknowledgements}
J.S.K.\ acknowledges funding from the Hightech Agenda Bayern, and L.M.\ and A.B.\ acknowledge funding from the German Federal Ministry of Research, Technology, and Space (BMFTR) through the Quantum Technologies programme (grant 13N17231, HoliQC2). A.B.\ further acknowledges support through the Munich Quantum Valley (MQV) Doctoral Fellowship, funded by the Bavarian state government through the Hightech Agenda Bayern.
We thank Craig Gidney (Google Quantum AI) for contributing a faster single-string multiplication kernel based on the bitwise phase counters of Stim~\cite{Gidney.2021-StimFastStabilizer}, and Ludmila Botelho and Roberto Di Remigio Eik{\aa}s (Algorithmiq) for reorganizing the build system, development environment, wheel builds, and Python bindings of \PE. We thank Konstantin Golovkin and the PauLie team for including \PE\ in their public performance comparison~\cite{QPauLie.2026-PaulistringComparison} and for feedback motivating faster string construction and coefficient-free Pauli strings. We further thank Florian Kr\"otz (LMU Munich) for discussions on high-performance Pauli arithmetic, Philipp Greiner for supporting the development of the Lie-closure routines, Moritz Willmann (Fraunhofer IPA) for reporting installation issues, and Oliver H\"uttenhofer for fruitful discussions.
We gratefully acknowledge the computational resources provided by the LiCCA HPC cluster at the University of Augsburg, co-funded by the Deutsche Forschungsgemeinschaft (DFG, German Research Foundation) -- Project-ID 499211671.

\bibliography{literature}

\revtexseparator
\appendix
\section*{Appendix}
\appendixtoc

\section{Pauli-string arithmetic}
\label{app:pauli_arithmetic}

\subsection{Binary multiplication and phase reconstruction}
\label{app:pauli_phase_reconstruction}

Our phase reconstruction algorithm in~\cref{eq:pauli_multiplication} adapts the bitwise-counter algorithm developed by C.~Gidney for Stim~\cite[Fig.~2]{Gidney.2021-StimFastStabilizer} to the binary encoding of \PE. For two Pauli strings $P,Q\in\mathcal P_n$ with signatures $v(P)=(\boldsymbol x,\boldsymbol y)$ and $v(Q)=(\boldsymbol x',\boldsymbol y')$, the product $PQ=\omega(P,Q)R$ has the output signature
\begin{equation}
    v(R)=(\boldsymbol x'',\boldsymbol y'')
    =(\boldsymbol x\oplus\boldsymbol x',
      \boldsymbol y\oplus\boldsymbol y').
\end{equation}
The remaining phase $\omega(P,Q)$ is determined by the local multiplication rules summarized via:
\begin{equation}
\begin{array}{c|c|c|c}
(P_j,Q_j)&(x_j,y_j)&(x_j',y_j')&\omega_j\\
\hline
(X,Y)&(1,0)&(0,1)&+i\\
(Y,Z)&(0,1)&(1,1)&+i\\
(Z,X)&(1,1)&(1,0)&+i\\
(Y,X)&(0,1)&(1,0)&-i\\
(Z,Y)&(1,1)&(0,1)&-i\\
(X,Z)&(1,0)&(1,1)&-i
\end{array}
\label{eq:pauli_local_phase_table}
\end{equation}

We process the binary signatures in $L_{\mathrm{word}}=\lceil n/w\rceil$ words of $w$ bits, padding unused positions with zeros. For each word $k=0,\ldots,L_{\mathrm{word}}-1$, we construct the auxiliary bit masks
\begin{equation}
\begin{aligned}
    \boldsymbol u_k
        &=\boldsymbol x_k\land\boldsymbol y_k',\\
    \boldsymbol m_k
        &=(\boldsymbol x_k'\land\boldsymbol y_k)
          \oplus\boldsymbol u_k,\\
    \boldsymbol d_k
        &=\boldsymbol x_k''\oplus\boldsymbol y_k''
          \oplus\boldsymbol u_k,
\end{aligned}
\label{eq:pauli_phase_word_masks}
\end{equation}
where $\boldsymbol m_k$ identifies locally anticommuting Pauli factors and $\boldsymbol d_k$ distinguishes their phases according to~\cref{eq:pauli_local_phase_table}. At bit position $r$, the local phase exponent is
\begin{equation}
    \epsilon_{k,r}=m_{k,r}(2d_{k,r}-1)
    \in\{-1,0,1\}.
    \label{eq:pauli_local_phase_exponent}
\end{equation}
This local exponent coincides with the phase function $g$ of the stabilizer tableau algorithm of Aaronson and Gottesman~\cite{Aaronson.2004-ImprovedSimulationStabilizer}, here evaluated for $w$ sites at once through bitwise operations. 
To accumulate these exponents modulo four without performing population counts for each word, we use two $w$-bit counter registers initialized to zero, $\boldsymbol c_1^{(0)}=\boldsymbol c_2^{(0)}=\boldsymbol 0$, and updated according to
\begin{equation}
\begin{aligned}
    \boldsymbol c_2^{(k+1)}
        &=\boldsymbol c_2^{(k)}\oplus
          \bigl[(\boldsymbol c_1^{(k)}\oplus\boldsymbol d_k)
          \land\boldsymbol m_k\bigr],\\
    \boldsymbol c_1^{(k+1)}
        &=\boldsymbol c_1^{(k)}\oplus\boldsymbol m_k.
\end{aligned}
\label{eq:F1F2}
\end{equation}
Each bit position $r$ acts as an independent counter across successive words, satisfying the invariant
\begin{equation}
    2c_{2,r}^{(k)}-c_{1,r}^{(k)}
    \equiv\sum_{s=0}^{k-1}\epsilon_{s,r}\pmod4.
    \label{eq:pauli_counter_invariant}
\end{equation}
Indeed, if $m_{k,r}=0$, both counter bits remain unchanged. Otherwise, the updates in~\cref{eq:F1F2} change the represented residue by $2d_{k,r}-1$ modulo four, establishing the invariant by induction. Summing over all bit positions then yields the global phase through only two population counts,
\begin{equation}
\begin{aligned}
    \widehat e(P,Q)
        &=2\operatorname{popcount}
          \bigl(\boldsymbol c_2^{(L_{\mathrm{word}})}\bigr)\\
        &\quad-\operatorname{popcount}
          \bigl(\boldsymbol c_1^{(L_{\mathrm{word}})}\bigr),\\
    e(P,Q)&=\widehat e(P,Q)\bmod4,\\
    \omega(P,Q)&=i^{e(P,Q)}.
\end{aligned}
\label{eq:pauli_phase_reconstruction}
\end{equation}
This reproduces the phase exponent $e(P,Q)\equiv\nu_+(P,Q)-\nu_-(P,Q)\pmod4$ introduced in~\cref{eq:pauli_phase}.

The negative contribution of $\boldsymbol c_1$ in~\cref{eq:pauli_phase_reconstruction} distinguishes our implementation from the original Stim algorithm~\cite[Fig.~2]{Gidney.2021-StimFastStabilizer} and follows from the different binary encodings, related through $(\boldsymbol\xi,\boldsymbol\zeta)=(\boldsymbol x\oplus\boldsymbol y,\boldsymbol y)$, where $(\boldsymbol\xi,\boldsymbol\zeta)$ denotes the conventional $X/Z$ signature.

In practice, phase accumulation is restricted to words present in both operands, since missing words represent identities. The remaining output words are copied from the longer operand. A complete example of the counter updates and phase reconstruction is illustrated in~\cref{fig:pauli_phase_counter}.

\begin{figure}[t]
    \centering
    \begin{tikzpicture}[
        show background rectangle,
        background rectangle/.style={
            fill=black!3,
            rounded corners=5pt
        },
        font=\footnotesize,
        stage/.style={
            font=\footnotesize\bfseries,
            anchor=west
        },
        lab/.style={
            font=\scriptsize,
            text=black!65
        },
        qb/.style={
            draw=blue!55!black,
            fill=blue!8,
            rounded corners=1pt,
            minimum width=0.54cm,
            minimum height=0.45cm,
            inner sep=0pt
        },
        state/.style={
            draw=teal!55!black,
            fill=teal!5,
            rounded corners=2pt,
            minimum width=1.75cm,
            minimum height=1.08cm,
            align=center
        },
        result/.style={
            rounded corners=2pt,
            minimum width=3.0cm,
            minimum height=0.65cm,
            align=center,
            inner sep=4pt
        },
        arr/.style={
            -{Stealth[length=1.6mm]},
            line width=0.5pt,
            black!65
        }
    ]

    \node[stage] at (0,0) {1.\ Local phase contributions};

    \node[lab] at (2.1,-0.43) {word 0};
    \node[lab] at (4.6,-0.43) {word 1};

    \node at (0.45,-0.95) {$P$};
    \node at (0.45,-1.52) {$Q$};
    \node at (0.45,-2.0) {$\epsilon$};

    \foreach \x/\p/\q/\phase in {
        1.7/X/Y/{+1},
        2.5/X/Z/{-1},
        4.2/X/Y/{+1},
        5.0/I/Z/0
    } {
        \node[qb] at (\x,-0.95) {\p};
        \node[qb] at (\x,-1.52) {\q};
        \node[lab] at (\x,-2.0) {$\phase$};
    }

    \node[stage] at (0,-2.55)
        {2.\ Accumulate in two counter registers};

    \node[state] (c0) at (1.05,-3.52)
        {$\boldsymbol c_1=(0,0)$\\
         $\boldsymbol c_2=(0,0)$};

    \node[state] (c1) at (3.55,-3.52)
        {$\boldsymbol c_1=(1,1)$\\
         $\boldsymbol c_2=(1,0)$};

    \node[state] (c2) at (6.05,-3.52)
        {$\boldsymbol c_1=(0,1)$\\
         $\boldsymbol c_2=(1,0)$};

    \draw[arr] (c0) -- (c1);
    \draw[arr] (c1) -- (c2);

    \node[lab] at (1.05,-4.22) {initial};
    \node[lab] at (3.55,-4.22) {after word 0};
    \node[lab] at (6.05,-4.22) {after word 1};

    \node[stage] at (0,-4.8)
        {3.\ Recover the phase and commutator};

    \node at (3.55,-5.48) {
        $\begin{aligned}
            e&\equiv
            2\operatorname{popcount}(\boldsymbol c_2)
            -\operatorname{popcount}(\boldsymbol c_1)\\
             &\equiv 2\cdot1-1
             \equiv 1\pmod4.
        \end{aligned}$
    };
    \node[stage] at (0,-6.38) {$\boldsymbol{\Rightarrow}$};
    
    \node[
        result,
        draw=black!12,
        fill=white
    ] at (2.15,-6.38) {$PQ=i\,ZYZZ$};

    \node[
        result,
        draw=black!12,
        fill=white
    ] at (5.25,-6.38) {$[P,Q]=2i\,ZYZZ$};

    \end{tikzpicture}
    \caption{Bitwise phase reconstruction for the product of $P=XXXI$ and $Q=YZYZ$, illustrated using two-bit words ($w=2$). 1.~The local phase exponents are grouped into words according to~\cref{eq:pauli_local_phase_exponent}. 2.~Two counter registers accumulate the exponents modulo four, with each tuple representing two independent counter lanes (\cref{eq:F1F2}). 3.~Two population counts recover the global exponent $e(P,Q)=1$ and hence the product $PQ=iZYZZ$ (\cref{eq:pauli_phase_reconstruction}). The odd exponent also establishes anticommutation, yielding $[P,Q]=2iZYZZ$.}
    \label{fig:pauli_phase_counter}
\end{figure}

\subsection{Commutation and computational complexity}
\label{app:pauli_commutation}

Since $\boldsymbol m_k$ identifies the locally anticommuting Pauli factors, the commutation parity is obtained through
\begin{equation}
\begin{aligned}
    \chi(P,Q)
        &=\operatorname{popcount}\!\left(
          \bigoplus_{k=0}^{L_{\mathrm{word}}-1}
          \boldsymbol m_k\right)\bmod2\\
        &=\operatorname{popcount}\!\left(
          \boldsymbol c_1^{(L_{\mathrm{word}})}\right)\bmod2\\
        &=e(P,Q)\bmod2,
\end{aligned}
\label{eq:pauli_word_commutation_parity}
\end{equation}
where the last two identities follow from~\cref{eq:F1F2,eq:pauli_phase_reconstruction}. For commutator evaluation, we first determine this parity using a single XOR accumulator. Only if $\chi(P,Q)=1$ do we evaluate the product and double its coefficient, according to~\cref{eq:pauli_commutator}. The coefficient-free interface omits this factor of two, as discussed in~\cref{app:software_conventions}.

For the packed binary representation with machine-word width $w=64$, each word requires a constant number of bitwise operations. Multiplication and commutator evaluation therefore both have complexity $\mathcal O(L_{\mathrm{word}})$, with $\mathcal O(1)$ auxiliary words for phase accumulation and $\mathcal O(L_{\mathrm{word}})$ storage for the output signature. The corresponding algorithms for sparse site storage are developed in~\cref{app:sparse_pauli_arithmetic}.

\subsection{Prefix-partitioned arithmetic of Pauli sums}
\label{app:pauli_sum_product}

\paragraph{Signature projection and output routing.}

Consider two nonempty Pauli sums $A=\sum_{P\in S_A}a_PP$ and $B=\sum_{Q\in S_B}b_QQ$ with complex coefficients. To obtain a compact representation of their product signatures, we first identify the signature coordinates that vary within at least one of the input sums through the masks
\begin{equation}
\begin{aligned}
    V_A&=\left(\bigvee_{P\in S_A}v(P)\right)
          \oplus\left(\bigwedge_{P\in S_A}v(P)\right),\\
    V_B&=\left(\bigvee_{Q\in S_B}v(Q)\right)
          \oplus\left(\bigwedge_{Q\in S_B}v(Q)\right),\\
    V&=V_A\lor V_B,
\end{aligned}
\label{eq:pauli_varying_coordinates}
\end{equation}
where the Boolean operations act coordinatewise and shorter signatures are padded with zeros. Let $s=\operatorname{popcount}(V)$ and let $\pi_V:\mathbb F_2^{2n}\rightarrow\mathbb F_2^s$ denote the projection onto the retained coordinates, preserving their order. Since every omitted coordinate is constant within each input sum, its value in the product signature is fixed by the XOR of the two input constants. The projection therefore preserves both the uniqueness and the lexicographic order of all attainable output signatures. Furthermore, for $PQ=\omega(P,Q)R$, we have
\begin{equation}
    \pi_V(v(R))
    =\pi_V(v(P))\oplus\pi_V(v(Q)).
\end{equation}

Let $h_d(P)$ denote the first $d$ bits of $\pi_V(v(P))$. For prefixes $\alpha,\beta\in\mathbb F_2^d$, we define the subsets
\begin{equation}
\begin{aligned}
    S_{A,\alpha}^{(d)}
        &=\{P\in S_A:h_d(P)=\alpha\},\\
    S_{B,\beta}^{(d)}
        &=\{Q\in S_B:h_d(Q)=\beta\},
\end{aligned}
\end{equation}
which form contiguous ranges in the sorted input sums. The XOR relation in~\cref{eq:pauli_prefix_routing} then identifies the set of input pairs contributing to a given output prefix $\rho$ as the disjoint union
\begin{equation}
    \mathcal J_\rho^{(d)}
    =\bigcup_{\alpha\oplus\beta=\rho}
        S_{A,\alpha}^{(d)}\times S_{B,\beta}^{(d)}.
    \label{eq:pauli_prefix_blocks}
\end{equation}
Each Cartesian product is represented by its two input ranges, allowing us to partition the product space without evaluating individual Pauli products.

\paragraph{Adaptive refinement and accumulation.}

We recursively refine the partitions in~\cref{eq:pauli_prefix_blocks} by considering additional signature bits until their respective workloads are suitable for local coefficient accumulation. As illustrated in~\cref{fig:prefix-partitioned}, this allows partitions to terminate at different prefix depths. Recursive splitting on leading key bits, followed by concatenating the buckets in key order, follows the most-significant-digit-first principle of radix sorting~\cite[Sec.~5.2.5]{Knuth.1998-ArtComputerProgramming}. Here, however, the buckets consist of pairs of input ranges and are formed before any output key is constructed. Let $\mathcal T$ denote the resulting set of leaf partitions and $T_\tau$ the accumulated operator associated with leaf $\tau$. Since every input pair belongs to exactly one leaf and pairs producing identical output signatures follow the same refinement path, we obtain
\begin{equation}
\begin{aligned}
    AB&=\sum_{\tau\in\mathcal T}T_\tau,\\
    S_{T_\tau}\cap S_{T_{\tau'}}
        &=\varnothing
        \qquad(\tau\ne\tau'),
\end{aligned}
\label{eq:pauli_disjoint_leaf_results}
\end{equation}
where $S_{T_\tau}$ denotes the set of nonzero output labels in $T_\tau$. Consequently, coefficient accumulation can be performed independently within each partition, and the locally sorted results can be concatenated in prefix order without further merging.

Within each leaf, we accumulate coefficients by sorting and merging compact product keys or, for sufficiently small remaining key spaces, by direct table accumulation. Product phases are evaluated from the original signatures according to~\cref{app:pauli_phase_reconstruction}, while full output signatures are reconstructed only for surviving terms using the original signatures of a contributing pair. For parallel execution, independent partitions are grouped into work units with separate scratch storage and reserved output ranges. The decomposition into work units and the local accumulation order are independent of the number of threads.

\paragraph{Commutators and squares.}

The same partitioning can be used for commutators by restricting the contributions to anticommuting pairs and multiplying their coefficients by two, according to~\cref{eq:pauli_commutator}. Squaring a Pauli sum admits an additional reduction, since $P^2=I$ and anticommuting cross terms cancel. We obtain
\begin{equation}
\begin{aligned}
    A^2={}&\left(\sum_{P\in S_A}a_P^2\right)I\\
    &+2\!\sum_{\substack{P,Q\in S_A,\;P<Q\\
                        \chi(P,Q)=0}}
       a_Pa_Q\,PQ,
\end{aligned}
\label{eq:pauli_partitioned_square}
\end{equation}
where $<$ denotes the fixed signature order. It is thus sufficient to examine each unordered pair of distinct Pauli strings once and retain only the commuting pairs.

\paragraph{Complexity and memory.}

Constructing the varying-coordinate mask and projected input signatures requires $\mathcal O((M_A+M_B)L_{\mathrm{word}})$ work. General multiplication still evaluates $M_AM_B$ input pairs, requiring $\mathcal O(M_AM_BL_{\mathrm{word}})$ operations for their compact keys and phases, while reconstructing the surviving signatures requires $\mathcal O(K_{\mathrm{out}}L_{\mathrm{word}})$ additional work. For a leaf $\tau$ containing $m_\tau$ product records, comparison-based sorting has a cost bounded by
\begin{equation}
    \mathcal O\bigl(
        L_{\mathrm{key}}m_\tau\log(m_\tau+1)
    \bigr),
\end{equation}
where $L_{\mathrm{key}}=\max\{1,\lceil s/w\rceil\}$ denotes the number of machine words per compact key. Direct table accumulation instead scales with the number of candidate pairs and the size of the remaining key space. The overall cost therefore depends on the distribution of output signatures, the partitioning overhead, and the local accumulation strategy.

The same key space also bounds the number of distinct output terms within each partition. For a work unit $u$ containing $N_u$ candidate pairs with $s-d_u$ unresolved key bits, the implementation reserves
\begin{equation}
    C_u=
    \begin{cases}
        \min\{N_u,2^{s-d_u}\}, & s-d_u<63,\\
        N_u, & s-d_u\geq63
    \end{cases}
    \label{eq:pauli_partition_capacity}
\end{equation}
output rows, where the second case avoids overflow in evaluating the size of the key space. The reserved capacity $\sum_u C_u$ can exceed $K_{\mathrm{out}}$, with unused rows removed after accumulation. While prefix partitioning avoids the need for a global array of pairwise product records and a subsequent global merge, the reserved output capacity can still grow quadratically with the input sizes.

For numerical Pauli sums stored through packed binary signatures, the implementation selects between direct arithmetic, prefix partitioning, and the overlap-indexed commutator algorithm described in~\cref{app:pauli_overlap_commutator}, depending on the input sizes and estimated workloads. Symbolic coefficients are handled through separate coefficient-processing paths. For operators stored through their occupied sites, we employ the hash-partitioned arithmetic developed in~\cref{app:sparse_pauli_arithmetic}.

\subsection{Sparse site storage and arithmetic}
\label{app:sparse_pauli_arithmetic}

\paragraph{Storage of sparsely supported Pauli strings.}

To represent only the occupied sites of a Pauli string $P$, we associate each site $j\in\operatorname{supp}(P)$ and its local Pauli factor $P_j$ with the integer
\begin{equation}
    u_j(P)=4j+\ell(P_j),
    \qquad
    \ell(X)=1,\quad \ell(Y)=2,\quad \ell(Z)=3.
    \label{eq:sparse_pauli_site_encoding}
\end{equation}
Using zero-based qubit indices, each entry can be stored in a 32-bit unsigned integer, with the two least significant bits encoding the local Pauli factor and the remaining bits encoding the site index. The entries of each Pauli string are stored in ascending order of their site indices.

For linear combinations of Pauli strings, we concatenate the individual site lists into a contiguous array and store the coefficients and offsets delimiting the lists in two additional arrays. Using the notation introduced in~\cref{eq:pauli_sum_mean_weight}, the site entries require $4W_A$ bytes, together with $8(M_A+1)$ bytes for the offsets. By comparison, packed binary signatures require $16M_A\lceil s_A/64\rceil$ bytes. These estimates exclude the coefficients, container metadata, and additional allocated capacity. The storage requirements of an individual sparse string thus depend only on its own Pauli weight, allowing us to efficiently represent sums containing both low- and high-weight strings.

\paragraph{Elementary multiplication and commutation.}

To evaluate the product of two strings $P$ and $Q$, we simultaneously traverse their ordered site lists. At sites occupied by only one of the strings, we retain the corresponding entry. At sites occupied by both strings, we combine their local Pauli factors through the XOR rule of~\cref{eq:pauli_multiplication} and accumulate the corresponding contribution to the multiplication phase according to~\cref{eq:pauli_local_phase_table}. Equal local Pauli factors cancel to identity, in which case no entry is generated.

The resulting product is again an ordered site list with at most $\operatorname{wt}(P)+\operatorname{wt}(Q)$ entries. Since each input entry is visited at most once during the merge, the label-processing cost is $\mathcal O(1+\operatorname{wt}(P)+\operatorname{wt}(Q))$, as stated in~\cref{eq:sparse_pauli_elementary_complexity}. The multiplication phase and commutation parity can also be evaluated without constructing the output list, allowing us to form the product in commutator calculations only when the input strings anticommute.

\begin{ExampleBox}{Arithmetic on sparsely supported Pauli strings}
Consider the Pauli strings
\begin{equation*}
    P=X_0Z_{4000},
    \qquad
    Q=Y_0Z_{17}.
\end{equation*}
Their occupied sites are represented through the ordered lists
\begin{equation*}
    \bigl((0,X),(4000,Z)\bigr),
    \qquad
    \bigl((0,Y),(17,Z)\bigr).
\end{equation*}
Only the factors at site zero need to be multiplied, yielding
\begin{equation*}
    PQ=iZ_0Z_{17}Z_{4000},
    \qquad
    [P,Q]=2iZ_0Z_{17}Z_{4000}.
\end{equation*}
For the representation of $P$ through packed binary signatures, each of the two binary vectors requires 63 machine words of 64 bits, corresponding to 1008 bytes in total. Using~\cref{eq:sparse_pauli_site_encoding}, the two occupied sites require only 8 bytes, excluding the coefficient and container metadata.
\end{ExampleBox}

\paragraph{Hash-partitioned multiplication of Pauli sums.}

To multiply two Pauli sums $A$ and $B$, we group the input pairs into rows, each corresponding to the products of one term of $A$ with all terms of $B$. We process these rows in blocks and accumulate their contributions successively, avoiding the simultaneous construction of all $M_AM_B$ elementary products.

Within each block, the workers generate product labels and coefficients in separate temporary buffers. The leading bits of the hash of each output label determine the partition to which its contribution is assigned. Each partition occupies a separate segment of a hash table, allowing the contributions of different partitions to be accumulated independently and in parallel. Hash collisions are resolved by comparing the complete site lists. After processing all blocks, we discard labels with vanishing coefficients and collect the remaining terms in the contiguous storage format described above.

Summing the elementary merge costs over all input pairs gives
\begin{equation}
\begin{aligned}
    &\sum_{P\in S_A}\sum_{Q\in S_B}
        \bigl(\operatorname{wt}(P)+\operatorname{wt}(Q)\bigr)\\
    &\qquad=M_BW_A+M_AW_B.
\end{aligned}
\end{equation}
Including the constant work associated with each pair and the traversal of site lists during hashing and label comparisons, we obtain the expected label-processing cost
\begin{equation}
    \mathcal O\bigl(M_AM_B+M_BW_A+M_AW_B\bigr),
    \label{eq:sparse_pauli_product_complexity}
\end{equation}
assuming expected constant hash-table probe counts. Expressing the total weights through their respective means recovers~\cref{eq:sparse_pauli_sum_complexity}. For symbolic coefficients, the additional cost of multiplying and accumulating their polynomial representations must be taken into account.

While blockwise processing limits the number of elementary products generated simultaneously, the hash tables retain all distinct labels encountered during accumulation, including those whose coefficients eventually cancel. The memory requirements therefore also depend on the number of intermediate labels generated throughout the arithmetic.

The same accumulation scheme applies to commutators and operator squares, exploiting the respective cancellations of commuting and anticommuting input pairs discussed in~\cref{app:pauli_sum_product,app:pauli_overlap_commutator}.

\paragraph{Adaptive selection of storage formats.}

To select a suitable storage format for a given Pauli sum $A$, we compare the costs of processing its packed binary signatures and its lists of occupied sites. In a simple cost model, these costs scale with $M_As_A/64$ and $W_A$, respectively. Accounting for the additional work associated with processing individual site entries through an empirical factor of four, we obtain the approximate crossover condition
\begin{equation}
    4W_A\lesssim M_A\frac{s_A}{64}.
\end{equation}
Accordingly, our default selection rule at construction is
\begin{equation}
    s_A\geq256,
    \qquad
    \overline w_A\leq\frac{s_A}{256}.
    \label{eq:sparse_pauli_selection}
\end{equation}
The minimum span retains the efficient bitwise arithmetic for sufficiently short strings. Both thresholds can be configured through the Python interface. For a single Pauli string, the mean weight is replaced by its Pauli weight.

Since the Pauli weights and span of an operator can change throughout a calculation, we additionally assess whether sparse arithmetic is suitable for the anticipated output. Before multiplying two operators, at least one of which uses sparse site storage, we use
\begin{equation}
\begin{aligned}
    s_{\mathrm{est}}&=\max(s_A,s_B),\\
    \overline w_{\mathrm{est}}
        &=\min(\overline w_A+\overline w_B,s_{\mathrm{est}})
\end{aligned}
    \label{eq:sparse_pauli_product_prediction}
\end{equation}
to estimate the span and mean weight of the generated products. If these quantities satisfy the sparse retention criterion, the arithmetic is performed in the sparse site representation. Otherwise, we convert the sparse operands to packed binary signatures and employ the corresponding arithmetic algorithms.

Since $\operatorname{wt}(PQ)\leq\operatorname{wt}(P)+\operatorname{wt}(Q)$, the estimated mean weight bounds that of the unmerged pairwise products. Accumulation, cancellation, and the selection of anticommuting pairs can, however, change the mean weight of the retained terms. We therefore reevaluate the storage criterion after sparse arithmetic and convert the result to packed binary signatures if necessary. Zero and identity results may retain sparse storage.

Automatic storage selection is implemented in the Python constructors and arithmetic dispatcher for weighted Pauli strings and sums. Arithmetic between two operands already stored through packed binary signatures retains the packed arithmetic path. Conversion between the two formats preserves the Pauli labels and coefficients, with its computational cost depending on the word span and occupied sites of the operators.

\subsection{Overlap-indexed commutators for sparse Pauli sums}
\label{app:pauli_overlap_commutator}

Consider two Pauli sums $A=\sum_{P\in S_A}a_PP$ and $B=\sum_{Q\in S_B}b_QQ$. Since Pauli strings with disjoint supports commute, only input pairs belonging to the set
\begin{equation}
    \mathcal J_{\mathrm{ov}}=
    \bigl\{(P,Q)\in S_A\times S_B:
    \operatorname{supp}(P)\cap\operatorname{supp}(Q)
    \ne\varnothing\bigr\}
\end{equation}
can contribute to the commutator. Denoting the number of such pairs by $N_{\mathrm{ov}}=|\mathcal J_{\mathrm{ov}}|$, we can restrict the double sum in~\cref{eq:pauli_sum_arithmetic} to
\begin{equation}
    [A,B]=2\!\sum_{\substack{(P,Q)\in\mathcal J_{\mathrm{ov}}\\
                           \chi(P,Q)=1}}
    a_Pb_Q\,PQ.
    \label{eq:overlap_commutator}
\end{equation}
The computational advantage arises when the overlapping pairs can be identified more efficiently than testing all $M_AM_B$ input pairs for anticommutation.

\paragraph{Identification of overlapping pairs.}

For each qubit $j$, we define the incidence numbers
\begin{equation}
\begin{aligned}
    \mu_{A,j}&=|\{P\in S_A:j\in\operatorname{supp}(P)\}|,\\
    \mu_{B,j}&=|\{Q\in S_B:j\in\operatorname{supp}(Q)\}|,
\end{aligned}
\end{equation}
which count the terms of each operator acting nontrivially on that qubit. Summing their products over all qubits gives
\begin{equation}
\begin{aligned}
    C_{\mathrm{ov}}
    &=\sum_{j=1}^{n}\mu_{A,j}\mu_{B,j}\\
    &=\sum_{P\in S_A}\sum_{Q\in S_B}
      |\operatorname{supp}(P)\cap\operatorname{supp}(Q)|
    \geq N_{\mathrm{ov}}.
\end{aligned}
\label{eq:overlap_incidence_count}
\end{equation}
Thus, $C_{\mathrm{ov}}$ counts overlapping pairs with multiplicity and provides an upper bound on their number without requiring their explicit enumeration.

To construct $\mathcal J_{\mathrm{ov}}$, we associate each qubit $j$ with the incidence list $\mathcal L_B(j)=\{Q\in S_B:j\in\operatorname{supp}(Q)\}$. The candidate partners of a given term $P\in S_A$ are then obtained from the union
\begin{equation}
    \mathcal N_B(P)=
    \bigcup_{j\in\operatorname{supp}(P)}\mathcal L_B(j),
    \label{eq:overlap_candidate_partners}
\end{equation}
and $\mathcal J_{\mathrm{ov}}=\bigcup_{P\in S_A}\{P\}\times\mathcal N_B(P)$. The implementation stores the incidence lists in a contiguous array of term indices, with offsets marking the entries associated with each qubit. A timestamp array prevents duplicate insertion of terms that overlap on several qubits.

The candidate lists are constructed in two passes, first counting the distinct partners of each input term and then filling the allocated array. Both passes can be parallelized over the terms of $A$. Sorting each candidate list by its original term indices preserves the relative order of contributions in an exhaustive pair traversal. Note that this corresponds to the symbolic phase of Gustavson's sparse matrix multiplication~\cite{Gustavson.1978-TwoFastAlgorithms} applied to the term--qubit incidence matrices of $A$ and $B$.

\paragraph{Computational complexity.}

Let $W_A$ and $W_B$ denote the total Pauli weights of the two input sums as defined in \cref{eq:pauli_sum_mean_weight}. Extracting the supports from their packed signatures requires $\mathcal O((M_A+M_B)L_{\mathrm{word}}+W_A+W_B)$ work, while computing the incidence numbers and constructing the index requires $\mathcal O(n+W_A+W_B)$ work. By~\cref{eq:overlap_incidence_count}, the enumeration of candidate partners visits $C_{\mathrm{ov}}$ incidence-list entries, including duplicates.

Writing $p_P=|\mathcal N_B(P)|$ and $p_{\max}=\max_{P\in S_A}p_P$, sorting the candidate lists requires
\begin{equation}
    \mathcal O\left(
        \sum_{P\in S_A}p_P\log(p_P+1)
    \right)
\end{equation}
operations, bounded by $\mathcal O(N_{\mathrm{ov}}\log(p_{\max}+1))$. The subsequent anticommutation tests are performed only for the $N_{\mathrm{ov}}$ distinct candidate pairs. Combining these contributions, the total cost before product formation and coefficient accumulation is bounded by
\begin{equation}
\begin{aligned}
    \mathcal O\bigl(
        &n+(M_A+M_B)L_{\mathrm{word}}
        +W_A+W_B+C_{\mathrm{ov}}\\
        &+N_{\mathrm{ov}}\log(p_{\max}+1)
        +N_{\mathrm{ov}}L_{\mathrm{word}}
    \bigr).
\end{aligned}
\label{eq:overlap_commutator_complexity}
\end{equation}
Forming the products of the $K_{\mathrm{raw}}$ anticommuting pairs requires a further $\mathcal O(K_{\mathrm{raw}}L_{\mathrm{word}})$ operations, followed by coefficient accumulation. By construction, $K_{\mathrm{out}}\leq K_{\mathrm{raw}}\leq N_{\mathrm{ov}}\leq C_{\mathrm{ov}}$.

When using the sparse site representation introduced in~\cref{app:sparse_pauli_arithmetic}, the supports of the input strings are directly available through their stored site lists. For the identified set of overlapping pairs, the elementary label-processing work is bounded by
\begin{equation}
    \mathcal O\left(
        N_{\mathrm{ov}}
        +\sum_{(P,Q)\in\mathcal J_{\mathrm{ov}}}
        \bigl(\operatorname{wt}(P)+\operatorname{wt}(Q)\bigr)
    \right).
\end{equation}
For local Hamiltonians of bounded Pauli weight and bounded term incidence per qubit, this contribution is $\mathcal O(n)$. The costs of constructing the incidence index, identifying distinct candidate pairs, and accumulating the resulting terms are accounted for separately.

\paragraph{Scaling for local Hamiltonians.}

Let $k$ bound the Pauli weights of the input terms and $\Delta_B=\max_j\mu_{B,j}$ denote the maximum number of terms of $B$ acting nontrivially on any qubit. The incidence count in~\cref{eq:overlap_incidence_count} then satisfies
\begin{equation}
    C_{\mathrm{ov}}
    \leq\Delta_B W_A
    \leq k\Delta_B M_A,
    \label{eq:overlap_local_bound}
\end{equation}
and similarly $p_{\max}\leq k\Delta_B$. For local Hamiltonians with $M_A,M_B=\mathcal O(n)$, bounded Pauli weight $k=\mathcal O(1)$, and bounded local incidence $\Delta_B=\mathcal O(1)$, we therefore have $N_{\mathrm{ov}}\leq C_{\mathrm{ov}}=\mathcal O(n)$. Including index construction and product formation, the cost prior to coefficient accumulation reduces from $\mathcal O(n^2L_{\mathrm{word}})$ for exhaustive pair processing to $\mathcal O(nL_{\mathrm{word}})$. The subsequent accumulation involves at most $\mathcal O(n)$ product records. This reduction in the number of candidate pairs is illustrated in~\cref{fig:overlap_indexed_commutator} for nearest-neighbor Hamiltonians on a chain.

\begin{figure}[t]
    \centering
    \begingroup
    \definecolor{exhaustive}{HTML}{4C88BD}
    \definecolor{overlap}{HTML}{2D9688}
    \begin{tikzpicture}[
      show background rectangle,
    background rectangle/.style={
        fill=black!3,
        rounded corners=5pt
    },
        x=0.27cm,
        y=0.27cm,
        font=\footnotesize,
        cell/.style={
            draw=#1!75!black,
            fill=#1!65,
            line width=0.25pt,
            rounded corners=0.7pt
        },
        inactive/.style={
            draw=black!12,
            fill=white,
            line width=0.25pt,
            rounded corners=0.7pt
        },
        lab/.style={font=\scriptsize, text=black!75},
        heading/.style={font=\footnotesize, align=center},
    ]

    \begin{scope}
        \node[heading] at (5,1.8) {\textbf{Exhaustive testing}};
        \node[lab] at (5,0.65) {Terms of $B$};
        \node[lab,rotate=90] at (-0.8,-5) {Terms of $A$};

        \foreach \r in {0,...,9}{
            \foreach \c in {0,...,9}{
                \draw[cell=exhaustive]
                    (\c+0.06,-\r-0.06)
                    rectangle
                    (\c+0.94,-\r-0.94);
            }
        }

        \node[lab] at (5,-10.8) {All input pairs};
        \node at (5,-12.0) {$(n-1)^2$ pairs};
        \node at (5,-13.3) {$\mathcal O(n^2)$ tests};
    \end{scope}

    \begin{scope}[xshift=3.9cm]
        \node[heading] at (5,1.8) {\textbf{Overlap-indexed testing}};
        \node[lab] at (5,0.65) {Terms of $B$};
        \node[lab,rotate=90] at (-0.8,-5) {Terms of $A$};

        \foreach \r in {0,...,9}{
            \foreach \c in {0,...,9}{
                \pgfmathtruncatemacro{\dist}{abs(\r-\c)}
                \ifnum\dist<2
                    \draw[cell=overlap]
                        (\c+0.06,-\r-0.06)
                        rectangle
                        (\c+0.94,-\r-0.94);
                \else
                    \draw[inactive]
                        (\c+0.06,-\r-0.06)
                        rectangle
                        (\c+0.94,-\r-0.94);
                \fi
            }
        }

        \node[lab] at (5,-10.8) {Overlapping pairs};
        \node at (5,-12.0) {$3n-5$ pairs};
        \node at (5,-13.3) {$\mathcal O(n)$ tests};
    \end{scope}

    \end{tikzpicture}
    \endgroup
    \caption{Reduction of candidate pairs through overlap indexing for nearest-neighbor Hamiltonians $A=\sum_{j=1}^{n-1}X_jX_{j+1}$ and $B=\sum_{j=1}^{n-1}Z_jZ_{j+1}$ on an open chain, illustrated for $n=11$. Each cell represents a pair of input Pauli strings, with colored cells indicating pairs subjected to an anticommutation test. While exhaustive evaluation tests all $(n-1)^2$ pairs, overlap indexing restricts the tests to the $3n-5$ pairs with overlapping supports. The number of tests thus scales linearly rather than quadratically with the system size, in accordance with the bound in~\cref{eq:overlap_local_bound}.}
    \label{fig:overlap_indexed_commutator}
\end{figure}

More generally, if the supports of the two input families are independently and uniformly distributed over subsets of $k$ qubits, the expected incidence count is
\begin{equation}
    \mathbb E[C_{\mathrm{ov}}]
    =\frac{M_AM_Bk^2}{n}.
\end{equation}
The overlap index therefore also reduces the expected number of candidate pairs for nonlocal low-weight operators distributed over sufficiently large registers. The reduction depends on the distribution of their supports, since bounded Pauli weight alone does not imply a small incidence count.

\paragraph{Adaptive selection and memory.}

The numerical commutator implementation selects the overlap-indexed kernel according to the estimated workload. For $M_AM_B\geq4096$, a preliminary sample estimates the fraction of anticommuting pairs. If this fraction is sufficiently small, the incidence count is evaluated and the index is constructed provided that
\begin{equation}
    16C_{\mathrm{ov}}<M_AM_B.
\end{equation}
The empirical factor accounts for the overhead of index construction and candidate enumeration. Otherwise, the implementation uses direct or prefix-partitioned commutator arithmetic.

The incidence index and candidate lists require $\mathcal O(n+W_B+M_A+N_{\mathrm{ov}})$ additional storage, with each worker maintaining a timestamp array of size $M_B$. Storage for the resulting $K_{\mathrm{raw}}$ product records and their subsequent accumulation is accounted for separately.

For sparse site storage, we similarly use the incidence count to select between overlap-indexed and exhaustive pair traversal, employing the criterion
\begin{equation}
    8C_{\mathrm{ov}}<M_AM_B.
\end{equation}
The respective empirical factors account for the computational overhead of index construction and candidate enumeration in the two storage formats.

\subsection{Targeted commutator coefficients}
\label{app:pauli_targeted}

For a prescribed output Pauli string $R\in\mathcal P_n$, the XOR rule uniquely determines the partner $Q_{P,R}$ of each input term $P\in S_A$ that can contribute to the coefficient of $R$,
\begin{equation}
    v(Q_{P,R})=v(P)\oplus v(R).
\end{equation}
Combining this relation with~\cref{eq:pauli_sum_arithmetic,eq:pauli_commutator} and setting $b_Q=0$ for $Q\notin S_B$, the targeted coefficient reduces to a single sum,
\begin{equation}
\begin{aligned}
    &\operatorname{coeff}_{\mathcal P_n}([A,B];R)\\
    &\quad=2\sum_{\substack{P\in S_A\\
                           \chi(P,Q_{P,R})=1}}
       a_Pb_{Q_{P,R}}\,\omega(P,Q_{P,R}).
\end{aligned}
\label{eq:targeted_pauli_commutator}
\end{equation}

We exploit this reduction by constructing a hash index of the larger input sum and traversing the smaller one, identifying at most one contributing partner for each term, in the manner of a hash join keyed by the partner signature~\cite{Graefe.1993-QueryEvaluationTechniques}. Hash collisions are resolved by comparing the full signatures, while the phase and anticommutation parity are evaluated directly from the input signatures without constructing the product. The original operand order is preserved in the phase evaluation to account for the antisymmetry of the commutator. The query extracts the coefficient of the elementary Pauli string $R$, irrespective of any scalar coefficient attached to the target object.

Assuming expected constant-time hash-table access, index construction requires $\mathcal O(\max(M_A,M_B)L_{\mathrm{word}})$ signature work, while evaluating the sum in~\cref{eq:targeted_pauli_commutator}, or its equivalent with the operands interchanged, requires $\mathcal O(\min(M_A,M_B)L_{\mathrm{word}})$ expected work. Since our implementation constructs the index for each query, the overall expected complexity is
\begin{equation}
    \mathcal O\bigl((M_A+M_B)L_{\mathrm{word}}\bigr),
\end{equation}
with additional index storage linear in $\max(M_A,M_B)$. Thus, targeted evaluation reduces the expected signature-processing cost from quadratic to linear in the combined input size for comparably sized sums. These estimates assume fixed-cost numerical coefficient arithmetic; symbolic coefficients introduce additional expression-dependent costs.

For Pauli sums stored through their occupied sites, the same partner identity applies. We construct the partner labels through ordered-list merges and identify the contributing terms through a hash index of the larger input sum. Including the construction of this index, the expected label-processing cost is bounded by
\begin{equation}
\begin{aligned}
    \mathcal O\bigl(
        M_A+M_B+W_A+W_B
        +M_{\min}\operatorname{wt}(R)
    \bigr),
\end{aligned}
    \label{eq:sparse_pauli_targeted_complexity}
\end{equation}
where $M_{\min}=\min(M_A,M_B)$. For bounded input and target Pauli weights, this reduces to $\mathcal O(M_A+M_B)$, independently of the span of the input operators.

\section{Pauli-cycle and Pauli-orbit arithmetic}
\label{app:symmetry_arithmetic}

\subsection{Canonical identification of Pauli cycles}
\label{app:cycle_unification}

Since cyclically equivalent Pauli strings generate the same cycle (cf.~\cref{eq:pauli_cycle_def}), we assign them a unique canonical representative through the \texttt{unify()} routine. Instead of generating and comparing all $n$ translations, which requires $\mathcal O(nL_{\mathrm{word}})$ operations, we exploit the relative positions of the nonidentity Pauli factors.

For a Pauli string $P$ of weight $h=\mathrm{wt}(P)>0$, let $q_1<\cdots<q_h$ denote its occupied sites, with corresponding Pauli labels $\ell_j\in\{X,Y,Z\}$. Defining the cyclic distances $g_j=q_{j+1}-q_j$, with $q_{h+1}=q_1+n$, we associate $P$ with the sequence
\begin{equation}
    \bigl((g_1,\ell_1),\ldots,(g_h,\ell_h)\bigr),
    \quad
    \sum_{j=1}^{h}g_j=n.
\end{equation}
Cyclic translations preserve the distances and Pauli labels, changing only the starting point of this sequence. Conversely, the sequence determines the original Pauli string up to a cyclic translation. Canonical identification thus reduces to finding the lexicographically smallest cyclic rotation of the sequence, comparing distances first and Pauli labels second, with $X<Y<Z$.

We identify this rotation using a two-pointer minimum-rotation algorithm~\cite{Booth.1980-LexicographicallyLeastCircular,Shiloach.1981-FastCanonizationCircular} in $\mathcal O(h)$ operations and reconstruct the corresponding Pauli string by placing its first nonidentity factor at qubit index zero. Including the extraction of the occupied sites from the packed signature and the reconstruction of the representative, the overall complexity is
\begin{equation}
    T_{\mathrm{unify}}(P)
    =\mathcal O(L_{\mathrm{word}}+\mathrm{wt}(P)).
    \label{eq:cycle_unification_complexity}
\end{equation}
The identity string is already canonical and requires no rotation. The complete procedure is illustrated in~\cref{fig:unify}.

\begin{figure}[t]
  \centering
  \begin{tikzpicture}[
    show background rectangle,
    background rectangle/.style={
        fill=black!3,
        rounded corners=5pt
    },
      font=\footnotesize,
      qb/.style={draw=#1!75!black, fill=#1!25, line width=0.4pt, rounded corners=1pt,
                 minimum width=0.66cm, minimum height=0.55cm, inner sep=0pt},
      ident/.style={draw=black!30, fill=white, line width=0.4pt, rounded corners=1pt,
                 minimum width=0.66cm, minimum height=0.55cm, inner sep=0pt, text=black!40},
      item/.style={draw=#1!75!black, fill=#1!25, line width=0.4pt, rounded corners=1pt,
                   minimum width=1.3cm, minimum height=0.55cm, inner sep=0pt},
      lab/.style={font=\scriptsize, text=black!65},
      stage/.style={font=\footnotesize\bfseries, anchor=west},
      gap/.style={-{Stealth[length=1.5mm]}, line width=0.5pt, black!60},
      arr/.style={-{Stealth[length=1.8mm]}, line width=0.5pt, black!60},
    ]
    \newcommand{\qx}[1]{0.75*#1+0.4}

    \node[stage] at (0, 0) {1.\ Extract cyclic distances};
    \foreach \q in {0,...,7} {
      \node[ident] (s\q) at ({\qx{\q}}, -1.2) {I};
      \node[lab] at ({\qx{\q}}, -1.7) {\q};
    }
    \foreach \q/\p/\c in {0/Z/pZ, 2/X/pX, 4/Y/pY} \node[qb=\c] (s\q) at ({\qx{\q}}, -1.2) {\p};
    \draw[gap] (s0.north) to[bend left=50] node[lab, above] {2} (s2.north);
    \draw[gap] (s2.north) to[bend left=50] node[lab, above] {2} (s4.north);
    \draw[gap] (s4.north) to[bend left=50] node[lab, above] {4} ({\qx{7}+0.55}, -1.0);

    \node[stage] at (0, -2.3) {2.\ Find minimum rotation};
    \node[item=pZ] (i1) at (1.25, -2.95) {(2, Z)};
    \node[item=pX, line width=1.2pt] (i2) at (3.0, -2.95) {(2, X)};
    \node[item=pY] (i3) at (4.75, -2.95) {(4, Y)};
    \draw[arr] (i1) -- (i2);
    \draw[arr] (i2) -- (i3);
    \draw[arr] (i3.south) -- ++(0, -0.22) -| (i1.south);
    \node[lab, align=center] at (3.0, -4.0)
      {Minimum rotation starts at (2, X)\\since $(2,X)<(2,Z)<(4,Y)$};

    \node[stage] at (0, -4.75) {3.\ Get canonical representative};
    \foreach \q in {0,...,7} {
      \node[ident] (u\q) at ({\qx{\q}}, -6.2) {I};
      \node[lab] at ({\qx{\q}}, -6.7) {\q};
    }
    \foreach \q/\p/\c in {0/X/pX, 2/Y/pY, 6/Z/pZ} \node[qb=\c] (u\q) at ({\qx{\q}}, -6.2) {\p};
    \draw[gap] (u0.north) to[bend left=50] node[lab, above] {2} (u2.north);
    \draw[gap] (u2.north) to[bend left=40] node[lab, above] {4} (u6.north);
  \end{tikzpicture}
\caption{Canonical identification of the Pauli cycle generated by $P=ZIXIYIII$ on eight qubits. 1.~The occupied sites and their cyclic distances define the sequence $((2,Z),(2,X),(4,Y))$. 2.~Its lexicographically smallest cyclic rotation starts at $(2,X)$, since equal distances are ordered according to the Pauli labels $X<Y<Z$. 3.~Reconstructing the string with its first nonidentity factor at qubit index zero yields the canonical representative $XIYIIIZI$. Every cyclic translation of $P$ yields the same representative through the algorithm described in~\cref{app:cycle_unification}.}
\label{fig:unify}
\end{figure}

\subsection{Cyclic averaging and arithmetic}
\label{app:cycle_arithmetic}

The arithmetic identities in~\cref{eq:pauli_cycle_arithmetic} follow from the invariance of the cyclic average under translations. Introducing the relative shift $r=t-s\pmod n$ in the product of two cycles gives
\begin{equation}
\begin{aligned}
    C_PC_Q
    &=\frac{1}{n^2}\sum_{s,t=0}^{n-1}P^{(s)}Q^{(t)}\\
    &=\frac{1}{n^2}\sum_{s,r=0}^{n-1}(PQ^{(r)})^{(s)}\\
    &=\frac{1}{n}\sum_{r=0}^{n-1}C_{PQ^{(r)}}.
\end{aligned}
\label{eq:cycle_product_derivation}
\end{equation}
The same argument applies to commutators,
\begin{equation}
    [C_P,C_Q]
    =\frac{1}{n}\sum_{r=0}^{n-1}C_{[P,Q^{(r)}]},
    \label{eq:cycle_commutator_derivation}
\end{equation}
as previously established in the anti-Hermitian cycle convention of Ref.~\cite{Barligea.2026-EnablingLieAlgebraicClassical}. Both identities reduce the $n^2$ pairs of translated strings to $n$ relative-shift classes, as illustrated in~\cref{fig:cycle_relative_shifts}.

\paragraph{Computational complexity.}

Each relative shift requires elementary Pauli-string arithmetic and canonical identification of the resulting cycle. Since the weight of a Pauli product is bounded by the sum of the input weights, the overall cost is
\begin{equation}
    \mathcal O\!\left(
        n[L_{\mathrm{word}}+\mathrm{wt}(P)+\mathrm{wt}(Q)]
    \right),
\end{equation}
assuming expected constant-time indexed accumulation (see~\cref{app:cycle_unification} for the canonicalization algorithm). At fixed machine-word width, this gives the general $\mathcal O(n^2)$ bound.

For commutators, a necessary condition for a nonzero contribution at relative shift $r$ is
\begin{equation}
    \operatorname{supp}(P)\cap
    \operatorname{supp}(Q^{(r)})\ne\varnothing.
\end{equation}
Each occupied site $j$ of $P$ and $k$ of $Q$ determines exactly one such shift through $r=j-k\pmod n$. The number of distinct candidate shifts is therefore bounded by $\min\{n,\mathrm{wt}(P)\mathrm{wt}(Q)\}$. For bounded input weights, both the enumeration of candidate shifts and the number of elementary commutators remain bounded as $n$ increases, reducing the overall complexity to $\mathcal O(n)$~\cite{Barligea.2026-EnablingLieAlgebraicClassical}. This restriction does not apply to multiplication, since products of disjoint-support strings remain nonzero.

\paragraph{Targeted commutator coefficients.}

Let $R_r$ denote the unweighted Pauli string satisfying ${PQ^{(r)}=\omega(P,Q^{(r)})R_r}$. Combining~\cref{eq:cycle_commutator_derivation,eq:pauli_commutator}, the coefficient of a prescribed cycle $C_R$ in the averaged cycle basis $\mathcal C_n$ is
\begin{equation}
\begin{aligned}
    &\operatorname{coeff}_{\mathcal C_n}([C_P,C_Q];[R])\\
    &\quad=\frac{2}{n}
      \sum_{\substack{0\leq r<n\\
          \chi(P,Q^{(r)})=1\\
          R_r\sim_{\mathrm{cyc}}R}}
      \omega(P,Q^{(r)}),
\end{aligned}
\label{eq:cycle_targeted_coefficient}
\end{equation}
where $[R]$ denotes the cyclic equivalence class of $R$. Only anticommuting relative shifts whose products generate the target cycle contribute to this sum. Targeted evaluation thus avoids constructing and accumulating the remaining output cycles, while retaining the general $\mathcal O(n^2)$ and bounded-weight $\mathcal O(n)$ complexity bounds.

\begin{figure}[t]
    \centering
    \begingroup
    \definecolor{cycX}{HTML}{2D9688}
    \definecolor{cycY}{HTML}{CE7992}
    \definecolor{cycZ}{HTML}{7B5CB2}
    \definecolor{cycW}{HTML}{4C88BD}
    \resizebox{\columnwidth}{!}{%
    \begin{tikzpicture}[
        font=\footnotesize,
        show background rectangle,
        background rectangle/.style={
            fill=black!3,
            rounded corners=5pt
        },
        lab/.style={
            font=\scriptsize,
            text=black!70
        },
        entry/.style={
            draw=black!25,
            rounded corners=1.5pt,
            minimum width=0.49cm,
            minimum height=0.48cm,
            inner sep=0pt
        },
        group/.style={
            draw=black!20,
            rounded corners=2pt,
            minimum width=3.1cm,
            minimum height=0.48cm,
            font=\scriptsize,
            inner sep=2pt
        },
        result/.style={
            draw=black!30,
            fill=white,
            rounded corners=2pt,
            minimum width=3.65cm,
            minimum height=0.95cm,
            align=center,
            inner sep=3pt
        },
        arr/.style={
            -{Stealth[length=1.8mm]},
            line width=0.6pt,
            black!65
        }
    ]

        \node at (1.9,0.32) {\textbf{Pairs of translations}};
        \node at (5.95,0.32) {\textbf{Sum within each class}};

        \foreach \t in {0,...,3}{
            \node[lab] at ({1.15+0.58*\t},-0.34)
                {$Q^{(\t)}$};
        }

        \foreach \s in {0,...,3}{
            \node[lab] at (0.44,{-0.95-0.58*\s})
                {$P^{(\s)}$};
        }

        \foreach \r/\clr in {
            0/cycX,1/cycY,2/cycZ,3/cycW%
        }{
            \foreach \s in {0,...,3}{
                \pgfmathtruncatemacro{\t}{mod(\s+\r,4)}
                \node[entry,fill=\clr!26]
                    at ({1.15+0.58*\t},
                        {-0.95-0.58*\s})
                    {$\r$};
            }

            \node[group,fill=\clr!26]
                at (5.95,{-0.95-0.58*\r})
                {$r=\r:\quad 4C_{PQ^{(\r)}}$};
        }

        \draw[arr] (3.27,-1.82) -- (4.02,-1.82);
        \node[lab] at (3.64,-1.47) {group};

        \node at (4.0,-3.43)
            {$\displaystyle
              C_PC_Q=\frac14\sum_{r=0}^{3}C_{PQ^{(r)}}$};

        \draw[black!18,line width=0.4pt]
            (0,-3.95) -- (8.0,-3.95);

        \node at (4.0,-4.25)
            {\textbf{Example:} $P=XIII$, $Q=YIII$};

        \node[result] (local) at (1.85,-5.15)
            {$[P,Q^{(0)}]=2i\,ZIII$\\
             $[P,Q^{(r)}]=0,\quad r\ne0$};

        \node[result] (final) at (6.15,-5.15)
            {$\displaystyle
              [C_P,C_Q]=\frac{i}{2}C_{ZIII}$};

        \draw[arr] (local.east) -- (final.west);

    \end{tikzpicture}%
    }
    \endgroup
    \caption{Reduction of cycle arithmetic to relative shifts for $n=4$. Each matrix entry represents a product $P^{(s)}Q^{(t)}$, with colors and labels identifying its relative shift $r=t-s\pmod 4$. Summing the four products within each class gives $4C_{PQ^{(r)}}$, reducing the original double sum to the single sum in~\cref{eq:cycle_product_derivation}. For the commutator of the cycles generated by $P=XIII$ and $Q=YIII$, only the class $r=0$ contributes, yielding $[C_P,C_Q]=\frac{i}{2}C_{ZIII}$ according to~\cref{eq:cycle_commutator_derivation}.}
    \label{fig:cycle_relative_shifts}
\end{figure}

\subsection{Permutation-orbit arithmetic}
\label{app:pauli_orbit_product}

Analogously to the cyclic averaging identity in~\cref{eq:cycle_product_derivation}, the permutation twirl satisfies
\begin{equation}
\begin{aligned}
    \mathcal S(P)\mathcal S(Q)
    &=\frac{1}{(n!)^2}
      \sum_{\sigma,\tau\in S_n}
      P^{(\sigma)}Q^{(\tau)}\\
    &=\frac{1}{n!}\sum_{\pi\in S_n}
      \mathcal S\!\left(PQ^{(\pi)}\right),
\end{aligned}
\label{eq:permutation_product_identity}
\end{equation}
where $P^{(\pi)}=U_\pi P U_\pi^\dagger$. Rather than enumerating the $n!$ relative permutations, we classify the pairs of Pauli strings $P\in\Omega_\alpha$ and $Q\in\Omega_\beta$ by their contingency tables
\begin{equation}
    \eta_{ab}=\bigl|\{j:(P_j,Q_j)=(a,b)\}\bigr|,
    \label{eq:orbit_contingency_table}
\end{equation}
for $a,b\in\{I,X,Y,Z\}$. Let $\mu_a(\alpha)$ denote the multiplicity of the Pauli factor $a$ in orbit $\alpha$. The feasible tables $\eta\in\mathcal T_{\alpha\beta}$ are precisely the nonnegative integer tables satisfying
\begin{equation}
    \sum_b\eta_{ab}=\mu_a(\alpha),
    \qquad
    \sum_a\eta_{ab}=\mu_b(\beta).
    \label{eq:orbit_contingency_constraints}
\end{equation}
By the local Pauli multiplication rules, each table determines an output orbit label $\gamma(\eta)=(\tilde p,\tilde q,\tilde r)$ through
\begin{equation}
\begin{aligned}
    \tilde p&=\eta_{XI}+\eta_{IX}+\eta_{YZ}+\eta_{ZY},\\
    \tilde q&=\eta_{YI}+\eta_{IY}+\eta_{ZX}+\eta_{XZ},\\
    \tilde r&=\eta_{ZI}+\eta_{IZ}+\eta_{XY}+\eta_{YX},
\end{aligned}
\label{eq:orbit_product_output}
\end{equation}
and a common multiplication phase
\begin{equation}
    \omega(\eta)=(-1)^{E(\eta)}i^{D(\eta)},
    \label{eq:orbit_product_phase}
\end{equation}
where $D(\eta)=\sum_{a,b\in\{X,Y,Z\},\,a\ne b}\eta_{ab}$ and $E(\eta)=\eta_{YX}+\eta_{ZY}+\eta_{XZ}$.

There are $W(\eta)=n!/\prod_{a,b}\eta_{ab}!$ ordered pairs realizing a given contingency table. Simultaneous permutation of both input strings preserves the table and acts transitively on the output orbit. Consequently, every string in $\Omega_{\gamma(\eta)}$ occurs with the same multiplicity, yielding
\begin{equation}
    \sum_{\substack{P\in\Omega_\alpha,\ Q\in\Omega_\beta\\
                    \eta(P,Q)=\eta}}
    PQ=W(\eta)\omega(\eta)B_{\gamma(\eta)}.
    \label{eq:orbit_fixed_table_sum}
\end{equation}
Dividing by $N_\alpha N_\beta$ and summing over the feasible tables gives the multiplication coefficients in~\cref{eq:pauli_orbit_product_coeff}. Since reversing the order of the Pauli strings introduces a sign $(-1)^{D(\eta)}$, the corresponding commutator coefficients are
\begin{equation}
    c_{\alpha\beta}^{\gamma}
    =\frac{2}{N_\alpha N_\beta}
      \sum_{\substack{\eta\in\mathcal T_{\alpha\beta}\\
                      \gamma(\eta)=\gamma\\
                      D(\eta)\ \mathrm{odd}}}
      W(\eta)\omega(\eta).
    \label{eq:pauli_orbit_commutator_coeff}
\end{equation}
This recovers the commutator construction of Ref.~\cite{Barligea.2026-EnablingLieAlgebraicClassical} in the Hermitian orbit convention used here. The grouping of Pauli-string pairs and their contributions is illustrated in~\cref{fig:orbit_overlap_patterns}.

\paragraph{Computational complexity and targeted coefficients.}

The sixteen entries of a contingency table are subject to seven independent marginal constraints in~\cref{eq:orbit_contingency_constraints}, leaving nine free entries, which can be chosen as the block indexed by $\{X,Y,Z\}^2$. Since each entry is bounded by $n$, at most $\mathcal O(n^9)$ candidate tables need to be considered for a product or commutator. With precomputed combinatorial factors and constant-time indexed accumulation, each feasible table contributes to one of $\binom{n+3}{3}=\mathcal O(n^3)$ possible output labels in constant arithmetic time.

For a prescribed output orbit $\gamma=(\tilde p,\tilde q,\tilde r)$, the three additional constraints in~\cref{eq:orbit_product_output} determine the diagonal entries $\eta_{XX}$, $\eta_{YY}$, and $\eta_{ZZ}$ from the six off-diagonal entries of the nonidentity block and the input marginals. Targeted evaluation of~\cref{eq:pauli_orbit_commutator_coeff} therefore requires at most $\mathcal O(n^6)$ candidate tables.

For targets of bounded Pauli weight, a stronger estimate follows by distributing each prescribed output multiplicity among its four possible local input pairs in~\cref{eq:orbit_product_output}. The remaining entries, corresponding to identity outputs, are uniquely determined by the input marginals. The number of candidate tables is thus bounded by
\begin{equation}
    \binom{\tilde p+3}{3}
    \binom{\tilde q+3}{3}
    \binom{\tilde r+3}{3},
    \label{eq:orbit_targeted_table_bound}
\end{equation}
which is independent of $n$ for fixed $\tilde p+\tilde q+\tilde r$~\cite{Barligea.2026-EnablingLieAlgebraicClassical}. The resulting $\mathcal O(1)$ bound applies to elementary orbit pairs after combinatorial precomputation. Arithmetic on orbit sums additionally depends on the numbers of input terms, according to~\cref{eq:general_targeted_arithmetic}. All estimates above assume fixed-cost numerical coefficient arithmetic.

\begin{figure}[htbp]
    \centering
    \begingroup

    \newcommand{\OrbitRow}[3]{%
        \node[lab] at (0.25,#1) {$#2$};
        \foreach \j/\pauli/\sty in {#3}{
            \node[\sty] at ({0.83+0.53*\j},#1) {\pauli};
        }
    }

    \newcommand{\OrbitTable}{%
        \node[lab] at (0.47,-2.48) {$\eta_{ab}$};
        \foreach \j/\pauli in {0/I,1/X,2/Y,3/Z}{
            \node[lab] at ({0.92+0.49*\j},-2.48) {\pauli};
            \node[lab] at (0.47,{-2.90-0.44*\j}) {\pauli};
        }
        \foreach \i in {0,...,3}{
            \foreach \j in {0,...,3}{
                \node[tabcell] at ({0.92+0.49*\j},{-2.90-0.44*\i}) {};
            }
        }
    }

    \begin{tikzpicture}[
        font=\footnotesize,
        show background rectangle,
        background rectangle/.style={
            fill=black!3,
            rounded corners=5pt
        },
        lab/.style={
            font=\footnotesize,
            text=black!70
        },
        heading/.style={
            font=\footnotesize\bfseries,
            align=center
        },
        tile/.style={
            draw=black!25,
            fill=white,
            rounded corners=1pt,
            minimum width=0.46cm,
            minimum height=0.43cm,
            inner sep=0pt
        },
        px/.style={tile,draw=pX!70!black,fill=pX!24},
        py/.style={tile,draw=pY!70!black,fill=pY!24},
        pz/.style={tile,draw=pZ!70!black,fill=pZ!24},
        tabcell/.style={
            draw=black!25,
            fill=white,
            rounded corners=0.7pt,
            minimum width=0.44cm,
            minimum height=0.39cm,
            inner sep=0pt
        },
        result/.style={
            draw=black!20,
            fill=white,
            rounded corners=2pt,
            minimum width=2.85cm,
            minimum height=0.60cm,
            inner sep=3pt
        }
    ]

    \begin{scope}
        \node[heading] at (1.60,0)
            {1.\ Same site};

        \OrbitRow{-0.60}{P}
            {0/X/px,1/I/tile,2/I/tile,3/I/tile}

        \OrbitRow{-1.13}{Q}
            {0/Y/py,1/I/tile,2/I/tile,3/I/tile}

        \OrbitRow{-1.66}{R}
            {0/Z/pz,1/I/tile,2/I/tile,3/I/tile}

        \node at (1.60,-2.08) {$PQ=iR$};

        \OrbitTable

        \node[tabcell,fill=black!8]
            at (0.92,-2.90) {3};

        \node[tabcell,fill=pZ!24]
            at (1.90,-3.34) {1};

        \node at (1.60,-4.80)
            {$W(\eta)=4,\quad\omega(\eta)=i$};

        \node[result] at (1.60,-5.42)
            {$\displaystyle\frac{i}{4}B_{0,0,1}$};
    \end{scope}

    \draw[black!15,line width=0.4pt]
        (3.40,0.22) -- (3.40,-5.75);

    \begin{scope}[xshift=3.60cm]
        \node[heading] at (1.60,0)
            {2.\ Different sites};

        \OrbitRow{-0.60}{P}
            {0/X/px,1/I/tile,2/I/tile,3/I/tile}

        \OrbitRow{-1.13}{Q}
            {0/I/tile,1/Y/py,2/I/tile,3/I/tile}

        \OrbitRow{-1.66}{R}
            {0/X/px,1/Y/py,2/I/tile,3/I/tile}

        \node at (1.60,-2.08) {$PQ=R$};

        \OrbitTable

        \node[tabcell,fill=black!8]
            at (0.92,-2.90) {2};

        \node[tabcell,fill=pX!24]
            at (0.92,-3.34) {1};

        \node[tabcell,fill=pY!24]
            at (1.90,-2.90) {1};

        \node at (1.60,-4.80)
            {$W(\eta)=12,\quad\omega(\eta)=1$};

        \node[result] at (1.60,-5.42)
            {$\displaystyle\frac34B_{1,1,0}$};
    \end{scope}

    \node at (3.40,-6.17)
        {$\displaystyle
          B_{1,0,0}B_{0,1,0}
          =\frac{i}{4}B_{0,0,1}
           +\frac34B_{1,1,0}$};

    \end{tikzpicture}%
    \endgroup
    \caption{Contingency-table multiplication of $B_{1,0,0}$ and $B_{0,1,0}$ for $n=4$. The sixteen ordered pairs of Pauli strings fall into two overlap patterns, represented by the pairs $(P,Q)$ shown above. The corresponding tables $\eta_{ab}$ count the local pairs of Pauli factors, with rows referring to $P$, columns to $Q$, and empty entries representing zeros. Each pattern determines an output orbit, represented here by a string $R$, together with a phase $\omega(\eta)$ and multiplicity $W(\eta)$. Dividing their contributions by the input normalization $N_{1,0,0}N_{0,1,0}=16$ yields the product according to~\cref{eq:pauli_orbit_product_coeff}. Only the first pattern contributes to the commutator (\cref{eq:pauli_orbit_commutator_coeff}).}
    \label{fig:orbit_overlap_patterns}
\end{figure}

\section{Shared algorithms and data structures}

\subsection{Mixed-representation arithmetic}
\label{app:mixed_representation_dispatch}

All implemented weighted operator families admit an expansion in the ordinary Pauli basis, providing a common representation for mixed arithmetic. The arithmetic dispatcher selects the appropriate kernel according to the operand representations and coefficient types. Products of individual Pauli strings retain their elementary representation, while arithmetic within the cycle or orbit family uses the corresponding sum representation. Other combinations are evaluated by converting the operands to \texttt{PauliStringSum}. Numerical coefficients are promoted to symbolic coefficients whenever one of the operands is symbolic. Within the ordinary Pauli family, the dispatcher additionally selects between packed binary signatures and sparse site storage according to the criteria described in~\cref{app:sparse_pauli_arithmetic}.

Conversion to an ordinary Pauli sum distributes the coefficients over the constituent Pauli strings according to~\cref{eq:pauli_cycle_def,eq:pauli_orbit_def}, taking into account the software normalization conventions in~\cref{app:software_conventions}. While an elementary cycle expands into at most $n$ distinct Pauli strings, an orbit expands into $N_{p,q,r}$ strings, a number that can grow exponentially with $n$. Mixed arithmetic involving permutation orbits may therefore incur substantial computational and memory costs that are absent when performing arithmetic directly in the orbit basis.

\subsection{Normalized inner-product evaluation}
\label{app:inner_product_evaluation}

For ordinary Pauli sums $A=\sum_{P\in S_A}a_PP$ and $B=\sum_{Q\in S_B}b_QQ$, orthogonality of the Pauli basis reduces the normalized Hilbert--Schmidt inner product to
\begin{equation}
    \langle A,B\rangle_\kappa
    =\sum_{P\in S_A\cap S_B}\overline{a_P}b_P.
    \label{eq:pauli_sum_inner_product_algorithm}
\end{equation}
For Pauli sums stored through packed binary signatures, we construct a hash index of the smaller input sum and traverse the larger one to identify their common Pauli labels. Including index construction, the expected signature-processing cost is $\mathcal{O}((M_A+M_B)L_{\mathrm{word}})$, assuming fixed-cost numerical coefficient arithmetic.

Within the cycle and orbit families, inner products are evaluated directly from the basis labels, coefficients, and normalization factors in~\cref{eq:pauli_cycle_inner_product,eq:pauli_orbit_inner_product}. For a Pauli sum $H=\sum_Q h_QQ$ and an elementary cycle or orbit defined on the same register, we similarly obtain
\begin{equation}
\begin{aligned}
    \langle C_P,H\rangle_\kappa
        &=\frac{1}{d_P}
          \sum_{Q\in\Omega_P^{\mathrm{cyc}}}h_Q,\\
    \langle B_\alpha,H\rangle_\kappa
        &=\frac{1}{N_\alpha}
          \sum_{Q\in\Omega_\alpha}h_Q,
\end{aligned}
\label{eq:mixed_symmetry_inner_products}
\end{equation}
where absent Pauli terms are assigned zero coefficients. In particular, the inner product with an orbit can be evaluated by traversing the terms of $H$ and identifying those with the prescribed Pauli multiplicities, without expanding the orbit itself. The implementation extends this procedure to orbit sums by indexing their weighted orbit labels. Other mixed inner products may require expansion of one or both operands.

All routines account for the software normalization conventions specified in~\cref{app:software_conventions}. Operators on different register sizes are compared by embedding them into a common register through identity factors, under which the normalized Hilbert--Schmidt inner product remains unchanged.

\subsection{Coefficient backends and software conventions}
\label{app:software_conventions}

In many applications of operator arithmetic, we wish to retain the dependence of an operator on a set of parameters throughout a calculation. For example, products and nested commutators of a Hamiltonian with polynomial coefficients again yield operators with polynomial coefficients, whose parameter dependence can subsequently be investigated through differentiation and substitution. Since these operations require only polynomial arithmetic, we replace the general-purpose symbolic-algebra backend \textsf{SymEngine}~\cite{SymEngine.2025-SymEngine}, used in earlier versions of \PE, with a native C++ implementation of sparse multivariate polynomials in version~1.0. We use \texttt{std::complex<double>} both for purely numerical operator coefficients and for the numerical weights of individual monomials in the symbolic representation. 

The polynomial representation follows a similar principle to that of the operator representations developed in~\cref{sec:operatorrepresentations}. We encode each monomial through the identifiers and nonnegative integer exponents of its constituent symbols and store only monomials with nonvanishing weights. Polynomial arithmetic can then be performed by combining these labels and accumulating their weights. For ordinary Pauli sums stored through packed binary signatures, we can further combine the monomial labels with the Pauli signatures into packed encodings whenever their respective sizes permit, allowing us to accumulate the contributions to both the operator expansion and its polynomial coefficients in a joint step. Since each polynomial owns its storage, the coefficient calculations can also be distributed across threads without managing shared symbolic expression trees. A synchronized registry is used separately to assign and identify symbols.

We support polynomial coefficients across all weighted operator representations, with symbolic weights of permutation orbits introduced through \texttt{PauliOrbitSum}. The combinatorial factors entering the orbit arithmetic remain numerical. If an operator depends on non-polynomial functions of its parameters, these functions can be represented by auxiliary symbols whose values are supplied upon numerical substitution. Their functional dependencies must then be accounted for externally when differentiating the operator.

For applications in which only the labels of Pauli operators and their commutation relations are relevant, we additionally provide a coefficient-free representation of Pauli strings. This is particularly useful for constructing dynamical Lie algebras generated by individual Pauli strings, as discussed in~\cref{sec:applications}, where the scalar coefficients of newly generated strings are irrelevant for determining the resulting basis. We therefore store collections of coefficient-free strings as ordered sets of distinct Pauli labels with stable insertion indices. Individual strings retain their discrete phase $i^k$, and for two anticommuting strings $P$ and $Q$, the commutator method returns their product $PQ$, omitting the factor of two in~\cref{eq:pauli_commutator}. For commuting strings, the method returns no result. When coefficient-free strings are combined with weighted operators, their phases are promoted to scalar coefficients and the ordinary weighted arithmetic applies.

When translating between the mathematical expressions derived in this work and their software representations, two normalization conventions need to be considered. In terms of the averaged Hermitian operator bases introduced in~\cref{sec:operatorrepresentations}, the elementary symmetry-adapted objects represent
\begin{equation}
\begin{aligned}
    \texttt{PauliCycle}(n,P,a)
        &\longleftrightarrow anC_P,\\
    \texttt{PauliOrbit}(n,p,q,r)
        &\longleftrightarrow iB_{p,q,r}.
\end{aligned}
\label{eq:software_basis_conventions}
\end{equation}
Accordingly, the coefficients of an expansion in the Hermitian cycle and orbit bases must be multiplied by $1/n$ and $-i$, respectively, when constructing the corresponding software objects. From the normalization of the mathematical bases, the squared normalized Hilbert--Schmidt norms of the software basis elements follow as
\begin{equation}
\begin{aligned}
    \left\|nC_P\right\|_\kappa^2
        &=\frac{n^2}{d_P},\\
    \left\|iB_\alpha\right\|_\kappa^2
        &=\frac{1}{N_\alpha}.
\end{aligned}
\end{equation}

The choice of an anti-Hermitian orbit basis $\widehat B_\alpha=iB_\alpha$ is particularly convenient for Lie-algebraic calculations. Expressing the product and commutator expansions of Eqs.~\eqref{eq:pauli_orbit_product_coeff} and \eqref{eq:pauli_orbit_commutator_coeff} in this basis yields
\begin{equation}
\begin{aligned}
    \widehat B_\alpha\widehat B_\beta
    &=i\sum_\gamma m_{\alpha\beta}^{\gamma}
      \widehat B_\gamma,\\
    [\widehat B_\alpha,\widehat B_\beta]
    &=i\sum_\gamma c_{\alpha\beta}^{\gamma}
      \widehat B_\gamma.
\end{aligned}
\label{eq:orbit_software_normalization}
\end{equation}
Thus, the coefficients returned by the orbit commutator directly correspond to the real structure constants $f_{\alpha\beta}^{\gamma}=ic_{\alpha\beta}^{\gamma}$ of the Hermitian bracket defined in~\cref{sec:framework}.

Finally, when converting between Pauli expansions and matrix representations, we associate qubit zero with the most significant computational-basis bit. During matrix decomposition, coefficients whose magnitudes do not exceed the optional \texttt{threshold} parameter of \texttt{PauliStringSum.from\_matrix()} are discarded, with a default threshold of zero. Since a Pauli sum does not explicitly store its ambient register size, information about trailing idle qubits may be lost during this conversion. To reconstruct the original matrix, users can preserve the positions of the active qubits through the option \texttt{ignore\_unused\_qubits=False} of \texttt{to\_matrix()} and account for any remaining idle qubits using the original register dimension.

\section{Benchmark methodology and supplementary results}
\label{app:benchmarks}

\subsection{Overview of related software}
\label{app:benchmark_software}

\Cref{tab:related_software} summarizes the software packages considered in our benchmarks. Although many of these implementations share the binary representation of Pauli strings introduced in~\cref{sec:operatorrepresentations}, they differ substantially in how they store operator sums, accumulate arithmetic contributions, and exploit the structure of the represented operators.

\begin{table*}[t]
    \centering
    \footnotesize
    \setlength{\tabcolsep}{4pt}
    \renewcommand{\arraystretch}{1.12}
    \newcommand{\tfull}{\checkmark}
    \newcommand{\tpart}{(\checkmark)}
    \newcommand{\tnone}{--}
    \begin{tabular}{@{}llll cccccc@{}}
        \toprule
        \textbf{Package}
        & \textbf{Version}
        & \textbf{Backend / interface}
        & \textbf{Exp.}
        & \textbf{PS} & \textbf{Bin} & \textbf{Sym} & \textbf{Par} & \textbf{SA} & \textbf{TC} \\
        \midrule
        \PE\ \textit{(This work)}
            & 1.0.0 & C++20 / Python & P N L S
            & \tfull & \tfull & \tfull & \tfull & \tfull & \tfull \\
        \midrule
        \multicolumn{10}{@{}l}{\textit{Specialized Pauli-arithmetic libraries}}\\
        PauLIB~\cite{Krotz.2026-PauLIBHighperformanceLibrary}
            & \texttt{cbdf68e} & C++20 / Python & P N
            & \tfull & \tfull & \tnone & \tfull & \tnone & \tnone \\
        BinSim~\cite{Li.2026-HighperformancePaulialgebraFramework}
            & \texttt{6fa8a4e} & Julia, C++/CUDA & N
            & \tpart & \tfull & \tnone & \tnone & \tnone & \tnone \\
        PauliStrings.jl~\cite{Loizeau.2025-QuantumManybodySimulations}
            & 1.10.1 & Julia & N S
            & \tfull & \tfull & \tfull & \tnone & \tpart & \tnone \\
        PauliStrings.py~\cite{Loizeau.2025-QuantumManybodySimulations}
            & 0.1.1 (\texttt{86c72af}) & Python + C++ & P
            & \tpart & \tfull & \tnone & \tnone & \tnone & \tnone \\
        PauliArray~\cite{Dion.2024-EfficientlyManipulatingPauli}
            & 0.0.1 (\texttt{e0e245a}) & Python / NumPy & P L
            & \tfull & \tpart & \tnone & \tnone & \tnone & \tnone \\
        zixy~\cite{Quantinuum.2026-Zixy}
            & 0.4.3 & Rust / Python & P N
            & \tfull & \tfull & \tpart & \tnone & \tnone & \tnone \\
        fast-pauli~\cite{Qognitive.2024-FastPauli}
            & 0.0.35 & C++20 / Python & P N
            & \tfull & \tnone & \tnone & \tfull & \tnone & \tnone \\
        fastfermion~\cite{Fawzi.2026-FastFermion}
            & 0.2.0 / \texttt{2dc495d} & C++ / Python & P N
            & \tfull & \tfull & \tnone & \tnone & \tnone & \tnone \\
        \midrule
        \multicolumn{10}{@{}l}{\textit{Operator classes of quantum SDKs and simulators}}\\
        Qiskit~\cite{JavadiAbhari.2024-QuantumComputingQiskit}
            & 2.6.2 / \texttt{cf8c5ae} & Rust, NumPy / Python & P N
            & \tfull & \tpart & \tfull & \tpart & \tnone & \tnone \\
        OpenFermion~\cite{McClean.2020-OpenFermionElectronicStructure}
            & 1.8.1 & Python & P
            & \tpart & \tnone & \tfull & \tnone & \tnone & \tnone \\
        PennyLane~\cite{Bergholm.2018-PennyLaneAutomaticDifferentiation}
            & 0.42.3 & Python & P
            & \tfull & \tnone & \tnone & \tnone & \tnone & \tnone \\
        Cirq~\cite{Cirq.2025-Cirq}
            & 1.5.0 & Python & P
            & \tfull & \tnone & \tfull & \tnone & \tnone & \tnone \\
        tket~\cite{Sivarajah.2021-TketRetargetableCompiler}
            & 2.18.5 / \texttt{dbf6e1a} & C++ / Python & P N
            & \tfull & \tnone & \tfull & \tnone & \tnone & \tnone \\
        Qulacs~\cite{Suzuki.2021-QulacsFastVersatile}
            & 0.6.14 / \texttt{66a0aaf} & C++ / Python & P N
            & \tfull & \tfull & \tnone & \tnone & \tnone & \tnone \\
        struqture~\cite{HQS.2026-Struqture}
            & 2.6.0 & Rust / Python & P N
            & \tfull & \tnone & \tfull & \tnone & \tnone & \tnone \\
        symmer~\cite{Ralli.2026-Symmer}
            & 0.0.10 & Python / NumPy & P
            & \tpart & \tpart & \tnone & \tnone & \tnone & \tnone \\
        \midrule
        \multicolumn{10}{@{}l}{\textit{Stabilizer and Lie-algebra tools}}\\
        Stim~\cite{Gidney.2021-StimFastStabilizer}
            & 1.16.0 / \texttt{ffe0d04} & C++ / Python & P N L
            & \tpart & \tfull & \tnone & \tnone & \tnone & \tnone \\
        QuantumClifford.jl~\cite{Krastanov.2026-QuantumCliffordJl}
            & 0.11.8 & Julia & N
            & \tpart & \tfull & \tnone & \tnone & \tnone & \tnone \\
        PauLie~\cite{Shaya.2026-PauLieFastClassification}
            & 0.2.3 & Python & P L
            & \tfull & \tfull & \tnone & \tnone & \tnone & \tnone \\
        $\gsim$~\cite{Barligea.2026-EnablingLieAlgebraicClassical}
            & 0.1.0 (\texttt{c190f36}) & Numba / Python & S
            & \tnone & \tnone & \tnone & \tnone & \tpart & \tpart \\
        \bottomrule
    \end{tabular}
    \caption{Comparison of the software packages considered in our benchmarks, including their versions, implementation languages, and arithmetic capabilities. Experiment suites (Exp.): P, arithmetic through Python interfaces; N, arithmetic through native interfaces; L, Pauli-string construction and pairwise arithmetic following Ref.~\cite{QPauLie.2026-PaulistringComparison}; S, symmetry-adapted arithmetic. Capabilities: PS, elementary Pauli strings and weighted Pauli sums; Bin, packed binary encoding; Sym, symbolic coefficients; Par, shared-memory parallel Pauli arithmetic; SA, symmetry-adapted operator arithmetic; TC, direct targeted commutator evaluation. Symbols indicate full (\checkmark), partial ((\checkmark)), or no (--) support for the respective capability.}
    \label{tab:related_software}
    \par\smallskip
    \begin{minipage}{\textwidth}
    \footnotesize
    \raggedright
    \textit{Software versions.}
    Where applicable, the Python release and native source revision are listed separately. The source revisions of \PE\ used in the individual benchmarks are documented in the benchmark repository referenced in~\cref{app:benchmark_methodology}.

    \smallskip
    \textit{Partial support.}
    PS: elementary strings without weighted sums (Stim, QuantumClifford.jl), or weighted sums with strings represented as one-term sums (BinSim, PauliStrings.py, OpenFermion, symmer).
    Bin: unpacked binary symplectic arrays (PauliArray, Qiskit \texttt{SparsePauliOp}, symmer).
    Sym: symbolic coefficients only in the Python layer, not in the Rust core (zixy).
    Par: parallel sorting within the canonicalization of \texttt{SparseObservable} only (Qiskit).
    SA: translation symmetry only (PauliStrings.jl), or cycle and orbit commutators without products ($\gsim$).
    TC: targeted structure constants for orbits only, within the adjoint-generator construction ($\gsim$).
    The $\gsim$ implementation uses \PE\ for elementary Pauli strings.
\end{minipage}
\end{table*}

Among the specialized libraries considered here, PauLIB~\cite{Krotz.2026-PauLIBHighperformanceLibrary} is most closely related to \PE\ in its approach to numerical Pauli arithmetic. Both libraries use packed binary signatures and flat storage for operator sums, with different strategies for generating and accumulating product terms. PauLIB employs a struct-of-arrays layout that facilitates vectorization across collections of Pauli strings, and combines the resulting terms through sorting and a subsequent accumulation step. In our implementation, we partition the pairs of input terms according to the prefixes of their output signatures, allowing us to accumulate contributions independently within each partition (see~\cref{app:pauli_sum_product}). Moreover, the development of PauLIB places particular emphasis on high-performance computing, including shared-memory parallelization with OpenMP~\cite{Dagum.1998-OpenMPIndustryStandard}, GPU arithmetic with CUDA~\cite{Nickolls.2008-ScalableParallelProgramming}, and distributed-memory arithmetic with MPI~\cite{MPIForum.2023-MPIMessagepassingInterface}. While our benchmarks are restricted to CPU arithmetic on a single node, these additional execution models are of interest for computations that exceed the memory or computational capacity of individual nodes.

Another closely related package is PauliStrings.jl~\cite{Loizeau.2025-QuantumManybodySimulations}, which was developed primarily for the study of quantum many-body dynamics using sparse Pauli expansions. It stores integer-encoded Pauli strings and their coefficients in hash-based dictionaries, which allow individual terms to be inserted, updated, or removed during arithmetic and truncation. The package provides routines for Heisenberg-picture time evolution and Krylov methods~\cite{Parker.2019-UniversalOperatorGrowth}, where controlling the growth of the Pauli expansion is a central computational task. Particularly relevant to our work is its representation of translation-invariant operators on lattices of arbitrary dimension, for which we consider the one-dimensional periodic case in our Pauli-cycle benchmarks. PauliStrings.jl also supports symbolic coefficients through package extensions for Symbolics.jl~\cite{Gowda.2021-HighperformanceSymbolicnumericsMultiple} and Mathematica, and its Python counterpart, PauliStrings.py~\cite{Loizeau.2025-QuantumManybodySimulations}, provides an implementation of ordinary Pauli arithmetic.

An alternative to operating on packed signatures is pursued by PauliArray~\cite{Dion.2024-EfficientlyManipulatingPauli}, which exploits NumPy~\cite{Harris.2020-ArrayProgrammingNumPy} broadcasting and vectorization to perform arithmetic on arrays of Pauli strings and operators. The array-oriented interface is particularly well suited to calculations involving collections of operators, such as the evaluation of commutators between a Hamiltonian and an operator pool in adaptive variational algorithms~\cite{Grimsley.2019-AdaptiveVariationalAlgorithma}. For products of large Pauli sums, however, the unpacked binary signatures and the intermediate arrays generated by broadcasting can lead to substantial memory requirements. Sort-based accumulation of packed Pauli signatures is also employed by BinSim~\cite{Li.2026-HighperformancePaulialgebraFramework}, where equal product terms are identified through radix sorting. The development of BinSim is motivated by quantum-chemistry applications, with dedicated algorithms for constructing qubit Hamiltonians and applying them to quantum states. Its operator--state arithmetic exploits common bit-flip patterns among Pauli strings and symmetry restrictions of the underlying state space.

The Rust library zixy~\cite{Quantinuum.2026-Zixy} follows a software design similar to that outlined in~\cref{sec:software_architecture,ssec:Extensibility}, separating operator components from their coefficient types and providing distinct containers for sets and linear combinations of operators. Its implementation extends beyond Pauli operators to fermionic operators and quantum states, and supports symbolic coefficients through SymPy~\cite{Meurer.2017-SymPySymbolicComputing} in the Python interface. Further specialized libraries include fast-pauli~\cite{Qognitive.2024-FastPauli}, which focuses on the parallel application of Pauli operators to batches of quantum states, and fastfermion~\cite{Fawzi.2026-FastFermion}, which provides arithmetic routines for the simulation of fermionic systems.

Many applications in quantum computing rely on the operator classes provided by general-purpose quantum software. Qiskit~\cite{JavadiAbhari.2024-QuantumComputingQiskit}, for instance, provides both a sparse representation through \texttt{SparseObservable} (i.e., “Qiskit 1” in our experiments) and an array-based binary symplectic representation through \texttt{SparsePauliOp} (i.e., “Qiskit 2” in our experiments). OpenFermion~\cite{McClean.2020-OpenFermionElectronicStructure} and PennyLane~\cite{Bergholm.2018-PennyLaneAutomaticDifferentiation} employ dictionary-based representations that store the nonidentity factors of individual Pauli strings. Symbolic operator coefficients are supported by several of these frameworks, including Qiskit, OpenFermion, Cirq~\cite{Cirq.2025-Cirq}, tket~\cite{Sivarajah.2021-TketRetargetableCompiler}, and struqture~\cite{HQS.2026-Struqture}. These implementations are relevant to our comparisons because they provide the arithmetic routines on which many existing applications are built.

For elementary Pauli-string arithmetic, Stim~\cite{Gidney.2021-StimFastStabilizer} and QuantumClifford.jl~\cite{Krastanov.2026-QuantumCliffordJl} provide reference implementations based on packed binary signatures and discrete Pauli phases. Both libraries were developed for stabilizer simulations~\cite{Gottesman.1998-HeisenbergRepresentationQuantum,Aaronson.2004-ImprovedSimulationStabilizer}, where arithmetic on individual Pauli strings is a frequent operation. Similar computational requirements arise in the classification of dynamical Lie algebras with Pauli generators~\cite{Wiersema.2024-ClassificationDynamicalLie,Kokcu.2024-ClassificationDynamicalLie,Aguilar.2024-FullClassificationPauli}, as implemented in PauLie~\cite{Shaya.2026-PauLieFastClassification}. 

For symmetry-adapted arithmetic, we also consider the earlier Python and Numba~\cite{Lam.2015-NumbaLLVMbasedPython} implementations of cycle and orbit commutators developed for Lie-algebraic simulation in Ref.~\cite{Barligea.2026-EnablingLieAlgebraicClassical} ($\gsim$). Since these implementations operate on the same symmetry-adapted bases as those developed in~\cref{ssec:PauliCycles,ssec:PauliOrbits}, they allow us to assess the computational improvements arising from the new arithmetic implementations independently of the advantages of the underlying operator representations. Related approaches to symmetry-adapted operator arithmetic include QCommute~\cite{Lychkovskiy.2026-QCommuteToolSymbolic}, which evaluates symbolic nested commutators of translation-invariant operators in the thermodynamic limit, and the merging of symmetry-equivalent Pauli strings in Pauli propagation~\cite{Teng.2025-LeveragingSymmetryMerging}. We include the latter in our symmetry-adapted benchmarks, using the accompanying SymMerge code~\cite{Teng.2026-SymMerge} together with the arithmetic routines of PauliPropagation.jl~\cite{Rudolph.2025-PauliPropagationComputational}.

\subsection{Benchmark configurations and methodology}
\label{app:benchmark_methodology}

The benchmarks were conducted on the LiCCA computing cluster at the University of Augsburg, using nodes equipped with AMD EPYC 7713 processors (Zen~3). We consider thread budgets of $T\in\{1,16,64\}$, restricting each benchmark process to the corresponding number of physical CPU cores. Measurements through the Python interfaces and the native C++, Rust, and Julia implementations are reported separately.

Unless specified otherwise, each runtime measurement comprises ten independently sampled input pairs with ten timing repetitions per pair, preceded by untimed warmup calls. We use identical input operators across the compared implementations and exclude input construction and numerical validation from the measured runtimes, while including arithmetic, required coefficient accumulation, and internal representation conversions. For sufficiently short operations, we measure batches of repeated arithmetic calls. We report arithmetic mean runtimes with error bars indicating the standard error of the mean across the ten input-level means.

The complete experimental configurations, software revisions, build settings, benchmark implementations, numerical validation records, and underlying measurement data are available at \url{https://github.com/adelina-b/pauliengine-benchmarks}. The repository additionally provides scripts for reproducing all benchmark figures and tables presented in this work, together with supplementary results covering the full range of investigated configurations.

\subsubsection{Ordinary Pauli arithmetic}
\label{app:benchmark_methodology_pauli}

We generate pairs of Pauli sums containing $M$ distinct strings on $n$ qubits, with numerical coefficients sampled uniformly from $[-1,1]$. For dense strings, we sample each local factor uniformly from $\{I,X,Y,Z\}$. For weight-four strings, we uniformly sample four distinct occupied sites and assign their nonidentity factors independently. The storage selection and arithmetic settings of the respective measured implementations are documented in the benchmark repository.

Where dedicated commutator routines are unavailable, we evaluate commutators through the difference of two products. For packages without native Pauli-sum representations, we implement the arithmetic using their elementary Pauli-string routines. The returned operators may differ in their degree of simplification, which we account for when comparing their numerical coefficients. Independent reference calculations are performed for elementary-string products and selected Pauli-sum configurations, primarily with up to 100 terms per operand. We apply a per-process memory limit of 96\,GiB and a runtime watchdog of 60 seconds without progress, reporting runtime statistics only for completed configurations that pass the applicable numerical checks.

\subsubsection{Symmetry-adapted arithmetic}
\label{app:benchmark_methodology_symmetry}

We perform all symmetry-adapted benchmarks on a single physical CPU core, considering pairs of elementary Pauli cycles and orbits with the input weights specified in the respective figures. All operators have unit coefficients in their software bases, with equivalent ordinary Pauli expansions constructed according to the normalization conventions of~\cref{app:software_conventions}. The measured arithmetic includes the canonicalization and normalization of the resulting operator expansions where required. Configurations exceeding computational resource limits or failing numerical checks are excluded.

For targeted commutator benchmarks, we sample targets from the nonvanishing components of independently computed reference commutators and additionally consider dense input orbits with prescribed nonvanishing weight-one target coefficients. For memory benchmarks, we perform one arithmetic operation per input pair in a fresh process, recording the resident set size (RSS) before and after operand construction, the peak RSS during arithmetic, and the RSS with both operands and the result retained. All reported memory measurements are relative to the initialized process and include memory-allocation overhead.

\subsection{Numerical validation}
\label{app:numerical_validation}

We validate the arithmetic implementations of \PE\ against reference calculations that are independent of the algorithms developed in~\cref{sec:operatorrepresentations}. For small quantum systems, we obtain reference results through explicit matrix arithmetic and, for symmetry-adapted operators, through arithmetic on their ordinary Pauli expansions. For each category of operations listed in~\cref{tab:implementation-validation}, we generate 100 seeded test instances per register size $n=1,\ldots,6$. We additionally test all pairs of Pauli strings with unit coefficients for $n\leq 3$ and all 36 ordered pairings of the six weighted operator representations.

For larger registers, we obtain reference results from independent implementations of the elementary Pauli multiplication rules and exact combinatorial calculations for symmetry-adapted operators. To verify the arithmetic across packed-storage boundaries, we test ordinary Pauli strings, sums, and cycles at register sizes immediately below, at, and above 64, 128, 256, and 512 qubits. For permutation orbits, we extend the validation to $n=170$. We test sparse site arithmetic on registers of up to $16{,}384$ qubits, including arithmetic between the two storage formats and conversion to packed binary signatures. To verify parallel arithmetic, we perform the Pauli-sum stress tests with thread settings of $T\in\{1,16,64\}$. All comparisons account for the normalization conventions described in~\cref{app:software_conventions}.

For each computed operator coefficient or scalar $c$, we require agreement with the corresponding reference value $c_{\mathrm{ref}}$ within the tolerance
\begin{equation}
    |c-c_{\mathrm{ref}}|
    \leq \epsilon_{\mathrm{abs}}
    +\epsilon_{\mathrm{rel}}\max(|c|,|c_{\mathrm{ref}}|),
\end{equation}
where $\epsilon_{\mathrm{abs}}=10^{-12}$ and $\epsilon_{\mathrm{rel}}=10^{-9}$. We validate symbolic arithmetic by comparing independently evaluated coefficients at five parameter assignments, including complex values. We further verify polynomial differentiation, conjugation, and partial substitution against SymPy~\cite{Meurer.2017-SymPySymbolicComputing}.

The validation results are summarized in~\cref{tab:implementation-validation}. All 115,971 numerical checks across 10,353 distinct test instances satisfy the prescribed tolerance. The largest absolute deviation across all tested operations is $4.55\times10^{-13}$, observed in symbolic arithmetic.

\begin{table*}[t]
    \centering
    \small
    \setlength{\tabcolsep}{3pt}
    \renewcommand{\arraystretch}{1.12}
    \begin{tabular*}{\textwidth}{
        @{\extracolsep{\fill}}
        >{\raggedright\arraybackslash}p{0.18\textwidth}
        >{\raggedright\arraybackslash}p{0.26\textwidth}
        r
        >{\raggedright\arraybackslash}p{0.18\textwidth}
        r
        r
        @{}
    }
        \toprule
        Operation
        & Independent reference
        & Instances
        & Qubits
        & \shortstack[r]{Max. abs.\\deviation}
        & \shortstack[r]{Failed\\checks} \\
        \midrule
        Pauli-string products and commutators
        & Explicit matrix arithmetic; Pauli rules (stress)
        & 5,138
        & 1--6, 63--65, 127--129, 255--257, 511--513, 1024, 4096, 8192, 16384
        & $4.58\times10^{-16}$
        & 0 \\
        
        Pauli-sum arithmetic
        & Explicit matrix arithmetic; Pauli rules (stress)
        & 1,372
        & 1--6, 63--65, 127--129, 255--257, 511--513, 1024, 2048, 8192, 16384
        & $4.31\times10^{-14}$
        & 0 \\
        
        Cycle arithmetic
        & Expanded Pauli sums; exact cyclic reference (stress)
        & 720
        & 1--6, 63--65, 127--129, 255--257, 511--513
        & $3.02\times10^{-15}$
        & 0 \\
        
        Orbit arithmetic
        & Expanded Pauli sums; exact integer reference (stress)
        & 680
        & 1--6, 8, 10, 12, 16, 32, 64, 128, 170
        & $9.04\times10^{-16}$
        & 0 \\
        
        Targeted commutators
        & Coefficients of independent full commutators
        & 600
        & 1--6
        & $1.83\times10^{-15}$
        & 0 \\
        
        Inner products
        & Normalized matrix traces
        & 600
        & 1--6
        & $7.11\times10^{-15}$
        & 0 \\
        
        Symbolic arithmetic
        & Independent coefficient evaluation; SymPy
        & 643
        & 1--6, 1024, 8192, 16384
        & $4.55\times10^{-13}$
        & 0 \\
        
        Mixed arithmetic
        & Ordinary Pauli expansions and matrices
        & 600
        & 1--6
        & $1.78\times10^{-15}$
        & 0 \\
        \bottomrule
    \end{tabular*}

\caption{Numerical validation of the operator arithmetic implemented in \PE~1.0 against independent reference calculations. We report the number of distinct test instances, the considered register sizes, the maximum absolute deviation from the reference values, and the number of numerical checks exceeding the tolerance specified in~\cref{app:numerical_validation}. Repeated evaluations with different thread settings constitute separate checks of the same test instance. Across 10,353 instances and 115,971 numerical checks, no deviations exceed the prescribed tolerance.}

    \label{tab:implementation-validation}
\end{table*}

The validation code, test configurations, and complete numerical results are available in our benchmark repository referenced in~\cref{app:benchmark_methodology}.

\subsection{Additional Pauli-arithmetic benchmarks}
\label{app:pauli_benchmarks}

The benchmarks presented in~\cref{ssec:benchmark_pauli} focus on single-threaded arithmetic through the Python interfaces of the considered software packages. To investigate the performance of the underlying arithmetic implementations and their parallelization, we additionally compare runtimes through the native C++, Rust, and Julia interfaces and consider thread budgets of up to 64 threads. Here, we present representative multiplication benchmarks for dense and low-weight Pauli sums, together with comparisons of elementary Pauli-string arithmetic. The complete set of measurements is available in the benchmark repository referenced in~\cref{app:benchmark_methodology}.

\begin{figure*}[htbp]
    \centering
    \includegraphics[width=1\linewidth]{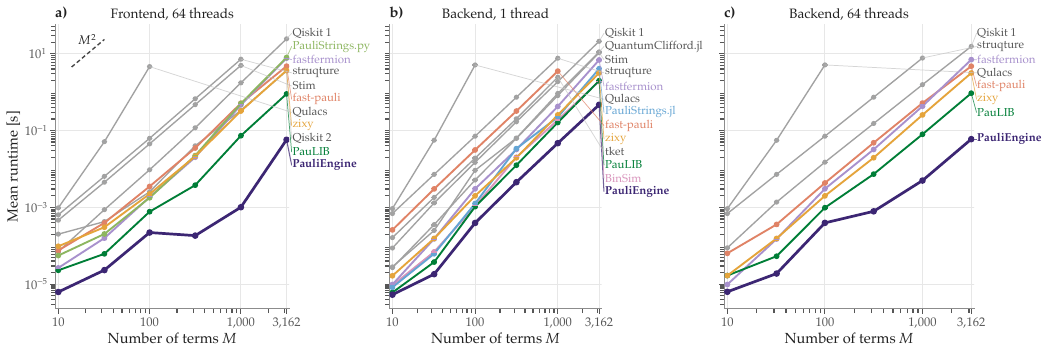}
    \caption{Runtime of products of dense Pauli sums, each containing $M$ randomly sampled Pauli strings on $n=64$ qubits, through (a) the Python interfaces with a thread budget of 64, (b) the native C++, Rust, and Julia interfaces with a single thread, and (c) the native interfaces with a thread budget of 64. Only the implementations benchmarked under the respective thread and interface configurations are shown. The dashed line indicates quadratic scaling with the number of terms $M$. Each data point represents the mean runtime across ten independent input pairs with ten timing repetitions each, with error bars indicating the standard error of the mean across the input instances. The experimental configurations and measurement procedures are detailed in~\cref{app:benchmark_methodology}.}
    \label{fig:dense-multiplication}
\end{figure*}

\Cref{fig:dense-multiplication} shows the dependence of multiplication runtimes on the number of terms in the input operators for three different interface and thread configurations. For dense operators, the number of generated product terms generally grows quadratically with $M$, and their accumulation constitutes a substantial part of the arithmetic. The adaptive output-prefix partitioning algorithm developed in~\cref{app:pauli_sum_product} reduces this cost by grouping contributions according to their output signatures before accumulation, thereby avoiding a global sorting step over all generated product terms.

The native single-threaded comparison in~\cref{fig:dense-multiplication}(b) shows that the performance advantage of \PE\ persists when the overhead associated with the Python interfaces is removed. Moreover, the independent accumulation of contributions within the output-prefix partitions allows us to distribute the arithmetic across multiple threads. For small operators, the overhead of parallel execution can exceed its computational benefit, and our implementation therefore uses a serial arithmetic routine below a prescribed workload threshold. The transition to parallel execution becomes visible in~\cref{fig:dense-multiplication}(a,c) as a reduced growth in runtime beyond approximately $M=100$.

The set of packages considered in the native and multithreaded comparisons differs from that of the single-threaded Python benchmarks, which reflects the availability of native interfaces and the thread configurations specified in~\cref{app:benchmark_methodology}. Among the compared implementations, PauLIB again achieves runtimes closest to those of \PE\ for the multiplication of dense Pauli sums.

\begin{figure*}[htbp]
    \centering
    \includegraphics[width=\linewidth]{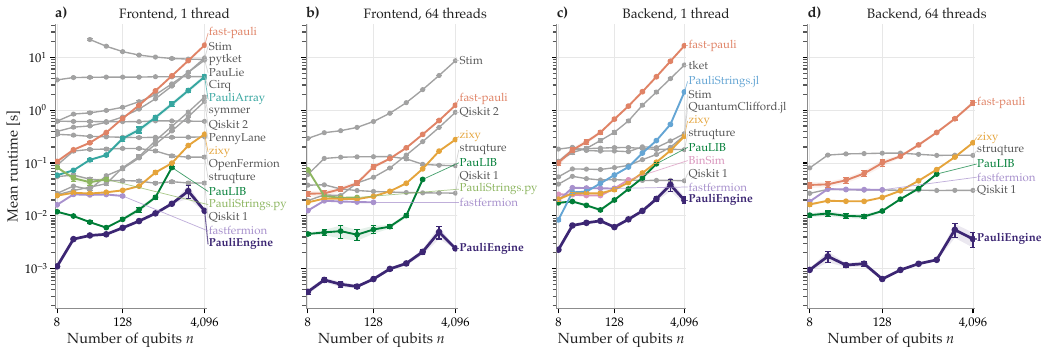}
    \caption{Runtime of products of Pauli sums, each containing $M=316$ randomly sampled Pauli strings of weight four, as a function of the register size $n$. We compare arithmetic through the Python interfaces with (a) a single thread and (b) a thread budget of 64, and through the native C++, Rust, and Julia interfaces with (c) a single thread and (d) a thread budget of 64. Only the implementations benchmarked under the respective thread and interface configurations are shown. Each data point represents the mean runtime across ten independent input pairs with ten timing repetitions each, with error bars indicating the standard error of the mean across the input instances. The experimental configurations and measurement procedures are detailed in~\cref{app:benchmark_methodology}.}
    \label{fig:lowweight-multiplication}
\end{figure*}

For operators composed of Pauli strings of fixed weight, the number of occupied sites per string remains independent of the register size, allowing us to reduce the computational cost of elementary products through the sparse site representation developed in~\cref{sssec:sparse_pauli_arithmetic}. The corresponding multiplication benchmarks are shown in~\cref{fig:lowweight-multiplication}. Implementations employing sparse representations of individual Pauli strings, such as Qiskit's \texttt{SparseObservable}, exhibit a comparatively weak dependence on the register size, while multiplication runtimes generally increase for implementations employing packed binary signatures.

In \PE, the transition between the two storage formats becomes visible through the decrease in runtime between $n=2048$ and $n=4096$ across all considered interface and thread configurations. This behavior agrees with the adaptive storage selection described in~\cref{app:sparse_pauli_arithmetic}, which allows us to retain efficient bitwise arithmetic on smaller registers and exploit sparse site arithmetic on sufficiently large registers. Together with the parallel hash-partitioned accumulation of the generated products, this yields the lowest multiplication runtimes among the compared implementations at large register sizes, with particularly pronounced advantages under parallel execution.

\begin{figure*}[htbp]
    \centering
    \includegraphics[width=1\linewidth]{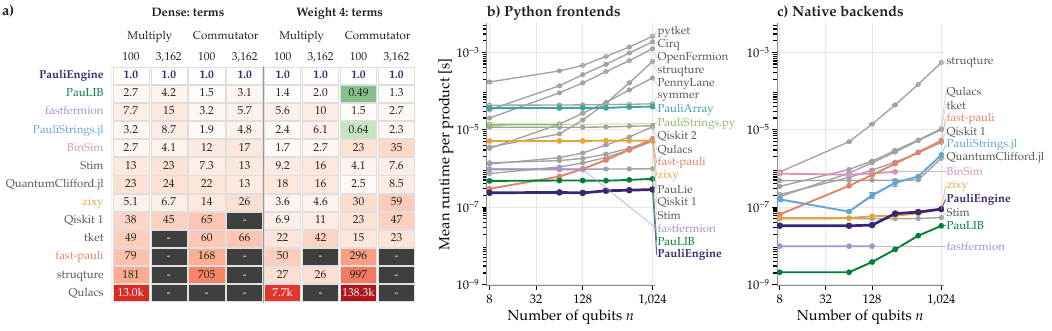}
    \caption{Comparison of native Pauli-sum arithmetic and elementary Pauli-string products. (a) Ratios of the single-threaded native runtimes of the respective implementations to those of \PE\ for products and commutators of dense and weight-four Pauli sums at $n=64$, evaluated at $M=100$ and $3162$ terms per operand. Values greater (smaller) than one indicate longer (shorter) runtimes than those of \PE, and dark-gray cells indicate configurations without a complete runtime measurement. The implementations are ordered by the geometric mean of their available runtime ratios. Panels (b) and (c) show the mean runtime per product of two randomly sampled Pauli strings with unit coefficients as a function of the register size $n$, through the Python and native interfaces, respectively, using a single thread. Elementary-string products are measured in batches according to the protocol described in~\cref{app:benchmark_methodology}.}
    \label{fig:native-factors-and-strings}
\end{figure*}

The relative runtimes of native product and commutator arithmetic in~\cref{fig:native-factors-and-strings}(a) exhibit a similar distribution to those obtained through the Python interfaces (see \cref{fig:frontend-factors}(a)). For commutators of weight-four operators with $M=100$, PauLIB and PauliStrings.jl achieve shorter runtimes than \PE, by factors of approximately two and 1.6, respectively. For larger operator sums, the advantages of our accumulation algorithms become more pronounced. These results show that the performance differences observed through the Python interfaces can largely be attributed to the underlying arithmetic implementations.

Finally, \cref{fig:native-factors-and-strings}(b,c) compares the runtimes of products of individual Pauli strings. In this case, the arithmetic requires only the computation of the output signature and its associated phase, without the subsequent accumulation of terms. For implementations based on packed binary signatures, the runtime therefore depends primarily on the number of machine words required to represent a string, as derived in~\cref{sec:operatorrepresentations}.

For the Python interfaces, \PE\ achieves the lowest runtimes across the considered register sizes, with a comparatively weak dependence on $n$. The native backend comparison shows a different ordering, with specialized implementations such as PauLIB and fastfermion achieving shorter runtimes than ours. At the level of elementary Pauli-string arithmetic, the native operations are sufficiently inexpensive that the overhead of invoking them through Python and constructing the corresponding result objects becomes a significant contribution to the measured runtime. For products of large operator sums, this overhead is incurred only once per arithmetic call, and the performance is instead governed primarily by the generation and accumulation of the individual Pauli contributions.

\subsection{Additional symmetry-adapted arithmetic benchmarks}
\label{app:symmetry_benchmarks}

\paragraph{Targeted commutator arithmetic.}

In applications where only individual coefficients of an operator expansion are required, the targeted arithmetic routines introduced in~\cref{sec:shared_operations} allow us to compute these coefficients without constructing the entire operator. For Pauli orbits, we showed in~\cref{app:pauli_orbit_product} that specifying the output label imposes additional constraints on the contingency tables contributing to the desired coefficient. In particular, when the Pauli weight of the target orbit is bounded, the number of admissible contingency tables remains bounded independently of the weights of the input operators and the register size.

To investigate the resulting computational advantage, we compare the direct evaluation of an individual commutator coefficient with the extraction of the same coefficient from the full commutator. For the dense and weight-two input orbits considered in~\cref{ssec:benchmark_symmetry}, we randomly select a target from the nonvanishing output components of the respective commutator. For identically vanishing commutators, we use $B_{1,0,0}$ as the target. We additionally construct pairs of dense input orbits whose labels differ by transferring one local Pauli factor between two Pauli types, such that their commutator contains a nonvanishing weight-one component. This allows us to investigate the targeted arithmetic for a fixed target weight while increasing the weights of the input operators.

\begin{figure}[htbp]
    \centering
    \includegraphics[width=\linewidth]{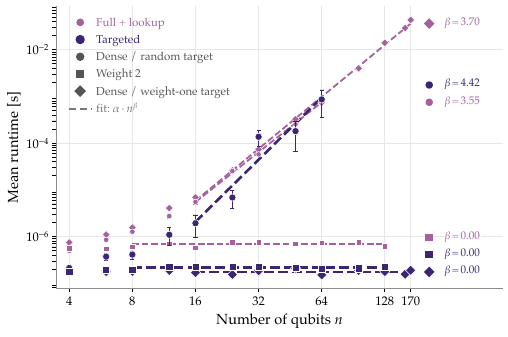}
    \caption{Runtime of evaluating individual Pauli-orbit commutator coefficients, either by computing the full commutator and subsequently extracting the desired coefficient (pink) or by directly computing the coefficient using the targeted arithmetic developed in~\cref{app:pauli_orbit_product} (purple). We consider dense input orbits with randomly selected targets (circles), input orbits of weight two (squares), and constructed pairs of dense input orbits with a weight-one target (diamonds). Dashed lines indicate power-law fits of the form $\alpha n^\beta$. Runtimes are measured using a single thread and ten calibrated batches of arithmetic calls for each of ten input pairs. Error bars indicate the standard error of the mean across the input pairs.}
    \label{fig:targeted-orbit-commutators}
\end{figure}

The resulting runtimes are shown in~\cref{fig:targeted-orbit-commutators}. For input orbits of weight two, both approaches exhibit essentially constant runtimes, with direct evaluation providing an additional reduction in the absolute runtime. For dense input orbits and randomly selected targets, the weights of the target orbits can themselves increase with the register size. In this case, the computational costs of both approaches grow polynomially, and their runtimes become comparable for larger registers.

The computational advantage of targeted arithmetic becomes particularly pronounced when the target weight remains fixed while the weights of the input orbits increase. While constructing the full commutator exhibits polynomial growth with a fitted exponent of $\beta=3.70$, the targeted evaluation of the weight-one coefficient exhibits constant runtime throughout the considered range of register sizes. At $n=170$, the direct evaluation of this coefficient is more than five orders of magnitude faster than constructing the entire commutator and extracting the coefficient.

\paragraph{Memory requirements.}

The runtime advantages of symmetry-adapted arithmetic are accompanied by substantial reductions in memory requirements. Following the measurement procedure described in~\cref{app:benchmark_methodology}, we distinguish the resident memory required to retain the input operators, the peak resident memory during arithmetic, and the resident memory with both the inputs and the resulting operator retained.

\begin{figure*}[htbp]
    \centering
    \includegraphics[width=\linewidth]{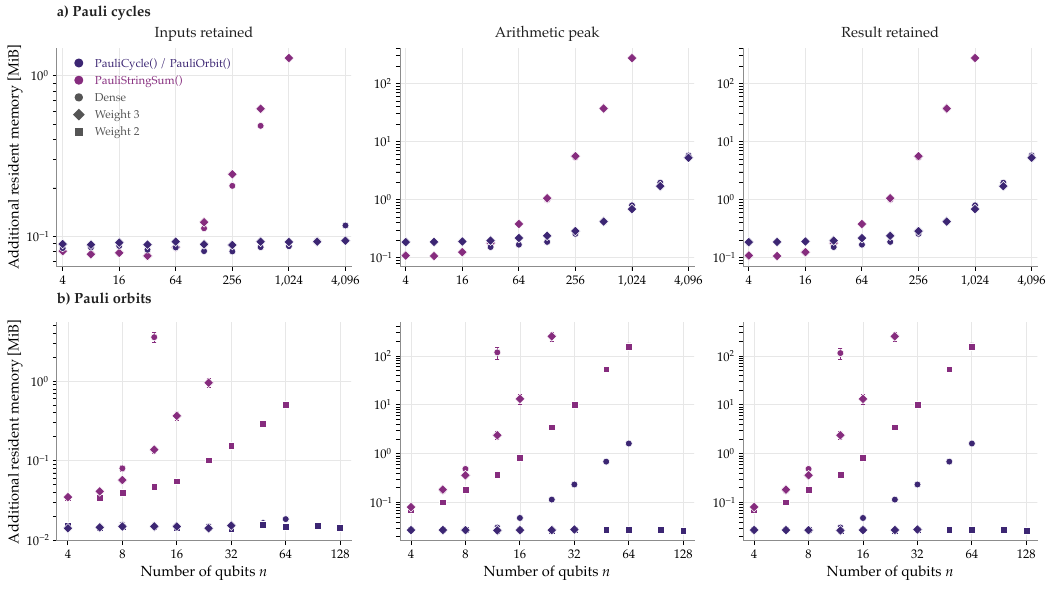}
    \caption{Memory requirements for multiplication in symmetry-adapted and ordinary Pauli bases. We compare the memory required for products of (a) Pauli cycles and (b) Pauli orbits with the corresponding products of their ordinary Pauli expansions. The input distributions correspond to those considered in~\cref{fig:cycles-commutators-t1,fig:orbits-commutators-t1}. The columns show the resident memory with both input operators retained before arithmetic, the peak resident memory during multiplication, and the resident memory with both the input and output operators retained after multiplication, respectively. All values are given relative to the baseline resident memory of the initialized process. Each data point represents the mean across ten independent process-level memory measurements, with error bars indicating the standard error of the mean.}
    \label{fig:combined-memory-multiplication}
\end{figure*}

\Cref{fig:combined-memory-multiplication} shows the memory requirements for products of Pauli cycles and orbits in the same input regimes considered in~\cref{ssec:benchmark_symmetry}. The memory required to store two elementary cycle or orbit operators exhibits only a weak dependence on the register size when using the symmetry-adapted representations. In contrast, constructing the corresponding ordinary Pauli expansions already requires storage of all distinct strings in the respective symmetry classes, resulting in a particularly pronounced increase in memory for dense Pauli orbits.

During multiplication, the difference between the representations becomes even more pronounced. For Pauli cycles, the product can contain a number of distinct cycles that increases with the register size, resulting in a corresponding increase in the memory requirements of the compressed output. The ordinary Pauli expansion of this output, however, requires storing the constituent Pauli strings of each cycle. For Pauli orbits, the number of possible output labels remains bounded when the weights of the input orbits are fixed, which is reflected in the nearly constant memory requirements during and after arithmetic. Even for dense orbit operators, the memory requirements of our symmetry-adapted arithmetic grow only moderately over the considered range of register sizes. Exploiting the symmetry of the input and output operators throughout the arithmetic thus allows us to manipulate operators whose ordinary Pauli expansions become prohibitively large, without constructing those expansions at any intermediate stage.

\section{Numerical demonstrations of applications}
\label{app:applications}

To complement the discussion in~\cref{sec:applications}, we investigate the performance of \PE\ within several of the computational workflows introduced there. In contrast to the arithmetic benchmarks of~\cref{sec:benchmarks}, we here measure the runtime of complete application-level calculations from prepared input operators. All measurements are performed on a single physical CPU core using the computational environment described in~\cref{app:benchmark_methodology}. Input preparation, conversion between software representations, and numerical validation are excluded from the reported runtimes. Depending on the runtime of an initial calibration call, we perform five, three, or one measured repetitions for each configuration. The complete experimental configurations, implementations, and validation results are provided in the accompanying benchmark repository.

\begin{figure*}[htbp]
    \centering
    \includegraphics[width=\textwidth]{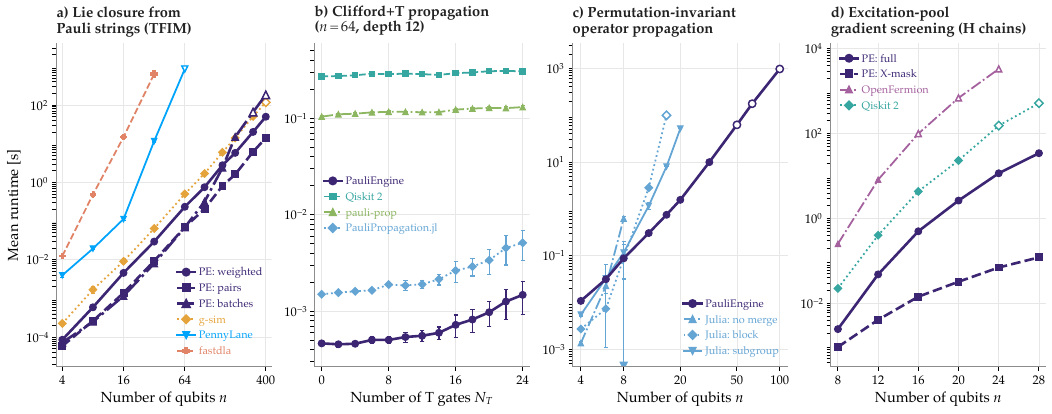}
    \caption{Runtime comparisons for four applications of quantum operator arithmetic. (a) Construction of the dynamical Lie algebra generated by the individual transverse-field and nearest-neighbor interaction terms of an open Ising chain, comparing weighted and coefficient-free arithmetic in \PE\ with existing Lie-closure implementations. (b) Exact Heisenberg propagation of a Pauli observable through 64-qubit circuits of depth 12 as a function of the number $N_T$ of non-Clifford rotations. (c) Permutation-invariant observable propagation under a product-formula evolution of the Lipkin--Meshkov--Glick model, comparing native Pauli-orbit arithmetic with ordinary Pauli propagation and different symmetry-merging strategies. (d) Evaluation of the complete excitation-pool gradient vector for linear hydrogen chains, comparing full commutator arithmetic with an implementation exploiting the computational-basis structure of the reference state (X-mask). All calculations use a single thread. Error bars indicate the standard deviation across repeated measurements in panels (a), (c), and (d), and the standard error of the mean across five circuit instances in panel (b). Hollow markers indicate configurations measured once, and the downward arrow in panel (c) indicates an error bar extending below the displayed logarithmic range.}
    \label{fig:application-overview}
\end{figure*}

\subsection{Construction of dynamical Lie algebras}
\label{app:applications_lie}

As a representative example of the Lie-algebraic preprocessing discussed in~\cref{ssec:applications_lie}, we consider the individual generators of the transverse-field Ising model (TFIM) on an open chain of $n$ qubits,
\begin{equation}
    \mathcal G
    =\{X_j\}_{j=1}^{n}
    \cup\{Z_jZ_{j+1}\}_{j=1}^{n-1},
    \label{eq:application_tfim_generators}
\end{equation}
which generate a dynamical Lie algebra of dimension $n(2n-1)$~\cite{Kokcu.2022-FixedDepthHamiltonian,Kazi.2025-AnalyzingQuantumApproximate}. Since each generator is an individual Pauli string, we can construct the entire Lie algebra by identifying the distinct Pauli labels generated through repeated commutation.

In~\cref{fig:application-overview}(a), we compare three implementations using \PE: a weighted implementation with hash-based identification of basis elements, a coefficient-free implementation evaluating individual commutators, and a coefficient-free implementation evaluating commutators between batches of newly generated basis elements and the generator set. We additionally consider the original Lie-closure implementation of $\gsim$~\cite{Barligea.2026-EnablingLieAlgebraicClassical} and the corresponding routines of PennyLane~\cite{Bergholm.2018-PennyLaneAutomaticDifferentiation} and fastdla~\cite{Iiyama.2025-FastNumericalGeneration}.

The coefficient-free implementation based on individual commutators achieves the shortest runtimes throughout the considered range of system sizes. In particular, it outperforms the batched implementation, illustrating that for this workload the direct generation and insertion of individual Pauli labels is preferable to constructing intermediate collections of commutator results. All three implementations using \PE\ successfully construct the complete Lie algebra for systems of up to 400 qubits, corresponding to 319,600 basis elements. The obtained bases are validated against the expected Lie-algebra dimension and, for smaller systems, independent span calculations.

\subsection{Observable propagation}
\label{app:applications_propagation}

\subsubsection{Exact Clifford and near-Clifford propagation}
\label{app:applications_clifford}

To investigate the performance of the Pauli-propagation algorithm described in~\cref{sssec:applications_clifford}, we generate circuits of depth 12 on 64 qubits. Each layer consists of randomly chosen single-qubit Pauli rotations of angle $\pm\pi/2$, followed by nearest-neighbor $ZZ$ rotations of angle $\pm\pi/2$ on alternating brickwork edges. Starting from the observable $Z_{33}$, we introduce non-Clifford rotations by halving the angles of $N_T$ randomly selected single-qubit rotations within its backward light cone. These rotations are equivalent to $T$ or $T^\dagger$ gates up to Clifford conjugation and global phases.

We propagate the observable according to~\cref{eq:applications_pauli_propagation} without truncating its Pauli expansion and evaluate the resulting operator with respect to the initial state $\ket{0}^{\otimes n}$. For \PE\ and Qiskit's \texttt{SparsePauliOp}~\cite{JavadiAbhari.2024-QuantumComputingQiskit}, we use the same propagation procedure with the respective operator-arithmetic backends, while for pauli-prop~\cite{IBM.2026-PauliProp} and PauliPropagation.jl~\cite{Rudolph.2025-PauliPropagationComputational} we employ their native propagation routines. For each $N_T\in\{0,2,\ldots,24\}$, we perform five timing repetitions for each of five circuit instances.

As shown in~\cref{fig:application-overview}(b), the implementation using \PE\ achieves the shortest runtimes across all considered non-Clifford gate counts. The increasing runtime with $N_T$ reflects the growth of the Pauli expansion through successive non-Clifford rotations. We verify agreement of the complete propagated Pauli expansions across the implementations, such that the observed runtime differences do not arise from different approximation or truncation strategies.

\subsubsection{Permutation-invariant operator propagation}
\label{app:applications_symmetric_propagation}

To illustrate the advantages of propagating observables directly in a symmetry-adapted representation, we consider a permutation-invariant Lipkin--Meshkov--Glick (LMG) model~\cite{Lipkin.1965-ValidityManybodyApproximation} with Hamiltonian components
\begin{equation}
    H_{XX}=\frac{1}{n}\sum_{i<j}X_iX_j,
    \qquad
    H_Z=\frac{1}{2}\sum_{i=1}^{n}Z_i.
    \label{eq:application_lmg_hamiltonian}
\end{equation}
We propagate the observable $\overline Z=n^{-1}\sum_i Z_i$ under a first-order product formula~\cite{Lloyd.1996-UniversalQuantumSimulators,Childs.2021-TheoryTrotterError},
\begin{equation}
    U(t)=
    \left(
        e^{-i\Delta t H_Z}
        e^{-i\Delta t H_{XX}}
    \right)^{N_{\mathrm{steps}}},
    \label{eq:application_lmg_evolution}
\end{equation}
with $\Delta t=0.1$ and $N_{\mathrm{steps}}=50$, and evaluate its expectation value with respect to $\ket{0}^{\otimes n}$.

Since both Hamiltonian components and the initial observable are permutation-invariant, we can perform the entire propagation in the Pauli-orbit basis. To this end, we evaluate the nested-commutator expansion of~\cref{eq:applications_heisenberg_series} for each Hamiltonian component using appropriately subdivided time steps and a total operator-norm remainder budget of $10^{-10}$, without truncating individual Pauli coefficients. We compare this approach with PauliPropagation.jl~\cite{Rudolph.2025-PauliPropagationComputational}, applying the same product-formula evolution through elementary Pauli rotations, either without symmetry merging or with the symmetry merging of Ref.~\cite{Teng.2025-LeveragingSymmetryMerging}, using the block- and subgroup-merging routines of its \texttt{symmetry} branch~\cite{SparqleSim.2026-PauliPropagationJl}. 

The advantage of retaining the permutation-invariant representation throughout the propagation becomes apparent as the register size increases [\cref{fig:application-overview}(c)]. While the different PauliPropagation.jl implementations reach system sizes of up to 8, 16, and 20 qubits, respectively, we can propagate the observable using \PE\ for systems of up to 100 qubits within the considered computational budget. We verify the resulting expectation values against an independent simulation in the permutation-symmetric spin-$n/2$ subspace, obtaining agreement within $10^{-9}$.

\subsection{Excitation-pool gradient screening}
\label{app:applications_variational}

As an example of the operator calculations arising in variational quantum chemistry, we investigate the excitation-pool gradient screening introduced in~\cref{ssec:applications_variational}. We consider linear hydrogen chains $\mathrm{H}_k$ with $k\in\{4,6,8,10,12,14\}$, an interatomic distance of $1\,\text{\AA}$, and the STO-3G basis~\cite{Hehre.1969-SelfconsistentMolecularorbitalMethods}. Using restricted Hartree--Fock orbitals computed with PySCF~\cite{Sun.2020-RecentDevelopmentsPySCF}, we construct the molecular Hamiltonian and a pool of spin-conserving occupied-to-virtual single and double excitations in Tequila~\cite{Kottmann.2021-TequilaPlatformRapid}, mapping the operators to $n=2k$ qubits through the Jordan--Wigner transformation. Across the considered systems, the number of Hamiltonian terms increases from 185 to 27,735, while the number of excitation operators increases from 26 to 3,381.

For each excitation generator $T_\mu$, we construct the anti-Hermitian generator $A_\mu=(T_\mu-T_\mu^\dagger)/2$ and evaluate the corresponding energy gradient from~\cref{eq:applications_variational_gradient} with respect to the Hartree--Fock reference state. We compare the full commutator arithmetic of \PE, OpenFermion~\cite{McClean.2020-OpenFermionElectronicStructure}, and Qiskit~\cite{JavadiAbhari.2024-QuantumComputingQiskit} with an additional implementation using \PE\ that exploits the computational-basis structure of the reference state.

Specifically, a Pauli string contributes to the expectation value with respect to a computational-basis state only if it consists exclusively of $I$ and $Z$ factors. Defining the bit-flip mask $\boldsymbol{\xi}(P)$ of a Pauli string $P$ to identify the sites occupied by $X$ or $Y$, the product of two strings $P$ and $Q$ can contribute only if
\begin{equation}
    \boldsymbol{\xi}(P)=\boldsymbol{\xi}(Q).
    \label{eq:application_gradient_mask}
\end{equation}
By indexing the Hamiltonian terms according to these masks, we can therefore restrict the commutator arithmetic to pairs of terms potentially contributing to the desired expectation values. Since this index can be reused across repeated screening calculations, we exclude its construction from the measured runtimes.

As shown in~\cref{fig:application-overview}(d), the full commutator implementation of \PE\ already yields a substantial runtime reduction compared with the other considered implementations. Exploiting the structure of the reference state provides a further improvement of more than two orders of magnitude for the largest molecular system. We verify agreement of the complete gradient vectors across the implementations within $10^{-10}$, including the identification of the excitation selected according to the gradients. This example illustrates how efficient general operator arithmetic can be combined with application-specific restrictions on the required output to further reduce the computational cost.

\end{document}